\documentclass[a4paper,11pt]{article}
\usepackage{jheppub} % for details on the use of the package, please
\usepackage[T1]{fontenc} % if needed

\usepackage{amsmath,amsfonts,amsbsy,amssymb,array,accents,dsfont}
\usepackage{enumerate,latexsym,graphicx}
\usepackage{xcolor,tikz}
\usepackage{tcolorbox}
\usepackage{mathtools}

\usepackage{subfigure}
\usepackage{stmaryrd}
\usepackage[compat=1.1.0]{tikz-feynman}

\newcommand{\beq}{\begin{equation}}
\newcommand{\eeq}{\end{equation}}
\newcommand{\bea}{\begin{eqnarray}}
\newcommand{\eea}{\end{eqnarray}}

\newcommand{\der}{\partial}

\newcommand{\nn}{\nonumber}
\newcommand{\<}{\langle}
\renewcommand{\>}{\rangle}
\renewcommand{\Re}{{\rm Re}}

\newcommand{\tdot}[1]{\dddot{#1}}
\newcommand{\fdot}[1]{\ddddot{#1}}

\usepackage{feynmp-auto}

\usepackage{braket}
\usepackage{tikz}
\usetikzlibrary{matrix,calc,positioning,decorations.markings,decorations.pathmorphing,decorations.pathreplacing}%tikzmark,
\usetikzlibrary{arrows,cd}
\usepackage{bbding}
\usetikzlibrary{positioning}
\tikzset{>=stealth}

\newcommand{\qqquad}{\;, \quad\qquad}  %more space, with a comma in between

\newcommand{\RR}{\mathbb{R}}

\newcommand{\Ff}{\mathcal{F}}

\newcommand{\R}{\ensuremath{\mathbb{R}}}

\newcommand{\w}{\omega}

\usepackage{color}

\usepackage[normalem]{ulem}  % \sout{old text} for strikeout

\definecolor{darkred}{rgb}{0.6,0,0}
\definecolor{darkblue}{rgb}{0,0,0.6}

\newcommand{\be}{\begin{equation}}
\newcommand{\ee}{\end{equation}}

\usepackage[bbgreekl]{mathbbol}
\usepackage{amsfonts}
\DeclareSymbolFontAlphabet{\mathbb}{AMSb} % This makes you use the "usual" mathbb font for upper case letters
\DeclareSymbolFontAlphabet{\mathbbl}{bbold} % Use this for mathbb lower case letters! You have two commands now and you do not overlap with the usual mathbb

\newcommand{\ww}{\boldsymbol{\w}}

\ExplSyntaxOn

\NewDocumentCommand{\pFq}{O{}mmmmm}
 {
  \group_begin:
  \keys_set:nn { hypergeometric } { #1 }
  \hypergeometric_print:nnnnn { #2 } { #3 } { #4 } { #5 } { #6 }
  \group_end:
 }
\NewDocumentCommand{\hypergeometricsetup}{m}
 {
  \keys_set:nn { hypergeometric } { #1 }
 }

\tl_new:N \l_hypergeometric_divider_tl
\tl_new:N \l_hypergeometric_left_tl
\tl_new:N \l_hypergeometric_right_tl

\keys_define:nn { hypergeometric }
 {
  symbol .tl_set:N = \l_hypergeometric_symbol_tl,
  symbol .initial:n = F,
  separator .tl_set:N = \l_hypergeometric_separator_tl,
  separator .initial:n = {},
  skip .tl_set:N = \l_hypergeometric_skip_tl,
  skip .initial:n = 8,
  divider .choice:,
  divider/semicolon .code:n = \tl_set:Nn \l_hypergeometric_divider_tl { \;; },
  divider/bar .code:n = \tl_set:Nn \l_hypergeometric_divider_tl { \;\middle|\; },
  divider .initial:n = semicolon,
  fences .choice:,
  fences/brack .code:n = 
   \tl_set:Nn \l_hypergeometric_left_tl {[}
   \tl_set:Nn \l_hypergeometric_right_tl {]},
  fences/parens .code:n = 
   \tl_set:Nn \l_hypergeometric_left_tl {(}
   \tl_set:Nn \l_hypergeometric_right_tl {)},
  fences .initial:n = brack,
 }

\cs_new_protected:Nn \hypergeometric_print:nnnnn
 {
  {} \sb {#1} \l_hypergeometric_symbol_tl \sb { #2 }
  \left\l_hypergeometric_left_tl
  \genfrac .. % no delimiters
           {0pt} % no line
           {} % default style
           { \__hypergeometric_process:n { #3 } } % numerator
           { \__hypergeometric_process:n { #4 } } % denominator
  \l_hypergeometric_divider_tl
  #5
  \right\l_hypergeometric_right_tl
 }

\cs_new_protected:Nn \__hypergeometric_process:n
 {
  \clist_use:nn { #1 }
   {
    {\l_hypergeometric_separator_tl}
    \mspace { \l_hypergeometric_skip_tl mu }
   }
 }

\ExplSyntaxOff

\title{\boldmath Spectral flow and the conformal block expansion for strings in  AdS$_3$  
}

\author[a%,b%d
]{Sergio Iguri}
\author[b,c]{, Nicolas Kovensky}
\author[a,d]{and Juli\'an H.~Toro}
\affiliation[a]{Instituto de Astronomía y Física del Espacio (IAFE), CONICET - Universidad de Buenos Aires, Facultad de Ciencias Exactas y Naturales,  Ciudad Universitaria, 1428 Buenos Aires, Argentina.}
\affiliation[b]{Institut de Physique Th\'eorique, Universit\'e Paris Saclay, CEA, CNRS, Orme des Merisiers, 91191 Gif-sur-Yvette CEDEX, France.}
\affiliation[c]{Instituto de F\'isica de La Plata - CONICET
Diagonal 113 e/ 63 y 64, 1900 - La Plata, Argentina.}
\affiliation[d]{Instituto de Investigaciones Matemáticas Luis A. Santaló (IMAS), CONICET - Universidad de Buenos Aires, Facultad de Ciencias Exactas y Naturales, Ciudad Universitaria, 1428 Buenos Aires, Argentina.}
\emailAdd{siguri@iafe.uba.ar}
\emailAdd{nicolas.kovensky@ipht.fr}
\emailAdd{jtoro@dm.uba.ar}

\abstract{ We present a detailed study of spectrally flowed four-point functions in the SL(2,$\R$) WZW model, focusing on their conformal block decomposition. Dei and Eberhardt conjectured a general formula relating these observables to their unflowed counterparts. Although the latter are not known in closed form, their conformal block expansion has been formally established. By combining this information with the integral transform that encodes the effect of spectral flow, we show how to describe a considerable number of $s$-channel exchanges, including cases with both flowed and unflowed intermediate states. For all such processes, we compute the normalization of the corresponding conformal blocks in terms of products of the recently derived flowed three-point functions with arbitrary spectral flow charges. Our results constitute a highly non-trivial consistency check, thus strongly supporting the aforementioned conjecture, and establishing its computational power. 
}

\newcommand{\of}[1]{\left(#1\right)}
\newcommand{\off}[1]{\left[#1\right]}

\begin{document} 
\hypergeometricsetup{
  separator={,},
  divider=bar,
}
\maketitle
\flushbottom

\section{Introduction}
\label{sec: intro}

String propagation in AdS$_3$ is one of the most studied realizations of the AdS/CFT correspondence, partly because it allows for a level of computational control that is perhaps unprecedented beyond the purely topological instances of the holographic duality in more than two bulk dimensions. Indeed, this applies to the  boundary theory, which is a two-dimensional CFT, believed to be closely related to a symmetric orbifold model,  but also to the worldsheet theory, which, in the pure NS-NS case, is given by a potentially solvable Wess-Zumino-Witten (WZW) model, thus giving access to the description of string dynamics  well beyond the supergravity limit. 

The bosonic sector of the worldsheet CFT is described by the WZW model based on the universal cover of SL(2,$\mathbb R$). Its distinctive feature is that, in addition to the {\em standard} representations, which are constructed as in any rational WZW model,  one must also incorporate their images under the so-called spectral flow automorphisms of the affine symmetry algebra. For instance, these are necessary to describe the classical configurations known as long strings in AdS$_3$.  Spectral flow thus plays a central role in generating a consistent physical spectrum in the SL(2,$\mathbb R$) WZW model \cite{Maldacena:2000hw}. In the supersymmetric case, it also allows for the construction of the tower of short string states forming the spacetime chiral ring \cite{Dabholkar:2007ey,Giribet:2007wp,Iguri:2022pbp,Iguri:2023khc}.

From the computational point of view, including vertex operators with non-zero spectral flow charges entails considerable technical difficulties for the calculation of the corresponding correlation functions \cite{Maldacena:2001km}. Correlators involving only unflowed states can be obtained simply by analytical extension from those of the Euclidean counterpart, that is, the H$_3^+$ coset model \cite{Teschner:1997ft,Teschner:1999ug}. This includes two- and three-point functions, which were computed exactly in analogy to the Liouville case \cite{Zamolodchikov:1995aa}, the OPE structure, and also four-point functions. The latter are not known in closed form, but they can be determined recursively from the leading order solution to the Knizhnik-Zamolodchikov (KZ) equation. This establishes their factorization properties at the primary level, by which we mean their description as a sum over the exchange of all intermediate states allowed by the fusion rules of the SL(2,$\R$) model. On the other hand, correlation functions with flowed insertions require much more intricate techniques. These originally involved an increase in the effective number of insertions due to the inclusion of the so-called spectral flow operators \cite{Fateev,Maldacena:2001km}, rendering the analysis substantially more challenging. 

At the level of the flowed four-point functions, which will be our main focus, this approach only led to concrete results for a handful of particular cases where all spectral flow charges are quite low. Let us stress that we are mostly interested in $x$-basis correlators, namely those written in the appropriate language for holographic considerations. This is because the complex variable $x$ parametrizes the asymptotic boundary of AdS$_3$. The only four-point functions which were analyzed in detail by using the techniques of \cite{Fateev,Maldacena:2001km} were the singly-flowed case  \cite{Minces:2005nb,Cagnacci:2015pka}, and, under some assumptions, the spectral flow conserving correlators \cite{Minces:2007td}. Some additional $m$-basis results were provided in \cite{Ribault:2005ms,Baron:2008qf,Iguri:2009cf,Giribet:2011xf}. Roughly speaking, the latter applies only to correlation functions for which all flowed insertions are located either at $x=0$ or at $x = \infty$. Determining the OPE structure for vertex operators belonging to the spectrally flowed sectors of the theory and the factorization properties of the corresponding four-point functions remains an important open problem. Solving it will likely provide crucial insights to understand the SL(2,$\R$) WZW model and, perhaps, even prove the AdS$_3$/CFT$_2$ holographic duality beyond the tensionless limit\footnote{The tensionless case was established in \cite{Eberhardt:2018ouy,Eberhardt:2019ywk,Eberhardt:2020bgq,Dei:2020zui,Dei:2023ivl}. A concrete proposal for the holographic CFT beyond the tensionless limit was put forward in \cite{Eberhardt:2021vsx,Dei:2022pkr} and further tested in \cite{Knighton:2023mhq,Knighton:2024qxd,Sriprachyakul:2024gyl} using free-field methods.}. 

Recently, an alternative technique based on the \textit{local} Ward identities was introduced in \cite{Eberhardt:2019ywk,Dei:2021xgh}. This allowed the authors to propose a closed formula for the three-point functions with arbitrary spectral flow charges. This was then proven in \cite{Iguri:2022eat,Bufalini:2022toj} by making use of the ramified holomorphic covering maps from the string worldsheet to the AdS$_3$ boundary. These allow one to bypass some of the main complications stemming from the OPEs between flowed vertex operators and the currents of the model. Covering maps are crucial in the tensionless limit, and remain quite useful beyond it. Moreover, a general integral expression relating four-point correlation functions in non-trivial spectral flow sectors to their unflowed cousins was conjectured in \cite{Dei:2021yom} by means of a comprehensive case-by-case analysis. This formula has not been proven in general, although it has been explicitly verified in several cases. It is thus natural to wonder if, and how, it can be used as a computational technique, for instance in order to study how these four-points functions factorize. %\SI{it is worth asking if this expression provides or not a new technique for computing correlators.} 

Let us stress that this question is a highly non-trivial one. On the one hand, the integral transform involved in defining the $x$-basis correlators from the perspective of \cite{Dei:2021xgh,Dei:2021yom} is quite challenging. It cannot be carried out in full generality even in the considerably simpler case of three-point functions. On the other hand, in the four-point case, the usefulness of the conjecture of \cite{Dei:2021yom} for the task at hand  relies on our knowledge of the corresponding \textit{unflowed} four-point functions. As discussed above, the only explicit expression for this kind of correlator is the leading order in the conformal block expansion in powers of the worldsheet cross-ratio obtained by Teschner in \cite{Teschner:1999ug}, together with the recursion relations that, in principle, define the following orders. Given that the expression conjectured in \cite{Dei:2021yom} mixes the spacetime and worldsheet cross-ratios with multiple integration variables in a complicated way, whether or not it is possible to recover a factorized expansion for the spectrally flowed four-point function at leading order, together with a consistent set of recursion relations for computing the subleading contributions to the conformal blocks, is yet to be understood.

In this paper, we address precisely these questions. Although the inherent complexity of four-point functions in the spectrally flowed sectors of the SL(2,$\R$) WZW model prevents us from carrying out the analysis in full generality, we work with arbitrary spectral flow charges for the external states and isolate a considerable number of non-trivial channels.  We explore channels involving both flowed and unflowed intermediate states, and prove that, in all cases, one obtains a consistent contribution to the factorized expansion of the full correlator. More precisely, we describe how the insertion of the factorized expression for the unflowed correlator in the integrand of the conjectured transform accounting for the effect of spectral flow leads to (1) the expected overall dependence on both the worldsheet and the boundary cross-ratios, (2) the leading order expression for the flowed conformal blocks, and (3) their appropriate normalization in terms of flowed structure constants. This involves the full complicated form of the latter observables, as derived in \cite{Dei:2021xgh,Bufalini:2022toj}. 
Our results constitute an involved consistency check based on the fundamental properties of the worldsheet CFT, which provides strong support for the conjecture of \cite{Dei:2021yom}, and further establishes its computational power. 

Regarding the derivation of the subleading contributions to the conformal blocks, we also expect the method based on the local Ward identities to allow for the determination of the corresponding recursions relations. We leave this calculation and that of the remaining factorization channels for future work.

%\smallskip

The paper is organized as follows. In Section \ref{Sec2} we review the SL(2,$\mathbb R$) WZW model, focusing on the role played by spectral flow in the construction of its spectrum, and its incidence while computing two- and three-point functions. We also review the known facts about the unflowed four-point functions \cite{Teschner:1999ug,Maldacena:2001km} and introduce the conjecture of \cite{Dei:2021yom} relating them to their spectrally flowed counterparts.

We then initiate our discussion of the factorization properties of flowed four-point functions in Section \ref{sec: unflowedcases} by analyzing in detail a number of sample cases, involving only vertex operators with either zero or small spectral flow charges.  The advantage of studying these specific situations is two-fold. First, it allows us to reproduce the results that were previously derived in 
\cite{Minces:2005nb,Minces:2007td,Cagnacci:2015pka,Giribet:2011xf} employing different methods, and compare their analysis with the one based on the local Ward identities. As we will show in the following sections, the latter is best suited for an extension to the general case. Moreover, the sample cases we consider here provide a simplified context for the detailed description of certain technical aspects of our approach, which will also play an important role in the rest of the paper. At the end of this section we further present a roadmap that will guide us through our study of the more general spectrally flowed four-point functions. 

In Section \ref{Sec4} we move past specific cases to study the factorization of four-point correlators with generic spectral flow assignments, first focusing on those admitting a well-defined $m$-basis limit. These are the so-called edge cases, in the language of \cite{Dei:2021xgh,Dei:2021yom,Bufalini:2022toj}, because the spectral flow charges nearly saturate the range allowed by the selection rules of the model, derived originally in \cite{Maldacena:2001km}. For the maximally spectral flow violating case, we also extend our analysis to the $x$-basis correlator, for which the integral transform associated with spectral flow can be carried out explicitly.   

Section \ref{Sec5} contains the main results of this paper. We discuss $x$-basis four-point functions with arbitrary spectral flow charges. We first show that the formula conjectured in \cite{Dei:2021yom} correctly accounts for the exchange of unflowed states along the $s$-channel, whenever they are allowed by the SL(2,$\R$) fusion rules. In other words, we show that using our knowledge of the unflowed conformal blocks, we can derive those associated with the exchange of such intermediate states between the different spectrally flowed vertex operators. We also obtain the corresponding normalizations, given by products of the relevant three-point functions, thus rendering such contributions consistent with the usual factorization ansatz. We then consider the exchange of flowed states and show that, for all channels isolated in the limit of small (worldsheet) cross-ratio, combining the formula of \cite{Dei:2021yom} with the results of \cite{Teschner:1999ug} for unflowed correlators leads to contributions to the  factorized four-point function that are again  consistent with the conformal symmetry constraints. 
Finally, in Section \ref{Sec7} we summarize our results, and discuss their impact for future investigations. Several useful identities are relegated to  Appendix \ref{sec: app A}.

%%%%%%%%%%%%%%%%%%%%%%%%%%%%%%%%%%%
\section{Review of the SL(2,$\R$) WZW model}
\label{Sec2}
%%%%%%%%%%%%%%%%%%%%%%%%%%%%%%%%%%%%
\subsection{Currents, spectrum and vertex operators} 
%%%%%%%%%%%%%%%%%%%%%%%%%%%%%

Let us start by introducing the basic ingredients of the SL$(2,\RR)$ WZW model at level $k>3$ \cite{Giveon:1998ns,Kutasov:1999xu,Maldacena:2000hw,Maldacena:2001km}. The symmetry algebra is characterized by the affine currents $J^a(z)$ with $a = 3,\pm$,  satisfying the following operator product expansions (OPEs), 
\begin{equation}
    J^a(z)J^b(w) \sim \frac{k\eta^{ab}/2}{(z-w)^2}+\frac{if^{ab}{}_{c}J^c(w)}{z-w}, \label{JJ OPE}
\end{equation}
where $\eta^{+-} =-2\eta^{33} =  2$, $f^{+-}_{\phantom{+-}3}=-2$ and $    f^{3+}_{\phantom{0+}+}=-
f^{3-}_{\phantom{0-}-}=1$.

Physical states come in three families. The long string states belong to the (flowed) continuous representations $C_{j}^{\alpha,\w}$ and have 
\begin{equation}
    j \, \in \, \frac{1}{2} + i \R 
    \,, \quad \alpha \in [0,1) 
    \,, \quad \w \in \mathbb{Z} \, .
\end{equation}
Here $j$ is the unflowed SL$(2,\RR)$ spin and $\w$ is the spectral flow charge. 
The short string states belong to the (flowed) discrete  representations ${\cal{D}}_j^{\pm,\w}$, which are of the highest/lowest-weight type and have 
\begin{equation}
    \frac{1}{2} < j < \frac{k-1}{2} \, ,
    \quad \w \in \mathbb{Z} \, . 
\end{equation}
Supergravity states belong to the unflowed discrete representations  ${\cal{D}}_j^{\pm} \equiv {\cal{D}}_j^{\pm,0}$. While such states can be understood by analytic continuation from the $H_3^+$ model studied in \cite{Teschner:1997ft,Teschner:1999ug}, the presence of spectrally flowed states is crucial for the consistency of the worldsheet spectrum. 

In the so-called $x$-basis, the vertex operators are denoted as $V_{jh}^{\w}(x,z)$. Here, $x$ and $z$ are the boundary and worldsheet complex coordinates,  respectively, while $h$ is the spacetime conformal weight. Incidentally, here we can restrict to $\w>0$ since states with negative $\w$ are encoded in the same vertex operators \cite{Maldacena:2001km}. Then, $h$ is defined by the $J_0^3$ eigenvalue $m$ before spectral flow, i.e.~$h=m + \frac{k}{2}\w$. For $\w=0$ we have $h = j$, hence we simply write $V_j(x,z)$. Here and in what follows we mostly omit the antiholomorphic variables, namely $\bar{x}, \bar{z}$ and $\bar{h}$. 

For strings propagating in an AdS$_3\times \mathcal M_{\rm int}$ geometry supported by pure NS-NS fluxes, the zero modes $J^a_0$ are identified with the global modes of the spacetime Virasoro algebra under the holographic duality, while the (AdS$_3$ part of the) worldsheet Virasoro algebra is obtained through the Sugawara construction, the corresponding central charge being $c=\frac{3k}{k-2}$. In contrast to the unflowed states, vertex operators $V_{jh}^{\w}(x,z)$ with $\w>0$ are not affine primaries, but they are Virasoro primaries of weight 
\begin{equation}
\label{def Delta w}
    \Delta_j = -\frac{j(j-1)}{k-2} - h\w + \frac{k}{4}\w^2\,.
\end{equation}
It should be noted that not all representations are actually independent. Indeed, in the continuous sector one has a reflection symmetry under $j \to 1-j$. For $\w=0$ this takes the form 
\begin{equation}
\label{reflection w=0}
V_{j}(x,z) = B(j)\int d^2x \, |x-x'|^{-4j} V_{1-j}(x',z) \, ,
\end{equation}
where the integral is performed over the full complex plane, and  
\begin{equation}
    B(j)=\frac{2j-1}{\pi}
    \frac{\Gamma[1-b^2(2j-1)]}{\Gamma[1+b^2(2j-1)]} \, \nu^{1-2j} \, , \quad  b^2 = (k-2)^{-1}
     \label{defBj} \,
\end{equation} 
with $\nu$ a free parameter of the theory which will be unimportant for us, 
while for $\w>0$ we have
\begin{equation}
\label{reflection w>0}
    V_{jh}^\w(x,z) = R(j,h,\w) V_{1-j,h}^\w(x,z) \, ,
\end{equation}
with 
\begin{equation}
\label{def N and R}
         R(j,h,\w) = \frac{ \pi \gamma \left(h-\frac{k}{2}\w+j\right) B(j)}{\gamma(2j) \gamma\left(h-\frac{k}{2}\w+1-j\right)} \, , \qquad 
        \gamma(x) = \frac{\Gamma(x)}{\Gamma(1-\bar{x})} \, .
\end{equation}
Moreover, in the discrete sector one has the so-called series identifications
\begin{equation}
\label{series-identif}
    V_{j,h=j + \frac{k}{2}\w}^\w(x,z) = {\cal{N}}(j) V_{\frac{k}{2}-j,h}^{\w+1}(x,z) \, , \qquad 
    {\cal{N}}(j) = \sqrt{\frac{B(j)}{B\left(
    \frac{k}{2}-j\right)}} \, .
\end{equation}

A set of linear combinations of flowed vertex operators which are particularly useful for the study of correlation functions is given by the so-called $y$-basis operators, related to the ones introduced above by the Mellin-type transform\footnote{Strictly speaking $\bar{h}$ is not necessarily the complex conjugate of $h$. In a slight abuse of notation, we will keep writing absolute values squared throughout the paper. } \cite{Dei:2021xgh,Iguri:2022eat}
\begin{equation}
    V_{jh}^{\w}(x,z) = 
    \int d^2y \,  \big|y^{j+\frac{k}{2}\w-h-1}\big|^2  
    V_j^\w (x,y,z)\,.
    \label{xtoybasis}
\end{equation}
In the $y$-basis, the defining OPEs are as follows. For the $J^+(z)$ current we have 
\begin{equation}
\label{J+Vw OPE}
   J^+(w)  V^{\w}_{j}(x,y,z)  =  \frac{\der_y 
    V_{j}^\w (x,y,z) }{(w-z)^{\w+1}} + 
   \sum_{n=2}^{\w} \frac{\left(J_{n-1}^+ 
    V_{j}^\w\right) (x,y,z) }{(w-z)^n}
    +\frac{\der_x 
    V_{j}^\w (x,y,z)  
      }{(w-z)}  + \cdots \, , 
\end{equation}
which includes a series of intermediate poles whose residues have no simple expression in terms of differential operators acting on the primaries (except for the $\w=0$ OPEs, obtained by taking $y \to x$). Their presence complicates the computation of correlation functions considerably. On the other hand, for 
\begin{equation}
    J^3(x,z) = J^3(z) - x J^+(z) \, , \qquad 
    J^-(x,z) = J^-(z) - 2 x J^3(z) + x^2 J^+(z) \, , 
\end{equation}
we have
\begin{equation}
\label{J3Vw OPE}
   J^3(x,w)  V^{\w}_{j}(x,y,z)  =  
   \frac{\left(y \der_y + j + \frac{k}{2}\w\right)
    V_{j}^\w (x,y,z)  
      }{(w-z)}  + \cdots \, , 
\end{equation}
and 
\begin{equation}
\label{J-Vw OPE}
   J^-(x,w)  V^{\w}_{j}(x,y,z)  =  (w-z)^{\w-1} \left(y^2\der_y + 2 j y\right)
    V_{j}^\w (x,y,z) + \cdots \, .  
\end{equation}
The vanishing of the first $\w-1$ terms in the expansion of the product $J^-(x,w)  V^{\w}_{j}(x,y,z)$ around $z=w$ imposes strong constraints on the correlation functions, known as \textit{local} Ward identities \cite{Eberhardt:2019ywk}. For $y$-basis operators the reflection symmetry reads 
\begin{equation}
\label{reflection ybasis}
V^\w_{j}(x,y,z) = B(j)\int d^2y' \, |y-y'|^{-4j} V^\w_{1-j}(x,y',z) \, .
\end{equation}

\subsection{Two- and three-point functions}
\label{sec: 2pt and 3pt functions}

The study of exact correlation functions in the bosonic SL$(2,\RR)$ model was initiated in \cite{Maldacena:2001km},  where the authors first derived the unflowed two- and three-point functions by analytic continuation from the H$_3^+$ results obtained in \cite{Teschner:1999ug}.  
For the flowed cases, important advances were achieved in recent years by studying the local Ward identities and making use of the $y$-basis representation \cite{Eberhardt:2019ywk,Dei:2021xgh,Dei:2021yom,Iguri:2022eat,Bufalini:2022toj,Iguri:2023khc}. 

As the auxiliary complex variable $y$ is somewhat unconventional from the CFT point of view, it will be useful to review the consequence of the global Ward identities for $y$-basis correlators \cite{Dei:2021xgh}. For two-point functions, one has 
\begin{equation}
    \langle V_{j_1 }^{\w_1}(x_1,y_1,z_1)V_{j_2 }^{\w_2}(x_2,y_2,z_2)\rangle =  
    \Bigg|
    \frac{x_{12}^{-h^0_1-h^0_2}}{
    z_{12}^{\Delta^0_1+\Delta^0_2}
    }
    \Bigg|^2
    \langle V_{j_1 }^{\w_1}\left(0,y_1 \frac{z_{12}^{\w_1}}{x_{12}},0\right)V_{j_2 }^{\w_2}\left(\infty,y_2 \frac{z_{12}^{\w_2}}{x_{12}},\infty\right)\rangle \, ,
\end{equation}
with 
\begin{equation}
    h_i^0 = j_i + \frac{k}{2} \w_i \, ,
    \qquad 
    \Delta_i^0 = -\frac{j_i(j_i-1)}{k-2}-j_i w_i - \frac{k}{2}\w_i^2  \, \label{h0andDelta0}, 
\end{equation}
and 
\begin{equation}
    V_j^\w(\infty,y,\infty) = \lim_{x,z \to \infty} |x|^{4h^0} |z|^{4\Delta_j^0} V_j^\w\left(x,y \frac{x^2}{z^{2\w}},z \right) \, .
\end{equation}
In this language, the two-point functions read \cite{Dei:2021xgh}
\begin{align}
    \langle V^{\w_1}_{j_1}(0,y_1,0)V^{\w_2}_{j_2}(\infty,y_1,\infty)\rangle = & \, \delta_{\w_1,\w_2}B(j_1)\delta(j_1-j_2) |1-y_1y_2|^{-4j_1}\\
    &+\delta_{\w_1,\w_2}\delta(j_1+j_2-1)|y_1|^{-2j_1}|y_2|^{-2j_2}\delta^{(2)}(1-y_1y_2) \, \, . \nn
\end{align}
By carrying out the integral transform in Eq.~\eqref{xtoybasis} one recovers the original $x$-basis results of \cite{Maldacena:2001km}, namely 
\begin{align}
    & 
    %\ddot{\mathcal{F}}^{j_1,j_2}_{\w}\equiv 
    \langle V^{\w_1}_{j_1 h_1}(0,0)V^{\w_2}_{j_2 h_2}(\infty,\infty)\rangle =  \delta_{\w_1,\w_2}\of{\delta(j_1-j_2)R(j_1,h_1,\w_1)
    +\delta(j_1+j_2-1)} \, .
\end{align}

%%%%%%%%%%%%%%%%%%%%%%%%%%%%%%%%
Three-point functions are more involved. The spectral flow charge is not a conserved quantity, although, as shown in \cite{Maldacena:2001km}, all non-zero $n$-point functions must satisfy 
\begin{equation}  
2\max(\w_i)\leq \sum_{i=1}^n \w_i + 1 \, .\label{Wselectionrules}
\end{equation} 
An exact expression for the $y$-basis three-point functions was first proposed in \cite{Dei:2021xgh} and later proven in \cite{Iguri:2022eat,Bufalini:2022toj}. The result depends on the parity of the total spectral flow charge $\w = \w_1+\w_2+\w_3$. Denoting 
\begin{equation}
   \tdot{\mathcal{F}}(y_i) \equiv  \left\langle V^{\w_1}_{j_1}(0, y_1, 0) \,  V^{\w_2}_{j_2}(1, y_2, 1) \, V^{\w_3}_{j_3}(\infty, y_3, \infty) \right\rangle \, , 
\end{equation}
one gets 
\begin{equation}
      \tdot{\mathcal{F}}(y_i) = \left\{
\begin{array}{cc}
     C_{-}(j_1,j_2,j_3) \Big| Z_{123}^{\frac{k}{2}-J} \prod_{i=1}^3 Z_i^{J-\frac{k}{2}-2j_i}\Big|^2 \,, & 
     \qquad {\rm for} \,\, \w \,\,  {\rm odd}\\[1ex]
     C_{+}(j_1,j_2,j_3) \Big| Z_{\emptyset}^{J-k} \prod_{i<\ell}^3 Z_{i\ell}^{J-2j_i-2j_\ell}\Big|^2 \,, & 
     \qquad {\rm for} \,\, \w \,\,  {\rm even}
\end{array}
      \right.
      \label{3pt-final}
\end{equation}
where $J=j_1+j_2+j_3$. 
Here, for any subset $I \subset \{ 1,2,3 \}$ we have introduced\footnote{We reserve the original notation  $(Z_I,Q_{\boldsymbol{\w}}) \to (X_I,P_{\boldsymbol{\w}})$ for the four-point functions, discussed below. }
\be 
Z_I(y_1,y_2,y_3)\equiv \sum_{i \in I:\ \varepsilon_i=\pm 1} Q_{\boldsymbol{\w}+\sum_{i \in I} \varepsilon_i e_{i}} \prod_{i\in I} y_i^{\frac{1-\varepsilon_i}{2}} \ . 
\label{Z_I-3pt}
\ee
with $\boldsymbol{\w}=(\w_1, \w_2, \w_3)$, $
e_1 = (1,0,0)$, $e_2 = (0,1,0)$, and $e_3 = (0,0,1)$. The $Z_I$ play the role of \textit{generalized differences}, and the numbers $Q_{\boldsymbol{\w}}$ are defined as  
\be 
Q_{\boldsymbol{\w}} = 0 \qquad \text{if} \qquad \sum_j \w_j < 2 \max_{i=1,2,3} \w_i \quad \text{or}\quad \sum_i \w_i \in 2\mathds{Z}+1 \, , 
\ee
while otherwise 
\be
Q_{\boldsymbol{\w}} =S_{\boldsymbol{\w}} \frac{G\left(\frac{-\w_1+\w_2+\w_3}{2} +1\right) G\left(\frac{\w_1-\w_2+\w_3}{2} +1\right) G\left(\frac{\w_1+\w_2-\w_3}{2} +1\right) G\left(\frac{\w_1+\w_2+\w_3}{2}+1\right)}{G(\w_1+1) G(\w_2+1) G(\w_3+1)}  \ .  
\label{Qw-definition}
\ee
with $G(n) = \prod_{i=1}^{n-1} \Gamma(i)$ the Barnes $G$ function and  $S_{\boldsymbol{\w}}$ a simple phase, see \cite{Dei:2021xgh}. Finally, the overall normalizations are set by the unflowed structure constants $C(j_1,j_2,j_3)$ as
\begin{equation}
\label{consdei}
    C_{+}(j_1,j_2,j_3) = 
    C(j_1,j_2,j_3) \,, \qquad 
    C_{-}(j_1,j_2,j_3) = 
    {\cal{N}}(j_3) C\left(j_1,j_2,\frac{k}{2}-j_3 \right) \, . 
\end{equation}  
Note that, although it is not obvious in this notation,  $C_{-}(j_1,j_2,j_3)$ is invariant under the exchanges $1 \leftrightarrow 3$ and $2 \leftrightarrow 3$, as it should. This follows from the explicit form of $C(j_1,j_2,j_3)$, which was analyed in detail in \cite{Teschner:1997ft,Teschner:1999ug,Maldacena:2001km}.
For more general insertion points one again makes use of the global Ward identities, which in this case imply
\begin{align}
\begin{aligned}
   &\langle V_{j_1 }^{\w_1}(x_1,y_1,z_1)V_{j_2 }^{\w_2}(x_2,y_2,z_2)V_{j_3 }^{\w_3}(x_3,y_3,z_3)\rangle =  \Bigg|\frac{x_{21}^{h^0_3-h^0_1-h^0_2}
    x_{31}^{h^0_2-h^0_1-h^0_3}
    x_{32}^{h^0_1-h^0_2-h^0_3}}{
    z_{21}^{\Delta^0_1+\Delta^0_2-\Delta^0_3}
    z_{31}^{\Delta^0_1+\Delta^0_3-\Delta^0_2}
    z_{32}^{\Delta^0_2+\Delta^0_3-\Delta^0_1}} \Bigg|^2
     \,  \\[1ex]
    &  \times     \left\< V_{j_1}^{\w_1} \left(0,y_1 \frac{x_{32} z_{21}^{\w_1}z_{31}^{\w_1}}{x_{21}x_{31}z_{32}^{\w_1}},0\right)V_{j_2}^{\w_2} \left(1,
    y_2 \frac{x_{31} z_{21}^{\w_2}z_{32}^{\w_2}}{x_{21}x_{32}z_{31}^{\w_2}}
    ,1\right)V_{j_3}^{\w_3}\left(\infty,
    y_3 \frac{x_{21} z_{31}^{\w_3}z_{32}^{\w_3}}{x_{31}x_{32}z_{21}^{\w_3}},\infty\right)\right\>.
\end{aligned}
\label{ybasisx1x2x3fixing}
\end{align}
It then follows from Eq.~\eqref{xtoybasis} that the $x$-basis three-point function is given by
\begin{align}
&\langle V_{j_1 h_1}^{\w_1}(x_1,z_1)V_{j_2h_2 }^{\w_2}(x_2,z_2)V_{j_3 h_3}^{\w_3}(x_3,z_3)\rangle  \\
   & \qquad \qquad = C_{\boldsymbol{\w}}(j_i,h_i) \, \Bigg|\frac{x_{21}^{h_3-h_1-h_2}
    x_{31}^{h_2-h_1-h_3}
    x_{32}^{h_1-h_2-h_3}}{z_{21}^{\Delta_1+\Delta_2-\Delta^{\w_3}_3}z_{31}^{\Delta^{\w_1}_1+\Delta^{\w_3}_3-\Delta^{\w_2}_2}z_{32}^{\Delta^{\w_2}_2+\Delta^{\w_3}_3-\Delta^{\w_1}_1}} \Bigg|^2 \, , \nn
\end{align}
with flowed structure constants 
\begin{equation}
\label{def Cw(ji,hi)}
    C_{\boldsymbol{\w}}(j_i,h_i)
\equiv \int \prod_{i=1}^{3}d^2y_i \, |y_i^{j_i-h_i+\frac{k}{2}\w_i-1}|^2 \tdot{\mathcal{F}}_{\boldsymbol{\w}}(y_i)    \, .
\end{equation} 

Finally, let us recall that something special happens in the so-called collision limit $x_2 \to x_1$ \cite{Dei:2021xgh,Bufalini:2022toj}. This corresponds to the limit \cite{Dei:2021xgh}
\begin{align}
\label{eq: Mobiusunfix}
    & \braket{
    V^{\w_1}_{j_1}(0,y_1,0)
    V^{\w_2}_{j_2}(x,y_2,1)
    V^{\w_3}_{j_3}(\infty,y_3,\infty)  } \nn\\
     & = \lim_{x\to 0} |x^{j_3-j_1 - j_2  + \frac{k}{2}(\w_3-\w_1-\w_2)}|^2
     \braket{
     V^{\w_1}_{j_1}\left(0,\frac{y_1}{x},0\right)
     V^{\w_2}_{j_2}\left(1,\frac{y_2}{x},1\right)
     V^{\w_3}_{j_3}\left(\infty,y_3 x,\infty \right)  } \, . 
\end{align}
Assuming for simplicity that $\w_3 \geq \w_{1,2}$, it turns out that only three-point functions satisfying $|\w_3 - \w_1 - \w_2| \leq 1$ remain non-vanishing. These $y$-basis correlators simplify considerably, giving a $y_i$-dependence of the form 
\begin{subequations}
\label{3ptcollision}
\begin{align}
& w_1+w_2-w_3=1:   \hspace{0.5cm} y_1^{-j_1+j_2+j_3-\frac{k}{2}}y_2^{j_1-j_2+j_3-\frac{k}{2}}\left(y_1+y_2+y_1y_2y_3\right)^{\frac{k}{2}-j_1-j_2-j_3}\ , \\[1ex]
& w_1+w_2-w_3=0: \hspace{0.5cm} (y_1-y_2)^{-j_1-j_2+j_3}(1+y_1y_3)^{-j_1+j_2-j_3}\left(1+y_2y_3\right)^{j_1-j_2-j_3}\ , \\
& w_1+w_2-w_3=-1: \hspace{0.5cm} y_3^{j_1+j_2-j_3-\frac{k}{2}}\left(1+y_1y_3+y_2y_3\right)^{\frac{k}{2}-j_1-j_2-j_3}\ .
\end{align}
\end{subequations}
Upon integrating over the $y_i$ as usual, one finds that these are the only three-point functions for which the charge conservation condition $h_3 = h_1 + h_2$ holds.

The details of the derivation leading to  Eqs.~\eqref{3pt-final} are beyond the scope of this brief review. However, let us provide some intuition for interpreting these formulas. As mentioned above, the local Ward identities derived from including additional current insertions in the flowed correlators and using the OPE \eqref{J-Vw OPE} give a series of linear constraints involving not only primary correlators and their $x$- and $y$-derivatives, but also descendant correlators. In principle, the total number of constraints allows to solve for these unknowns and, when the dust settles, one is left with three differential equations that must be satisfied by the Moebius-fixed $y$-basis primary correlators. In practice, however, this procedure quickly becomes rather cumbersome as the spectral flow charges are   increased. 

Luckily, at least for three-point functions there is a workaround to this problem. The motivation comes from the relation with the holographic CFT. It is known that, at $k=3$, the AdS$_3$ becomes string-size and the strings become tensionless \cite{Gaberdiel:2017oqg,Gaberdiel:2018rqv,Giribet:2018ada}. This leads to drastic simplifications of the worldsheet WZW model, which can be shown to be exactly dual to a symmetric orbifold CFT living on the boundary \cite{Eberhardt:2018ouy,Eberhardt:2019ywk,Eberhardt:2020bgq}. %, the corresponding seed theory being the sigma model on the internal manifold of the bulk configuration. 
In this setup the holomorphic covering maps routinely employed in the computation of two-dimensional  symmetric orbifold models \cite{Lunin:2000yv,Lunin:2000yv,Pakman:2009ab,Pakman:2009zz,Dei:2019iym} play a key role in computing the worldsheet correlators. It was shown in \cite{Eberhardt:2019ywk,Dei:2020zui,Dei:2023ivl} that the latter  localize on configurations where the worldsheet itself \textit{is} the appropriate covering surface of the AdS$_3$ boundary. Importantly, the spectral flow charge $\w$ of a given vertex operator is identified with the twist of its boundary avatar. 

Although the story is more complicated for $k>3$, where the dual theory is expected to be a suitable twist-two deformation of a somewhat different symmetric orbifold,
%, including a Liouville-type field associated with the bulk radial direction, 
\cite{Eberhardt:2021vsx,Dei:2022pkr} the covering maps remain very useful tools. For instance, the authors of \cite{Knighton:2023mhq,Knighton:2024qxd} were recently able to establish the holographic matching for a series of residues of $n$-point functions in the perturbative regime associated with poles in the space of complex (unflowed) spins by using the free-field description of the SL(2,$\R$) WZW model, valid at large radial distances. For the exact theory, which is the focus of the present work, the localisation property no longer holds. Nevertheless, the full solution for the case of $n=3$ presented above can be derived by combining the properties of certain covering maps with the symmetries of the model, including in particular the series identifications \eqref{series-identif} \cite{Dei:2021xgh,Iguri:2022eat,Bufalini:2022toj}. The relevant covering map data is encoded in the ratios of the $Q_{\boldsymbol{\w}}$ numbers defined in Eq.~\eqref{Qw-definition}. %The residues alluded to above then correspond to the locii \SI{??} in $y$-space where one or several of the $Z_I$ vanish (on top of those associated with poles of the unflowed structure constants). 

%%%%%%%%%%%%%%%%%%%%%%%%%%%%%

\subsection{Four-point functions and the KZ equation in the unflowed sector}
\label{sec: unflowed factorization}

The main focus of this paper is on four-point functions and their factorization properties. Let us first review some basic facts about the unflowed case, namely 
\begin{equation}
    \langle V_{j_1}(x_1,z_1)V_{j_2}(x_2,z_2)V_{j_3}(x_3,z_3)V_{j_4}(x_4,z_4)\rangle. \label{unflowed4pt}
\end{equation}
As usual, by using both worldsheet and spacetime conformal invariance one can fix the insertion points at $(z_1,z_2,z_3,z_4) = (0,1,\infty,z)$ and $(x_1,x_2,x_3,x_4) = (0,1,\infty,x)$ where
\begin{equation}
    z = \frac{z_{32}z_{14}}{z_{12}z_{34}}\qqquad x = \frac{x_{32}x_{14}}{x_{12}x_{34}} \, , 
\end{equation}
with $z_{ij}= z_i-z_j$ and $x_{ij}= x_i-x_j$. In other words, the four-point function \eqref{unflowed4pt} can be expressed as
\begin{align}
    &\langle V_{j_1}(x_1,z_1)V_{j_2}(x_2,z_2)V_{j_3}(x_3,z_3)V_{j_4}(x_4,z_4)\rangle= \nn\\
    & \qquad \qquad \Bigg|\frac{x_{12}^{-j_1-j_2+j_3-j_4}x_{13}^{-j_1+j_2-j_3+j_4}x_{23}^{j_1-j_2-j_3+j_4}x_{34}^{-2j_4}}{z_{12}^{\Delta_1+\Delta_2-\Delta_3+\Delta_4}z_{13}^{\Delta_1-\Delta_2+\Delta_3-\Delta_4}z_{23}^{-\Delta_1+\Delta_2+\Delta_3-\Delta_4}z_{34}^{2\Delta_4}} \Bigg|^2\fdot{\mathcal{F}}_{0}(x,z) \,,
\end{align}
where we have defined
\begin{equation}
    \fdot{\mathcal{F}}_{0}(x,z) \equiv \langle V_{j_1}(0,0)V_{j_2}(1,1)V_{j_3}(\infty,\infty)V_{j_4}(x,z)\rangle \, ,
\end{equation}
leaving its $j_i$-dependence implicit. 
The function $\fdot{\mathcal{F}}_{0}(x,z)$  must correspond to a (monodromy invariant) solution of the KZ  equation, which follows from the fact that 
\begin{equation}
    L_{-1} - \frac{1}{2(k-2)} \left(J^+_{-1} J^-_0 + J^-_{-1} J^+_0-
    2 J^3_{-1} J^3_0\right) 
\end{equation}
vanishes when acting on affine primaries. More precisely, this reads 
\begin{equation}
\label{KZ eq w=0}
    \left[\der_z - \frac{1}{k-2} \left(\frac{{\cal{P}}}{z}+\frac{{\cal{Q}}}{z-1}\right)\right] \fdot{\mathcal{F}}_{0}(x,z)=0 \, ,
\end{equation}
where 
\begin{subequations}
\label{def P and Q ops}
\begin{align}
 & \hspace{-0.1cm}   {\cal{P}} =
    x^2(x-1)\der_x^2 - \left[(\kappa-1)x + 2j_1  - 2 j_4 (x-1)\right]x \der_x - 2 j_4(j_1 + \kappa x) \, , 
    \\
& \hspace{-0.1cm}
    {\cal{Q}} = 
    -x(1-x)^2\der_x^2 + \left[(\kappa-1)(1-x) + 2j_2 + 2 j_4 x \right](1-x)\der_x - 2 j_4[j_2 + \kappa (1-x)] \, ,
\end{align}    
\end{subequations}
with $\kappa = j_3-j_1-j_2-j_3$. Beyond the usual singularities at $z=0,1,\infty$ and $x=0,1,\infty$, unflowed four-point functions have an additional one at  $z=x$ \cite{Maldacena:2001km}, 
\begin{equation}
    \fdot{\mathcal{F}}_{0}(x,z) \sim |x-z|^{2 (k-j_1-j_2-j_3-j_4)} \, .
\end{equation}
%This  occurs because at this point there exist a (genus zero) holomorphic map $\Gamma$ from the worldsheet to the AdS$_3$ boundary satisfying $\Gamma(z_i) = x_i$, namely, the identity map. \SI{Es estandar que se llame identity map?}

For unflowed vertex operators the OPE expansion can be derived by analytic continuation from that of the H$_3^+$ model \cite{Teschner:1999ug,Maldacena:2001km}. For states in the continuous representations (which are tachyonic for $\w=0$), one has 
\begin{equation}
\label{unflowed OPE}
V_{j_1}(x_1,z_1)V_{j_4}(x_4,z_4)  = \int_{\frac{1}{2} + i \R}dj \int_\mathbb{C} d^2x \frac{\big|z_{14}^{\Delta_j-\Delta_1-\Delta_4}\big|^2C(j_1,j_4,j) V_{1-j}(x,z_1)}{\big|x_{14}^{j_1+j_4-j}(x_4-x)^{j_4+j-j_1}(x-x_1)^{j+j_1-j_4}\big|^2} + \cdots\,, 
\end{equation}
up to descendant contributions weighted by larger powers of $z_{12}$. The appearance of the operator $V_{1-j}$ is related to the reflection symmetry \eqref{reflection w=0}. The analytic continuation of this OPE is required for exploring the short string sector, where 
$j_1$ or $j_4$ would take real values. Note that as $j_{12}$ is moved around in the complex plane \eqref{unflowed OPE} can pick up additional contributions coming from the poles of  $C(j_1,j_4,j)$ and the powers of $x_{ij}$ that may cross the integration contour. The main aspects of this procedure were discussed in \cite{Maldacena:2001km}.

By inserting the OPE  in Eq.~\eqref{unflowed4pt}, one obtains the decomposition of the unflowed four-point function in terms of conformal blocks. More explicitly, one gets 
\begin{equation}           
      \fdot{\mathcal{F}}_{0}(x,z) = \int_{\frac{1}{2}+i\RR}dj \, {\cal{C}}(j)|F_j(x,z)|^2. \label{unflowedblocks}
\end{equation}
where 
\begin{equation}
   {\cal{C}}(j)= \frac{C(j_1,j_4,j)C(j,j_2,j_3)}{B(j)} \,, 
\end{equation} 
and $F_j(x,z)$ has the $z$ expansion 
\begin{equation}
\label{unflowedzexp}
    F_j(x,z) = z^{\Delta_{j}-\Delta_{1}-\Delta_{4}} \sum_{n=0}^{\infty} f_{j,n}(x)z^{n} \, ,
\end{equation}
with  
\begin{equation}
f_{j,0}(x) = x^{j-j_1-j_4}\pFq{2}{1}{j-j_1+j_4,j-j_3+j_2}{2j}{x} \, .
    \label{unflowed 2F1}
\end{equation}
This is consistent with the fact that the theory has a Virasoro symmetry acting on the variable $x$ since this is the coordinate on the conformal boundary of AdS$_3$ where the holographic CFT is defined. From the worldsheet perspective, Eq.~\eqref{unflowed 2F1} can be derived directly by using \eqref{reflection w=0} to write $V_{1-j}(x,z)$ in terms of an integral involving $V_j(x')$ and using the identity 
\begin{align}
    & \int_{\frac{1}{2}+i \R}dj \, {\cal{C}}(j)\Bigg|\frac{\pi x^{j-j_1-j_4}}{2j-1} \pFq{2}{1}{j-j_1+j_4,j-j_3+j_2}{2j}{x}\Bigg|^2 =
    \label{2F1-integral-Teschner}\\
    & \int_{\frac{1}{2}+i \R}dj \, {\cal{C}}(j)\int d^2x' d^2x'' |x^{j-j_1-j_4}\of{x'-x''}^{2j-2}x'^{j_4-j_1-j}(x-x')^{j_1-j_4-j}(1-x'')^{j_3-j_2-j}|^2 \, . \nn
\end{align}
Of course, $f_{j,0}(x)$ is a leading order solution of the KZ equation \eqref{KZ eq w=0}. Although, in principle, higher orders $f_{j,n>0}(x)$ can be obtained recursively, no closed-form expression is known for $F_j(x,z)$. It is important to keep in mind that these formulae are only valid if the external spins satisfy 
\begin{equation}
{\rm Max}\left[|\Re(j_1-j_4)|,|\Re(1-j_1-j_4)|,|\Re(j_2-j_3)|,|\Re(1-j_2-j_3)|\right]<\frac{1}{2} \,\label{unflowedrange}.
\end{equation}
Moving away from this range implies that the correlator will pick up additional contributions coming from the exchange of unflowed discrete states and, as shown in  \cite{Maldacena:2001km}, in some cases also singly-flowed states.

It will also be necessary for our proposes to use the conformal block expansion of the unflowed four-point function in the limit $x,z\rightarrow 0$, with $x/z$ fixed. This can be extrapolated from \eqref{unflowedblocks} by using the identity \cite{Dei:2022pkr}
\begin{align}
    \fdot{\mathcal{F}}_{0}\of{x,z} =\mathcal{N}(j_1)\mathcal{N}(j_3) |x|^{-4j_4}|z|^{2j_4}\fdot{\mathcal{F}}_0\Big|_{\frac{k}{2}-j_1,\frac{k}{2}-j_3}\of{\frac{z}{x},z}\label{flipid} \, .
\end{align}
This implies that  
\begin{equation}
    \fdot{\mathcal{F}}_{0}(z x,z) = \int_{\frac{1}{2}+i\RR}dj \mathcal{N}(j_1)\mathcal{N}(j_3)\frac{C(\frac{k}{2}-j_1,j_4,j)C(j,j_2,\frac{k}{2}-j_3)}{B(j)}|G_j(x,z)|^2 \, , 
\end{equation}
where the function $G_j(x,z)$ can be expanded as
\begin{equation}
G_j(x,z) = z^{\Delta_{j}-\Delta_{1}-\Delta_{4}-j_4-j_1+\frac{k}{4}} x^{-2j_4}\sum_{n=0}^{\infty} g_{j,n}(x)z^{n} \, ,
\end{equation}
the leading order being
\begin{equation}
    G_j(x,z) = z^{\Delta_{j}-\Delta_{1}-\Delta_{4}-j_4-j_1+\frac{k}{4}}x^{\frac{k}{2}-j-j_1-j_4}\pFq{2}{1}{j-\frac{k}{2}+j_1+j_4,j-\frac{k}{2}+j_3+j_2}{2j}{\frac{1}{x}} \label{zxexpansion}\, .
\end{equation}
By virtue of Eq.~\eqref{flipid}, these expressions are valid as long as  
\begin{equation}
{\rm Max}\left[|\Re(j_1+j_4-\frac{k}{2})|,|\Re(1+j_{14}-\frac{k}{2})|,|\Re(j_2+j_3-\frac{k}{2})|,|\Re(1-j_{23}-\frac{k}{2})|\right]<\frac{1}{2} \, ,\label{flippedunflowedrange}
\end{equation}
where $j_{i\ell} = j_i - j_\ell$.

%%%%%%%%%%%%%%%%%%%%%%
\subsection{Conjecture for the flowed four-point functions}

Spectrally flowed four-point functions are notoriously complicated. Recently, an integral expression relating them with the corresponding unflowed correlators was proposed in \cite{Dei:2021yom}. One can divide this conjecture into two parts. First, using the global Ward identities, which in this case imply 
\begin{align}
&\langle V_{j_1}^{\w_1}(x_1,y_1,z_1)V_{j_2}^{\w_2}(x_2,y_2,z_2)V_{j_3}^{\w_3}(x_3,y_3,z_3)V_{j_4}^{\w_4}(x_4,y_4,z_4) \rangle \nonumber\\[1ex]
& = \Bigg|\frac{ x_{21}^{-h_1^0-h_2^0+h_3^0-h_4^0} x_{31}^{-h_1^0+h_2^0-h_3^0+h_4^0}  x_{32}^{h_1^0-h_2^0-h_3^0+h_4^0} x_{34}^{-2 h_4^0}}
{z_{21}^{-\Delta_1^0-\Delta_2^0+\Delta_3^0-\Delta_4^0} z_{31}^{-\Delta_1^0+\Delta_2^0-\Delta_3^0+\Delta_4^0}  z_{32}^{\Delta_1^0-\Delta_2^0-\Delta_3^0+\Delta_4^0} z_{34}^{-2 \Delta_4^0}} \Bigg|^2\nonumber\\
   &\qquad\!\times \Bigg \langle V_{j_1}^{\w_1} \left(0,\frac{y_1 \, x_{32} \, z_{21}^{\w_1} \, z_{31}^{\w_1} }{x_{21} \, x_{31} \, z_{32}^{\w_1}},0\right) V_{j_2}^{\w_2} \left(1,\frac{y_2 \, x_{31} \, z_{21}^{\w_2} \, z_{32}^{\w_2}}{x_{21} \, x_{32} \, z_{31}^{\w_2}},1\right) V_{j_3}^{\w_3}\left(\infty,\frac{y_3 \, x_{21} \, z_{31}^{\w_3} \, z_{32}^{\w_3}}{x_{31} \, x_{32} \, z_{21}^{\w_3}},\infty\right)\nonumber\\
   &\qquad\qquad\qquad\qquad V_{j_4}^{\w_4} \left(\frac{x_{32} \, x_{14}}{x_{12} \, x_{34}},\frac{y_4 \, x_{31} \, x_{32} \, z_{21}^{\w_4} \,  z_{34}^{2 \w_4}}{x_{21} \, x_{34}^2 \, z_{31}^{\w_4} \,  z_{32}^{\w_4}}, \frac{z_{32} \, z_{14}}{z_{12} \, z_{34}}\right)\Bigg\rangle\ ,\label{eq:global Ward identities solution 4pt function}
\end{align}
in order to focus on 
\begin{equation}
    \fdot{\mathcal{F}}_{\boldsymbol{\w}}(x,y_i,z) \equiv 
    \langle V_{j_1}^{\w_1}(0,y_1,0)V_{j_2}^{\w_2}(1,y_2,1)V_{j_3}^{\w_3}(\infty,y_3,\infty)V_{j_4}^{\w_4}(x,y_4,z) \rangle\,,  
\end{equation}
it was proposed that the $y$-basis differential equations for four-point functions with arbitrary spectral flow charges are solved by 
\begin{align}
    \fdot{\mathcal{F}}_{\boldsymbol{\w}}(x,y_i,z) = 
    %\int \prod_i^{4}d^2y_i \, & y_i^{j_i-h_i+\frac{k}{2}\w_i-1}  
    & 
     \big|X_\emptyset^{j_1+j_2+j_3+j_4-k}X_{12}^{-j_1-j_2+j_3-j_4} X_{13}^{-j_1+j_2-j_3+j_4}\nn\\[1ex]
    & \quad\times X_{23}^{j_1-j_2-j_3+j_4}X_{34}^{-2j_4}\big|^2\fdot{\mathcal{F}}_{+}\of{\frac{X_{14}X_{23}}{X_{12}X_{34}},z}\label{even4pt}
\end{align}
when the total spectral flow charge is even, and by 
\begin{align}
   \fdot{\mathcal{F}}_{\boldsymbol{\w}}(x,y_i,z)  =\mathcal{N}(j_3)
   %\int \prod_i^{4}d^2y_i \, & y_i^{j_i-h_i+\frac{k}{2}\w_i-1}  
   &\big| 
   X_{123}^{\frac{k}{2}-j_1-j_2-j_3-j_4} X_{1}^{-j_1+j_2+j_3+j_4-\frac{k}{2}}X_{2}^{j_1-j_2+j_3+j_4-\frac{k}{2}}\nn\\[1ex]
    &\quad \times X_{3}^{j_1+j_2-j_3+j_4-\frac{k}{2}}
    X_{4}^{-2j_4}\big|^2
    \fdot{\mathcal{F}}_{-}\of{\frac{X_{2}X_{134}}{X_{123}X_{4}},z}\label{odd4pt}
\end{align}
when it is odd. The corresponding $x$-basis correlators
\begin{equation}
    \fdot{\mathcal{F}}_{\boldsymbol{\w}}(x,z) \equiv \langle V^{\w_1}_{j_1h_1}(0,0)V^{\w_2}_{j_2h_2}(1,1)V^{\w_3}_{j_3h_3}(\infty,\infty)V^{\w_4}_{j_4h_4}(x,z)\rangle
\end{equation}are then obtained as 
\begin{equation}
\label{2.57}
    \fdot{\mathcal{F}}_{\boldsymbol{\w}}(x,z) = 
    \hspace{-0.15cm} \int \prod_{i=1}^{4}d^2y_i \,  \big|y_i^{j_i+\frac{k}{2}\w_i-h_i-1}\big|^2
    %\bar{y}_i^{j_i-\bar{h}_i+\frac{k}{2}\w_i-1}
    \fdot{\mathcal{F}}_{\boldsymbol{\w}}(x,y_i,z).
\end{equation}
In these expressions, the $X_I$ are the appropriate generalizations of the factors appearing in the three-point functions discussed above, namely the $Z_I$ defined in \eqref{Z_I-3pt}, while $ \fdot{\mathcal{F}}_{\pm}\of{X,z}$  are arbitrary functions of the worldsheet cross-ratio $z$ and of the relevant generalized cross-ratio, i.e.~either $X=\frac{X_{14}X_{23}}{X_{12}X_{34}}$ or $X= \frac{X_{2}X_{134}}{X_{123}X_{4}}$. The structure of this conjecture is thus similar in spirit to the three-point case. 
More precisely, the $X_I$ are still linear functions in the corresponding $y_i$, 
\begin{equation}
    X_{I} = z^{\frac{1}{2}\delta_{\{1,4\}\in I}}\off{(z-1)(-1)^{\w_1(\w_2+\w_3)+\w_4(\w_2+\w_3)}}^{\frac{1}{2}\delta_{\{2,4\}\in I}} 
    \hspace{-0.2cm}\sum_{i\in I: \epsilon_i = \pm}\hspace{-0.2cm}P_{\boldsymbol{\w}+\sum_{i\in I}\epsilon_i \hat{e}_i}\prod_{i\in I} y_i^{\frac{1-\epsilon_i}{2}} \,,
    \label{def XI 4pt}
\end{equation}
although now the coefficients $P_{\boldsymbol{\w}}=P_{\boldsymbol{\w}}(x,z)$ are  polynomials that depend on both $x$ and $z$. They are defined as 
\begin{align}
    P_{\boldsymbol{\w}} =& f(\boldsymbol{\w})\of{1-x}^{\frac{1}{2}s(\w_2+\w_4-\w_1-\w_3)}\of{1-z}^{\frac{1}{4}s((\w_1+\w_2-\w_3-\w_4)(\w_1+\w_4-\w_2-\w_3))-\frac{1}{2}\w_2\w_4}\nn\\
    &\times x^{\frac{1}{2}s(\w_1+\w_4-\w_2-\w_3)}z^{\frac{1}{4}s((\w_1+\w_2-\w_3-\w_4)(\w_2+\w_4-\w_1-\w_3))-\frac{1}{2}\w_1\w_4}\tilde{P}_{\boldsymbol{\w}}(x,z)
    \,, \label{pfunction} 
\end{align}
where $s(\w) = \w \Theta(\w)$, $\Theta$ being the Heaviside step function. On the other hand, the $\tilde{P}_{\boldsymbol{\w}}(x,z)$ come from the different covering maps relevant for a given four-point function, see \cite{Pakman:2009zz,Dei:2021yom}. A necessary condition for the existence of a rational function $\Gamma(z)$ with the appropriate branch points is 
\begin{equation}
    \sum_{i=1}^4 \w_i \, \in \, \mathbb{Z} \, , \quad \text{and}
    \quad \sum_{i=1}^4 \w_i > 2 \,{\rm max} \, \w_i \, .
\end{equation}
However, as opposed to the three-point case, this is not a sufficient condition. Indeed, once we further impose $\Gamma(z_i)=x_i$ for $i=1,2,3$ there is no freedom left, hence a covering map can only exist if a certain relation between the worldsheet and boundary insertion points holds, namely $\Gamma(z_4)=x_4$. The $\tilde{P}_{\boldsymbol{\w}}(x,z)$ are irreducible polynomials which vanish whenever this condition holds, namely $\tilde{P}_{\boldsymbol{\w}}(x,z) = \prod (z-\Gamma^{-1}(x))$, where the product runs over all possible covering maps\footnote{These are direct generalizations of the $Q_{\boldsymbol{\w}}$ functions appearing in the three-point functions. Although they are not known in closed form, an algorithmic procedure for computing them was given in \cite{Pakman:2009zz}.}. As it turns out, their degree in $z$ is given by the Hurwitz number 
\begin{equation}
    H_{\boldsymbol{\w}} = \frac{1}{2}\min_{i=1,2,3,4}\off{\w_i\of{\sum_{j=1}^4 \w_j - 2\w_i}}\, ,
\end{equation}
while their degree in $x$ is 
\begin{equation}
    \Lambda_{\boldsymbol{\w}} = \frac{1}{2}\off{\min(\w_1+\w_2,w_3+\w_4)-\max(|\w_1-\w_2|,|\w_3-\w_4|)} \,. \label{lambda}
\end{equation}
Moreover, they are normalized such that if $H_\w = 0$ then $\tilde{P}_{\boldsymbol{\w}}(x,z)= 1$, and if $H_\w < 0$ then $\tilde{P}_{\boldsymbol{\w}}(x,z) = 0 $. 

The second part of the conjecture concerns the nature of the functions $ \fdot{\mathcal{F}}_{\pm}\of{X,z}$. These are unconstrained by the local Ward identities. However, one still has to impose the KZ equation for the flowed correlator. The derivation of the corresponding differential equation is of course more involved than in the unflowed case because the vertex operators under consideration are not affine primaries. The authors of \cite{Dei:2021yom} argued that the structure of the prefactors in Eqs.~\eqref{even4pt} and \eqref{odd4pt} is such that, as a function of $z$ and of the generalized cross-ratio $X$, $\fdot{\mathcal{F}}_{+}$ satisfies precisely the KZ equation as the unflowed correlator with the same external spins, see \eqref{KZ eq w=0}. The same goes for $\fdot{\mathcal{F}}_{-}$ provided we flip $j_3 \to \frac{k}{2}-j_3$. It is therefore natural to identify these functions with the corresponding unflowed correlators, 
\begin{equation}
\label{def Fpm}
    \fdot{\mathcal{F}}_{+} = \fdot{\mathcal{F}}_{0} \,, \qquad 
    \fdot{\mathcal{F}}_{-} = \fdot{\mathcal{F}}_{0}|_{j_3 \to \frac{k}{2}-j_3} \, .
\end{equation}
Note that the unflowed structure constants play an analogous role in the context of flowed three-point functions, see Eqs.~\eqref{3pt-final} and \eqref{consdei}. In this way, the flowed four-point functions can be understood as integral transforms of the unflowed ones.

This proposal is based on a detailed case-by-case analysis of the local Ward identities, combined with a series of non-trivial consistency checks. More explicitly, it was shown that 1) it reduces to the known three-point functions in the $\w_4 = j_4 = 0$ limit, 2) it satisfies certain null vector constraints when they are present, 3) it is invariant under the pairwise exchange of the different insertions, 4) it is consistent with the reflection symmetry under $j_i \to 1-j_i$, and 5) it reproduces the relatively small number of previously known results \cite{Fateev,Maldacena:2001km,Ribault:2005ms,Giribet:2011xf,Minces:2005nb,Cagnacci:2015pka}. However, it remains to be proven. 

One of the main goals of this paper is to provide further support for this conjecture. We will do so by studying its factorization properties, i.e.~by considering several non-trivial exchange channels and showing that, in all cases, one obtains perfect agreement with the expected OPE decomposition for the SL(2,$\R$) WZW model, including spectrally flowed sectors.

%%%%%%%%%%%%%%%%%%%%%%%%%%%%%%%
%%%%%%%%%%%%%%%%%%%%%%%%%%%%%%%
\section{Sample cases for low spectral flow charges}
\label{sec: unflowedcases}

In this section we focus on a few low-lying cases in terms of the external spectral flow charges $\w_i$, with $i=1,2,3,4$, leaving a more general analysis for the following ones. These examples, namely the correlators with $\boldsymbol{\w}=(0,0,1,0)$, $\boldsymbol{\w}=(1,0,1,0)$, and $\boldsymbol{\w}=(0,0,2,0)$, are interesting for several reasons. First, the cases with a single flowed operator with either $\w_3=1$ or $\w_3=2$ are the relevant ones for the comparison with previous results from \cite{Minces:2005nb,Cagnacci:2015pka,Giribet:2011xf}, which were obtained by using different techniques. Second, in these cases, the expressions in Eqs.~\eqref{even4pt} and \eqref{odd4pt} can easily be established explicitly, as will be shown below, and they are simple enough to allow for a tractable and detailed analysis which will provide some useful intuition for the general case. 

\subsection{The $\boldsymbol{\w}=(0,0,1,0)$ case}

We consider the four-point function 
\begin{equation}
    \langle V_{j_1}(x_1,z_1)V_{j_2}(x_2,z_2)V_{j_3}^{1}(x_3,y_3,z_3)V_{j_4}(x_4,z_4) \rangle\,. 
\end{equation}
In this case, there is a single constraint equation \cite{Eberhardt:2019ywk}, namely that 
\begin{align}
\label{3.2}
&\langle J^-(x_3,z)V_{j_1}(x_1,z_1)V_{j_2}(x_2,z_2)V_{j_3}^{1}(x_3,y_3,z_3)V_{j_4}(x_4,z_4) \rangle \nn \\[1ex]
& = (y_3^2\der_{y_3} + 2j_3 y_3)
\langle V_{j_1}(x_1,z_1)V_{j_2}(x_2,z_2)V_{j_3}^{1}(x_3,y_3,z_3)V_{j_4}(x_4,z_4) \rangle + {\cal{O}}(z-z_3) \, .
\end{align}
By inserting the OPEs \eqref{J+Vw OPE}-\eqref{J-Vw OPE} in Eq.~\eqref{3.2}, and fixing $x_1=z_1=0$, $x_2=z_2=1$ and $x_3 = z_3 \to \infty$, this translates into the following differential equation for the Moebius-fixed $y$-basis correlator:
\begin{equation}
    \left\{y_3(y_3-1) \der_{y_3} + (x-z) \der_x +  j_1+j_2+j_3(2 y_3-1)+ j_4 - \frac{k}{2} \right\} \fdot{\mathcal{F}}_{(0010)}(x,y_3,z) = 0 \, .
\end{equation}
The general solution takes the form 
\begin{equation}
\label{0010 ybasis solution 1}
    \fdot{\mathcal{F}}_{(0010)}(x,y_3,z) =\big|y_3^{j_1+j_2 - j_3 +j_4 + \frac{k}{2}}(1-y_3)^{\frac{k}{2}-j_1-j_2-j_3-j_4}\big|^2 F(X,z) \, ,
\end{equation}
where
\begin{equation}
X = \frac{xy_3-z}{y_3-1}
\end{equation}
is the appropriate generalized cross-ratio for this particular case, and $F(X,z)$ is an arbitrary function. On the other hand, as  $\w_4=0$, the KZ equation for the vertex $V_{j_4}$ is derived analogously to the unflowed case. However, one picks up an additional contribution from the double pole in the  OPE of the current with $V_{j_3}^1$. Making use of \eqref{0010 ybasis solution 1}, one finds that the resulting constraint on $F(X,z)$ takes exactly the same form as in Eqs.~\eqref{KZ eq w=0} and \eqref{def P and Q ops}, with the replacements $x \to X$ and $j_3 \to \frac{k}{2}-j_3$. Transforming back to the $x$-basis, and using the series identifications to fix the overall normalization as in \cite{Iguri:2022eat,Bufalini:2022toj}, we conclude that 
\begin{align}
\label{0010}
    \fdot{\mathcal{F}}_{(0010)}(x,z) = \mathcal{N}(j_3)\int d^2y_3 \big| y_3^{-h_3+j_1+j_2+j_4-1}(1-y_3)^{\frac{k}{2}-j_1-j_2-j_3-j_4}\big|^2\fdot{\mathcal{F}}_-\of{X,z}\,,
\end{align}
$\fdot{\mathcal{F}}_-(X,z)$ being the unflowed four-point function with $j_3 \to \frac{k}{2}-j_3$ as in \eqref{def Fpm}. This is the expression conjectured in \cite{Dei:2021yom}.

\begin{figure}[h!]
    \centering
    \includegraphics[width=0.4\linewidth]{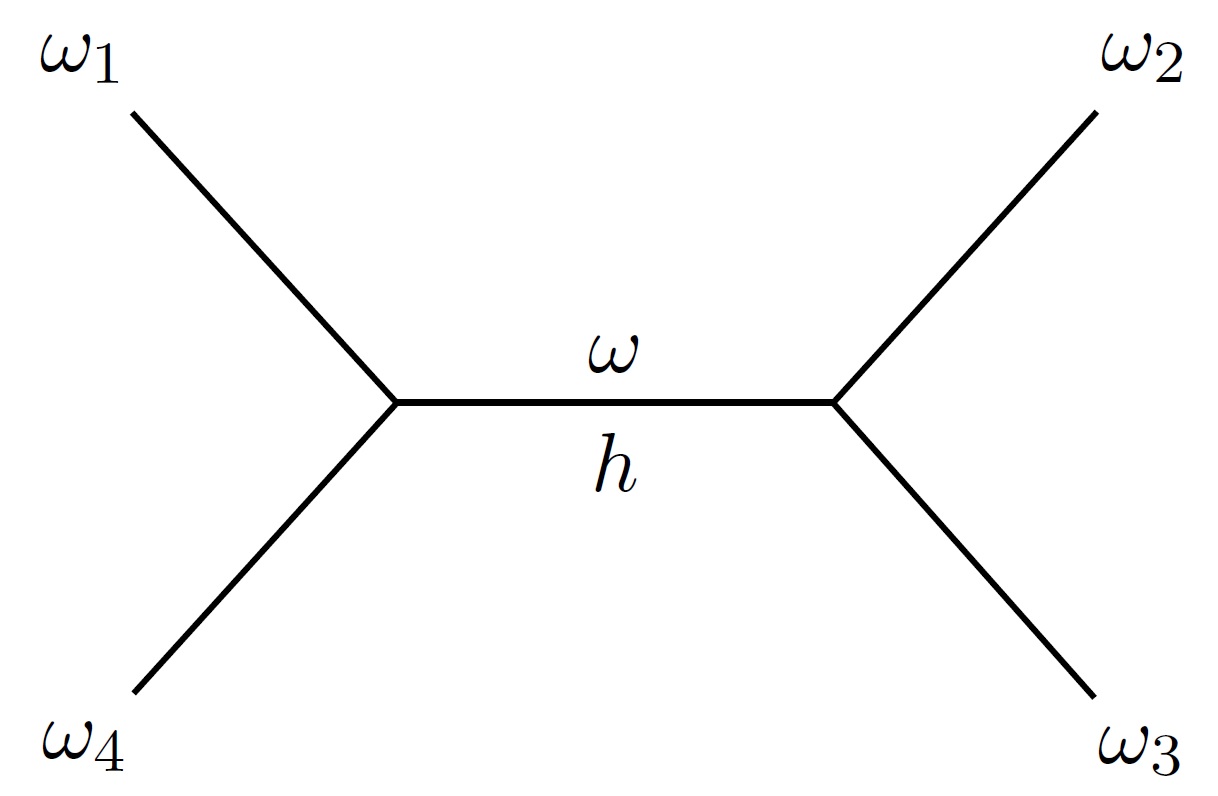}
    \caption{Schematic description of the exchange of states with spectral flow charge $\w$ and chage $h$ along the $14\to23$ channel.}
    \label{fig: Basic diagram}
\end{figure}

We now discuss the factorization properties of the correlator \eqref{0010}. 
The spectral flow selection rules for three-point functions in Eq.~\eqref{Wselectionrules} imply that this should involve two different intermediate channels for processes of the form $14\rightarrow 32$: the exchanged state can only have 
$\w=0$  or $\w=1$, see Fig.~\ref{fig: Basic diagram}. 
% \begin{figure}[h!]
%     \centering
%     \feynmandiagram [horizontal=a to b,large] {
%   i1 [particle=\(\w_1\)]--  a --  i2[particle=\(\w_4\)],
%   a -- [edge label=\(\w\),edge label'=\(h\)] b,
%   f1[particle=\(\w_2\)] --  b -- f2[particle=\(\w_3\)],
% };\caption{Schematic description of the exchange of states with spectral flow charge $\w$ and chage $h$ along the $14\to23$ channel.}
%     \label{fig: Basic diagram}
% \end{figure}
We expect to be able to study these processes by combining our knowledge of the factorized expansion of the \textit{unflowed} four-point function appearing in the integrand of \eqref{0010} with the integral transform that captures the effect of spectral flow. In other words, we aim for the conformal block decomposition of the spectrally flowed correlator by means of the corresponding decomposition of the unflowed one. 

Now, the scanning of exchange channels presents several subtleties. For instance, it strongly depends on the values of the external spins $j_i$. We shall bypass this problem by restricting our study to spin configurations reducing to Teschner strip as in the Euclidean model \cite{Teschner:1999ug} (up to the appropriate flips $j_i \to \frac{k}{2}-j$), for which a closed analytical expression for the four-point function is actually available. These domains only allow for the propagation of intermediate channels lying in the continuous series.  The emergence of  intermediate discrete states could be obtained after a proper analytical continuation \cite{Teschner:1999ug,Maldacena:2001km}, which is beyond the scope of this paper. Another factor that allows tuning new exchange states, in particular, channels in non-trivial spectral flow sectors, is related to the behaviour of the generalized  cross-ratio $X$ as $z$ approaches zero and, indirectly, with the interchange of the $y$-integrals and the $z\rightarrow 0$ limit while determining conformal blocks. Indeed, we shall see that different rescalings of $y_i$ depending on $x$ and $z$ while integrating \eqref{2.57} will lead us to recognize different intermediate channels, even when the external spins remain fixed.

Let us be more precise. When the external spins $j_i$ satisfy the constraint 
\begin{equation}
{\rm Max}\left[|\Re(j_1-j_4)|,|\Re(1-j_1-j_4)|,|\Re(j_2+j_3-\frac{k}{2})|,|\Re(1-j_2+j_3-\frac{k}{2})|\right]<\frac{1}{2} \,,  \label{0010j3range}
\end{equation}
the unflowed four-point function appearing in the integrand of \eqref{0010} admits a factorized expression of the form
\begin{equation}           
      \fdot{\mathcal{F}}_{-}\of{X,z} = \int_{\frac{1}{2}+i\RR}dj \mathcal{C}(j)F_j\of{X,z} \,, 
\end{equation}
where 
\begin{equation}
    \mathcal{C}(j) \equiv \frac{C(j_1,j_4,j)C\of{j,j_2,\frac{k}{2}-j_3}}{B(j)} \, .
\end{equation}
If we consider the limit $z \to 0$, while keeping $x$ and $y_3$ fixed, the generalized cross-ratio goes as  
\begin{equation}
    X \rightarrow \frac{x y_3}{y_3-1} \, .
\end{equation}
Eq.~\eqref{unflowedzexp} then shows that, at leading order in $z$, we can replace 
\begin{equation}
    F_j\of{X,z} \to z^{\Delta_j^{(0)}-\Delta_{1}-\Delta_{4}}\of{ \frac{x y_3}{y_3-1}}^{j-j_1-j_4} 
    \hspace{-0.5cm}
    \pFq{2}{1}{j-j_1+j_4,j+j_3+j_2-\frac{k}{2}}{2j}{\frac{x y_3}{y_3-1}}\,, 
\end{equation}
Here, we have introduced the notation $\Delta_j^{(0)}$ for the conformal dimension of the intermediate state at $\w=0$, which depends only on $j$. We obtain a contribution of the form 
\begin{align} 
    & \fdot{\mathcal{F}}_{(0010)}(x,z) \sim \mathcal{N}(j_3)\int d^2y_3 \int_{\frac{1}{2}+i\RR}dj \mathcal{C}(j) \Bigg| y_3^{-h_3+j_1+j_2+j_4-1}(1-y_3)^{\frac{k}{2}-j_1-j_2-j_3-j_4} \nn \\
    &\quad \times z^{\Delta_j^{(0)}-\Delta_{1}-\Delta_{4}}\of{\frac{xy_3}{y_3-1}}^{j-j_1-j_4}\pFq{2}{1}{j-j_1+j_4,j+j_3+j_2-\frac{k}{2}}{2j}{ \frac{xy_3}{y_3-1}} \Bigg|^2
    \label{00103F2Integral} \, , 
\end{align}
which indeed corresponds to the exchange of an unflowed state, as can be read off from the overall powers of $x$ and $z$. 

We would now like to compute the integral over $y_3$. The subtleties related to the exchange of the $j$ and $y_3$ integrals in \eqref{00103F2Integral} can be analyzed along the lines of  \cite{Maldacena:2001km}. Although this could lead to extra contributions associated to specific values of $j$, we expect these to be associated to either descendant contributions or two-particle states. Since we are interested in primary exchanges, we will ignore such contributions from now on. Using the complex integrals provided in Appendix \ref{sec: App A - Integrals}, we get   
\begin{align}
    \fdot{\mathcal{F}}_{(0010)}(x,z) &\sim 
    \int_{\frac{1}{2}+i\RR}dj \, \mathcal{C}(j) \frac{ \mathcal{N}(j_3) \gamma(h_3+j_3-\frac{k}{2})\gamma(j+j_2-h_3)}{\gamma(j+j_2+j_3-\frac{k}{2})}
 \\
    &\times \Bigg| z^{\Delta_j^{(0)}-\Delta_{1}-\Delta_{4}}x^{j-j_1-j_4}
    \pFq{2}{1}{j-j_1+j_4,j-h_3+j_2}{2j}{x}  \Bigg|^2\, . \nn
\end{align}
We find that the factors $\mathcal{C}(j)$ and $\mathcal{N}(j_3)$ combine with the $\gamma$-functions precisely in the right way to give the product of structure constants associated to the relevant flowed three-point functions, i.e.~\cite{Maldacena:2001km}
\begin{equation}
    \mathcal{N}(j_3) \mathcal{C}(j) \frac{\gamma(h_3+j_3-\frac{k}{2})\gamma(j+j_2-h_3)}{\gamma\of{j+j_2+j_3-\frac{k}{2}}} = \frac{C(j_1,j_4,j)C_{(001)}(j,j_2,j_3,h_3)}{B(j)}\, .
\end{equation}
%Moreover, we have obtained a hypergeometric function evaluated at the spacetime cross-ratio $x$ where, in comparison with the unflowed case, $j_3$ was replaced by $h_3$, see Eq.~\eqref{unflowed 2F1}. This had to be the case, as $h_3$ is the actual spacetime spin (or boundary holomorphic weight) for the corresponding flowed state. 
This shows that the structure of the factorized expression is analogous to that of the unflowed case. Indeed, on top of the structure constants we have obtained the \textit{flowed} conformal block at leading order in $z$, that is, the hypergeometric function appearing in \eqref{unflowedzexp} but with $j_3$ replaced by $h_3$, which is the spacetime spin (or boundary weight) of the flowed vertex. 

 This kind of flowed four-point function was studied previously in \cite{Minces:2005nb,Cagnacci:2015pka}. Their approach was quite different, as it was based on the spectral flow operator introduced in \cite{Fateev} and the subsequent definition of singly-flowed $x$-basis operators introduced in \cite{Maldacena:2001km}\footnote{This was recently generalized to arbitrary $\w$ in \cite{Iguri:2022eat}.}. Nevertheless, the authors obtained a KZ equation and recursion relations which are, of course, equivalent to those we have just derived in the $y$-basis language. Our result, including the region of validity given in Eq.~\eqref{0010j3range}, matches the combined results of \cite{Minces:2005nb,Cagnacci:2015pka}. 

%%%%%

Now, according to the spectral flow selection rules, the four-point function under consideration can also receive contributions from the exchange of intermediate states with $\w=1$. 
%%%%%%%%%%%%%%%%%%
In order to pick up a contribution of this type we turn back to Eq.~\eqref{0010} and perform the change of variables $
    y_3 \rightarrow z y_3/x$. This choice will be justified in Sec.~\eqref{sec: lessons} below. 
Keeping the new $y_3$ fixed,  the generalized cross-ratio now behaves at small $z$ as 
\begin{equation}
        X \rightarrow z (1-y_3) \, .
\end{equation}
As a consequence, the leading order expression of the unflowed correlator must be read off from  Eq.~\eqref{zxexpansion}. As long the external spins lie in the range
\begin{equation}
{\rm Max}\left[|\Re(j_1+j_4-\frac{k}{2})|,|\Re(1-j_4+j_1-\frac{k}{2})|,|\Re(j_2-j_3)|,|\Re(1-j_2-j_3)|\right]<\frac{1}{2} \label{0010j1range}\,,
\end{equation} 
this leads to  
\begin{align}
     & \fdot{\mathcal{F}}_{(0010)}(x,z) \sim \mathcal{N}(j_1) \int d^2y_3 \int_{\frac{1}{2}+i\R} dj \, \mathcal{C}(j) \Bigg|z^{\Delta_j^{(0)}-h_3+j_2+\frac{k}{4}-\Delta_{1}-\Delta_{4}}x^{h_3-j_2-j_1-j_4}
     \label{0010j1fact} \\
    &\qquad \times y_3^{-h_3+j_1+j_2+j_4-1}\of{1-y_3}^{\frac{k}{2}-j-j_1-j_4}\pFq{2}{1}{j_1+j_4+j-\frac{k}{2},j-j_3+j_2}{2j}{\frac{1}{1-y_3}} \Bigg|^2\nn,
\end{align}
now with
\begin{equation}
    \mathcal{C}(j) = \frac{C(\frac{k}{2}-j_1,j_4,j)C(j,j_2,j_3)}{B(j)} \, .
\end{equation}

We thus find that the $x$-dependence simplifies considerably as we only get an overall power. 
For this to describe a consistent contribution to the factorization expansion associated with a singly-flowed exchange, this should be of the form $x^{h-j_1-j_4}$, with $h$ the spacetime spin of the intermediate state. Hence, in this strict $z\to 0$ limit, we appear to only be capturing the process corresponding to $h = h_3-j_2$. This interpretation is further supported by the power of $z$ in \eqref{0010j1fact} since  
\begin{equation}
\Delta_{j}\Big|_{h=h_3-j_2}^{\w=1} = 
    \Delta_j^{(0)}-h_3+j_2+\frac{k}{4}-\Delta_{1}-\Delta_{4} \, ,
\end{equation}
as can be seen from Eq.~\eqref{def Delta w}. We will discuss why  this specific channel has been isolated later on, when analyzing the general case in Sec.~\eqref{sec: lessons}. For now, we focus on showing that, at least for this particular contribution, the four-point function factorizes correctly, in the sense that the relevant flowed structure constants emerge directly from \eqref{0010j1fact}.   

For this, we first note that the hypergeometric function appearing in \eqref{0010j1fact} must have a different origin from the one that appeared so far, since it is not evaluated at $x$ and is independent of the spacetime spins $h_i$. In order to show how it comes about, we proceed by reverse engineering. In analogy with the unflowed case, we expect to find certain factors corresponding to
\begin{equation}
\label{(001)(101)/(11)}
\frac{C_{(001)}(j_1,j_4,j,h)C_{(101)}(j,j_2,j_3,h,h_3)}{R(j,h,1)}\, , 
\end{equation}
where the denominator stands for the relevant $\w=1$ propagator in Eq.~\eqref{def N and R}.  
%The overall normalization is already present in \eqref{0010j1fact}. On the other hand, 
Since $h=h_3-j_2$,  we can combine the explicit integral expressions given in Sec. \ref{sec: 2pt and 3pt functions} for the flowed structure constants with the identity \eqref{3ptproductId} to get 
\begin{align}
\eqref{(001)(101)/(11)} &= \int d^2y \, d^2y_3 \,| y^{-2j} y_3^{j_3-h_3+\frac{k}{2}-1}|^2 \\
 &\quad \times\braket{V_{j_1}(0,0)V_{j_4}(1,1)V^1_{j,h_3-j_2}(\infty,y^{-1},\infty)}\braket{V^{1}_{1-j}(0,y,0)V_{j_2}(0,1)V^{1}_{j_3}(\infty,y_3,\infty)}\nn 
\\
& = \mathcal{N}(j_1)C\of{\frac{k}{2}-j_1,j_4,j}C(1-j,j_2,j_3) \nn \\
& \quad \times \int d^2y d^2y_3 \,| y_3^{j_3-h_3+\frac{k}{2}-1} (1-y)^{\frac{k}{2}-j_1-j_4-j}y^{j_3+j-j_2-1}(1-yy_3)^{j_2+j-j_3-1}|^2 \, . \nn
\end{align}
After the change of variable $y\rightarrow y_3^{-1}(1-y)$, the $y$-integral becomes of the hypergeometric type, giving 
\begin{align}
&\eqref{(001)(101)/(11)} = \mathcal{N}(j_1)C\of{\frac{k}{2}-j_1,j_4,j}\frac{\pi\gamma(j-j_3+j_2)\gamma(j+j_3-j_2)C(1-j,j_2,j_3)}{\gamma(2j)  } \\
    &\times \int d^2y_3 \, \Bigg|y_3^{j_1+j_4-h-1}(1-y_3)^{\frac{k}{2}-j-j_1-j_4}\pFq{2}{1}{j+j_1+j_4-\frac{k}{2},j-j_3+j_2}{2j}{\frac{1}{1-y_3}}
    \Bigg|^2,\nn
\end{align}
This exactly reproduces the contribution to the four-point function under consideration given in Eq.~\eqref{0010j1fact}, by means of the identity 
\begin{equation}
    \frac{\pi\gamma(j-j_3+j_2)\gamma(j+j_3-j_2)C(1-j,j_2,j_3)}{\gamma(2j)  } \label{Creflection}=
     \frac{C(j,j_2,j_3)}{B(j)}\,,
\end{equation}
which follows directly from the reflection symmetry  in the unflowed sector  \eqref{reflection w=0}.

\subsection{The $\boldsymbol{\w}=(1,0,1,0)$ case}

In this case the four point function is given by
\begin{align}
    \fdot{\mathcal{F}}_{(1010)}(x,z) &= z^{j_4}\int d^2y_1\, d^2y_3 \, | y_1^{\alpha_1-1}y_3^{\alpha_3-1}(1-y_1)^{-j_1-j_2+j_3-j_4}(1-y_3)^{j_1-j_2-j_3+j_4} \nn\\
    &\times(z-xy_3)^{-2j_4}(1-y_1y_3)^{-j_1+j_2-j_3+j_4}|^2\fdot{\mathcal{F}}_+\off{\frac{x-zy_1}{1-y_1}\frac{1-y_3}{z-xy_3},z}\label{1010},
\end{align}
where we have used the shorthand $\alpha_i = j_i+\frac{k}{2}\w_i-h_i$. The structure of the integrand follows from the two independent local Ward identities, which imply that the $y$-basis correlator $\fdot{\mathcal{F}}_{(1010)}(x,y_1,y_3,z)$ is annihilated by the differential operators 
\begin{equation}
y_3(y_3-1) \der_{y_3} + (y_1-1)\der_{y_1}+ (x-z) \der_x + j_1+j_2+j_3(2 y_3-1)+ j_4\, , 
\end{equation}
and 
\begin{equation}
y_1(y_1-1) \der_{y_1} + (y_3-1)\der_{y_3}+ \frac{x(x-z)}{z} \der_x + j_1 (2 y_1-1)+j_2+j_3+ j_4\left(\frac{2x}{z}-1\right)\, . 
\end{equation}
Combined with the KZ equation, this implies that $\fdot{\mathcal{F}}_{+}(X,z)$ should be identified with the corresponding unflowed correlator, as it satisfies the same differential equation in terms of the generalized cross-ratio 
\begin{equation}
    X = \frac{x-zy_1}{1-y_1}\frac{1-y_3}{z-xy_3} \, .
\end{equation}
%%%%%%
We now show that, as in the $\boldsymbol{\w}=(0,0,1,0)$ case,  considering different regimes for the external spins and taking the $z\to 0$ limit accordingly helps us isolate the different possibilities for the spectral flow charge of the intermediate state, which in this case can be either $\w=0$, $\w=1$ or $\w=2$. 

For the singly-flowed channel we simply take the small $z$ limit while keeping $y_{1,3}$ (and also $x$) fixed, hence $X \to \frac{y_3-1}{(1-y_1)y_3}$. Assuming that the external spins are in the range \eqref{unflowedrange} and inserting the leading order expression for the unflowed conformal block, we obtain 
\begin{align}
\label{F4 (1010) w=1}
   & \fdot{\mathcal{F}}_{(1010)}(x,z)  \sim \int d^2y_1d^2y_3 \int_{\frac{1}{2}+i\R} dj \mathcal{C}(j) \Bigg| z^{\Delta_{j}^{(0)}-\Delta_{1}^{(0)}-\Delta_{4}+j_4}x^{-2j_4}  y_3^{\alpha_3+j_1-j_4-j-1} (1-y_1)^{j_3-j_2-j} \nn \\
    &\times  y_1^{\alpha_1-1} (1-y_3)^{j-j_2-j_3}(1-y_1y_3)^{-j_1+j_2-j_3+j_4}\pFq{2}{1}{j-j_1+j_4,j-j_3+j_2}{2j}{\frac{y_3-1}{(1-y_1)y_3}} \Bigg|^2 \, ,
\end{align}
with $\mathcal{C}(j) = B(j)^{-1}C(j_1,j_4,j)C(j,j_2,j_3)$. We now find a power-law dependence in $x$ that is factorized from the $y_i$-integrals, and is compatible with the exchange of a flowed state with $h=h_1-j_4$. The power of $z$ supports this conclusion since, for $\w=1$, we have 
\begin{equation}
    \Delta_{j}^{(0)}-\Delta_{1}^{(0)}-\Delta_{4}+j_4 = \Delta_j^{(0)} - (h_1-j_4)+\frac{k}{4}-\Delta_{1}-\Delta_{4} = \Delta_j - \Delta_{1}-\Delta_{4} \, .
\end{equation}
The factorization along this particular channel is then consistent iff the remaining factors, including the $y_i$-integrals, give precisely the normalization 
\begin{equation}
\label{(101)(101)/(11)}
\frac{C_{(101)}(j_1,j_4,j,h_1,h)C_{(101)}(j,j_2,j_3,h,h_3)}{R(j,h,1)}\, . 
\end{equation}
Proceeding as in the previous section, we use the identities of Sec.~\ref{Useful properties} to rewrite this as 
\begin{align}
\eqref{(101)(101)/(11)} &= 
C(j_1,j_4,j)C(1-j,j_2,j_3)
    \int d^2y \, d^2y_1 \, d^2y_3 \, |y_1^{\alpha_1-1} y_3^{\alpha_3-1} \\
    &\times (1-y_3)^{1-j-j_2-j_3}  (y-y_1)^{j_4-j-j_1}(1-y)^{j+j_3-j_2-1}(1-y_3y)^{j_2+j-j_3-1}|^2 \nn \, , 
\end{align}
with $\alpha_i = j_i + \frac{k}{2}-h_i$. The integral in the auxiliary variable $y$ can be performed as in the previous section. It is simplified by changing variables  $y\rightarrow y_3^{-1}[1-(1-y_3)y]$, and the result is proportional to a hypergeometric function evaluated at $\frac{y_3-1}{(1-y_1)y_3}$. After using again \eqref{Creflection}, we recover precisely the expression on the RHS of Eq.~\eqref{F4 (1010) w=1}, as expected.

As it turns out, in this case yet another contribution corresponding to the exchange of a state with $\w=1$ can be obtained from \eqref{1010}. This can be seen by first taking $(y_1,y_3)\rightarrow (z^{-1}y_1,zy_3)$ and only then taking $z\to 0$ with the new $y_{1,3}$ fixed. A similar analysis shows that this corresponds to an intermediate spacetime spin of the form $h = h_3-j_2$.

It is also possible to isolate the unflowed channel by changing variables to $y_1 \rightarrow xy_1/z$. After doing so, the generalized cross-ratio behaves at small $z$ as 
\begin{equation}
    X\to \frac{z (1-y_1)(1-y_3)}{x y_1 y_3},
\end{equation}
and therefore we must use the expansion \eqref{zxexpansion} for the unflowed correlator. This leads to a contribution of the form 
\begin{align}
    \fdot{\mathcal{F}}_{(1010)}(x,z) & \sim \int d^2y_1 \, d^2y_3 \int_{\frac{1}{2}+i\R} dj \, \mathcal{C}(j) \Bigg|x^{j-h_1-j_4}z^{\Delta_{j}-\Delta_1-\Delta_{4}}  \label{1010exp2} \\
    & \quad \times y_1^{j+j_4-h_1-1} y_3^{j+j_2-h_3-1}(1-y_1)^{\frac{k}{2}-j-j_1-j_4}(1-y_3)^{\frac{k}{2}-j-j_2-j_3} \nn \\
    & \quad \times \pFq{2}{1}{j_1+j_4+j-\frac{k}{2},j_3+j_2+j-\frac{k}{2}}{2j}{\frac{x y_1y_3}{(1-y_1)(1-y_3)}}\Bigg|^2 , \nn
\end{align}
where now 
\begin{equation}
    \mathcal{C}(j) = \frac{\mathcal{N}(j_1)C(\frac{k}{2}-j_1,j_4,j)\mathcal{N}(j_3)C(j,j_2,\frac{k}{2}-j_3)}{B(j)}\,.
\end{equation}
Taking $y_i \rightarrow y_i/(y_i+1)$ gives
an integral that can be carried out in analogy to what was done around  \eqref{00103F2Integral}, leading to
\begin{align}
    \fdot{\mathcal{F}}_{(1010)}(x,z) &\sim  \int_{\frac{1}{2}+i\R} dj \, \mathcal{C}(j)  \frac{\gamma(j+j_2-h_3)\gamma(j_3+h_3-\frac{k}{2})}{\gamma(j_3+j_2+j-\frac{k}{2})}
    \frac{\gamma(j+j_4-h_1)\gamma(j_1+h_1-\frac{k}{2})}{\gamma(j_1+j_4+j-\frac{k}{2})} \nn \\
    & \quad \times \Bigg|  x^{j-h_1-j_4}z^{\Delta_{j}-\Delta_1-\Delta_{4}}
    \pFq{2}{1}{j-h_1+j_4,j-h_3+j_2}{2j}{x} \Bigg|^2
    \, .
\end{align}
The expression \eqref{1010exp2} is valid as long as the external spins are in the strip
\begin{equation}
{\rm Max}\left[|\Re(j_1+j_4-\frac{k}{2})|,|\Re(1+j_{14}-\frac{k}{2})|,|\Re(j_2+j_3-\frac{k}{2}),|\Re(1-j_{23}-\frac{k}{2})|\right]<\frac{1}{2} \label{1010j1j3range}\,.
\end{equation} 
Our result is thus consistent with the expected factorization structure for the process associated to the exchange of an unflowed state thanks to the identity 
\begin{align}
    & {\mathcal{C}}\of{j} \frac{\gamma(j+j_4-h_1)\gamma(j_1+h_1-\frac{k}{2})\gamma(j+j_2-h_3)\gamma(j_3+h_3-\frac{k}{2})}{\gamma(j_1+j_4+j-\frac{k}{2})\gamma(j_3+j_2+j-\frac{k}{2})}\nn \\
    & \qquad\qquad\qquad\qquad\qquad\qquad\qquad = \frac{C_{(100)}(j_1,j_4,j,h_1)C_{(001)}(j,j_2,j_3,h_3)}{B(j)} \, .
\end{align}

Finally, we note that the $\w=2$ channel sits at the edge of the selection rules for three-point functions, since $\w=\w_1+\w_4+1=\w_3+\w_2+1$. For reasons that will become clear below, we were only able to isolate this flowed channel in the spacetime collision limit $x\rightarrow 0$, which corresponds to taking $h=h_1+j_4$. Evaluating \eqref{1010} at $x=0$, rescaling $y_1 \rightarrow z y_1$, and using \eqref{zxexpansion}, we get a leading order behavior given by 
\begin{align}
    \fdot{\mathcal{F}}_{(1010)}(0,z) \sim&  \int_{\frac{1}{2}+i\R} dj \mathcal{N}(j_1)\mathcal{N}(j_3)\frac{C(\frac{k}{2}-j_1,j_4,j)C(j,j_2,\frac{k}{2}-j_3)}{B(j)} |z^{\Delta_j-\Delta_{1}-\Delta_{4}}|^2\\
    \times&\int d^2y_1 d^2y_3 \, \Bigg|y_1^{k-j-h_1-j_4-1}y_3^{\alpha_3-1}(1-y_3)^{\frac{k}{2}-j-j_2-j_3}\nn\\
    \times& \pFq{2}{1}{j_1+j_4+j-\frac{k}{2},j_3+j_2+j-\frac{k}{2}}{2j}{\frac{1}{y_1(y_3-1)}} \Bigg|^2\nn,
\end{align}
where the intermediate conformal weight is given by
\begin{align}
    \Delta_j = \Delta_j^{(0)}-2(h_1+j_4)+k.
\end{align}
The overall power of $z$ is compatible with a process where the intermediate channel has $\w=2$ and the aforementioned spacetime weight. For the decomposition to be consistent, the remaining factors should match 
\begin{equation}
\label{(102)(201)/(22)}
\frac{C_{(102)}(j_1,j_4,j,h_1,h)C_{(201)}(j,j_2,j_3,h,h_3)}{R(j,h,2)}\, , 
\end{equation}
which, thanks to the identity \eqref{3ptproductId}, can be written as 
\begin{align}
    &\mathcal{N}(j_1)\mathcal{N}(j_3)C\of{\frac{k}{2}-j_1,j_4,1-j}C\of{j,j_2,\frac{k}{2}-j_3}\int d^2y_1d^2y_3 d^2y \\
    & \times |y_1^{j_1-h_1+\frac{k}{2}-1}y_3^{j_3-h_3+\frac{k}{2}-1} y^{j_1+j_4+j-\frac{k}{2}-1}\of{1-y_1y}^{\frac{k}{2}-j_1-j_4+j-1}\of{1+y-y_3}^{\frac{k}{2}-j-j_2-j_3}|^2 \, . \nn
\end{align}
This is seen to be true by taking $y\rightarrow y/y_1$, performing the integral over the new $y$, and using 
\eqref{Creflection} once again. 

%%%%%%%%%%%%%%%%%%%%%%%%%%%
\subsection{The $\ww = (0,0,2,0)$ case}
\label{sec: (0020)}

For completeness we also consider the correlator with  $\ww = (0,0,2,0)$. This corresponds to the simplest configuration lying at the edge of the range allowed by the selection rules \eqref{Wselectionrules}. The four-point function is given by 
\begin{align}
    \fdot{\mathcal{F}}_{(0020)}(x,z)= f(z)\int d^2y_3 |y_3^{j_1+j_2+j_4-h_3-1}\off{(1-z)z+y_3(z-x)}^{k-j_1-j_2-j_3-j_4}|^2 \, .
    \label{(0020)}
\end{align}
Note that the derivation of this formula from the general expression in Eq.~\eqref{even4pt} is a bit subtle \cite{Dei:2021yom}. This will be discussed more generally in Sec.~\ref{sec: edge case} below. In particular, the definition of the function $f(z)$, which is  independent of $y_i$ and $x$, is provided in Eq.~\eqref{def f(z)}. 

For the particular case at hand, the local Ward identities give two constraints. Although in principle one of them would be used to solve for the unknown correlator corresponding to the insertion of $(J^+_1V_{j_3}^2)(x_3,z_3)$, this actually drops out. One finds that the correlator is annihilated by the differential operator  
\begin{equation}
   y_3 \der_{y_3} +  (z-x)\der_x + k - j_1 - j_2+j_3-j_4  \,, 
\end{equation}
hence the general solution is given by the integrand in Eq.~\eqref{(0020)} times an arbitrary function $F(y_3 (x-z),z)$. Since we also know that at large $y_3$ this must scale as $y_3^{-2j_3}$ \cite{Bufalini:2022toj}, we can fix $F(y_3 (x-z),z) = f(z)$. The KZ equation then fixes $f(z)$ in terms of the unflowed correlator. 

As it turns out, correlators saturating the bound on spectral flow are also simpler to deal with in terms of their factorization properties. Indeed, unlike in the previous cases, the integral in \eqref{(0020)} can be carried out explicitly  even before taking the small $z$ limit. After rescaling $y_3 \rightarrow z(1-z) y_3/(z-x)$, we obtain
\begin{equation}
    \fdot{\mathcal{F}}_{(0020)}(x,z)= f(z)\Bigg|\frac{\off{(1-z)z}^{k-h_3-j_3}}{(z-x)^{j_1+j_2+j_4-h_3}}\Bigg|^2 \frac{\pi\gamma(j_1+j_2+j_4-h_3)\gamma(j_3+h_3-k)}{\gamma(j_1+j_2+j_3+j_4-k)}.
\end{equation}
As follows from Eqs.~\eqref{def f(z)} and \eqref{zxexpansion}, at small $z$ the function $f(z)$ behaves as 
\begin{align}
    f(z) &\sim \int_{\frac{1}{2}+i\R} dj \mathcal{N}(j_1)\mathcal{N}(j_3)\frac{C\of{\frac{k}{2}-j_1,j_4,j}C\of{j,j_2,\frac{k}{2}-j_3}}{B(j)}\nn\\
    &\hspace{1cm}\times |z^{\Delta_j-\Delta_{1}-\Delta_{4}+j_2+j_3-\frac{3k}{4}}|^2\frac{\gamma(2j)\gamma(j_1+j_2+j_3+j_4-k)}{\gamma(j_1+j_4+j-\frac{k}{2})\gamma(j_2+j_3+j-\frac{k}{2})}
    \label{edgeflimit} \, .
\end{align}
Therefore, we get 
\begin{align}
    \fdot{\mathcal{F}}_{(0020)}(x,z)\sim& \int_{\frac{1}{2}+i\R} dj \mathcal{N}(j_1)\mathcal{N}(j_3)\frac{C\of{\frac{k}{2}-j_1,j_4,j}C\of{j,j_2,\frac{k}{2}-j_3}}{B(j)}\label{0020lim}\\
    &\hspace{-0.5cm} \times |z^{\Delta_j-\Delta_{1}-\Delta_{4}+j_2-h_3+\frac{k}{4}}x^{h_3-j_2-j_1-j_4}|^2\frac{\pi\gamma(j_1+j_2+j_4-h_3)\gamma(j_3+h_3-k)\gamma(2j)}{\gamma(j_1+j_4+j-\frac{k}{2})\gamma(j_2+j_3+j-\frac{k}{2})}.\nn
\end{align}
As usual, the overall powers of $x$ and $z$ allow us to recognize the intermediate channel in question. It describes the
exchange of a state with $\w = 1$, which is indeed the only channel allowed by the selection rules \eqref{Wselectionrules}, and $h = h_3-j_2$. The numerical constants then combine to give the relevant product of flowed structure constants,  
\begin{equation}
\label{(001)(102)/(11)}
 \frac{C_{(001)}(j_1,j_4,j,h)C_{(102)}(j,j_2,j_3,h,h_3)}{R(j,h,1)}\, , 
\end{equation}
which can be found in \cite{Maldacena:2001km,Cagnacci:2013ufa}.
%%%%%%%%%%%%%%%%%%%%%%%%%%%%%%%

\subsection{Lessons for the general case}
\label{sec: lessons}

Let us try and build on the results obtained in the previous sections and provide some intuition for the more general cases. On general grounds, worldsheet conformal invariance combined with the spectrum of the SL(2,$\R$) WZW model implies that four-point functions should factorize (in the 14 $\to$ 23 channel) schematically as follows: 
\begin{equation}
    \fdot{\Ff}_{\boldsymbol{\w}}(x,z) = \sum_{\rm \Delta} \frac{C_{1,4,\Delta}C_{\Delta,2,3}}{B_{\Delta}} |F_\Delta(x,z)|^2 \, . 
\end{equation}
Here we sum over all possible primary intermediate states of conformal dimension $\Delta$, while $F_\Delta(x,z)$ represents the corresponding conformal block, which also accounts for the descendant contributions. Let us emphasize that we are discussing Virasoro primaries, as opposed to the affine ones, hence we must include the spectrally flowed operators.
The $z$-independent factors $C_{ij\Delta}$ and $B_\Delta$ then capture the relevant structure constants and inverse propagator. In our context the sum over $\Delta$ is actually a sum over the spectral flow charge $\w$ of the exchanged state, together with a double integral over its flowed and unflowed spins, namely  $h$ and $j$, respectively.  

When the external $\w_i$ vanish and the corresponding $j_i$ are sufficiently closed to those of the continuous representations, only continuous unflowed states are exchanged, hence the integral over $j$ is along the contour $\frac{1}{2}+i \R$ \cite{Teschner:1999ug,Maldacena:2001km}. The above examples, and, more generally, the conjecture of \cite{Dei:2021yom}, suggest that, for some appropriately defined range of values for the external spins $j_i$, which does not necessarily coincide with Teschner's strip \eqref{unflowedrange}, one has 
\begin{align}
\label{general factorization}
    &\fdot{\Ff}_{\boldsymbol{\w}}(x,z) = \int_{\frac{1}{2}+i \R} dj \Bigg[
    \frac{C_{(\w_1,\w_4,0)}(j_1,j_4,j,h_1,h_4)C_{(0,\w_2,\w_3)}(j,j_2,j_3,h_2,h_3)}{B(j)} |F^{0}_{j}(x,z)|^2\\
    & \qquad \qquad   + \sum_{\w > 0} \sum_{h} \frac{C_{(\w_1,\w_4,\w)}(j_1,j_4,j,h_1,h_4,h)C_{(\w,\w_2,\w_3)}(j,j_2,j_3,h,h_2,h_3)}{R(j,h,\w)} |F^{\ww}_{jh}(x,z)|^2 \Bigg]\, , \nn
\end{align}
with $B(j)$, $R(j,h,\w)$ and $C_{\boldsymbol{\w}}(j_i,h_i)$ defined in Eqs.~\eqref{def N and R} and \eqref{def Cw(ji,hi)}. Here we have separated the contributions from the exchange of unflowed states from the flowed ones because for the latter there is an additional sum over the allowed values of $h$ (and $\bar{h}$). At leading order in $z$ the conformal blocks must be of the form\footnote{Strictly speaking, Eq.~\eqref{general block small z} is valid only up to potential monodromy projections removing the contribution of the shadow operator \cite{Simmons-Duffin:2012juh}. In all cases considered in this paper this projection trivializes. }
\begin{align}
\label{general block small z}
    |F^{\ww}_{jh}(x,z)|^2 & = |z^{\Delta-\Delta_1-\Delta_4} x^{h-h_1-h_4}|^2 \\
& \hspace{-1cm}\times \frac{(2h-1)^2}{\pi^2}\int d^2x' d^2x'' |\of{x'-x''}^{2h-2}x'^{h_4-h_1-h}(x-x')^{h_1-h_4-h}(1-x'')^{h_3-h_2-h}|^2 \, , \nn
\end{align}
where for $\w=0$ we take $h=j$. As in the unflowed case the integral Eq.~\eqref{general block small z} is of the hypergeometric type, albeit with all $j_i$ replaced by the corresponding $h_i$. 

The main obstacle for showing that the proposal of \cite{Dei:2021yom} factorizes in this way is that the complicated expressions in \eqref{even4pt} and \eqref{odd4pt} do not allow us to obtain the four-point functions in closed form. The reason is two-fold. On the one hand, they provide integral expressions for the former in terms of the $y_i$-variables, but the structure of the generalized differences $X_I$ and cross-ratio $X$ where these variables appear is quite involved, so much so that even in the three-point function case (where the Moebius-fixed unflowed correlator is simply a constant) it has not been possible to compute these integrals in closed form. On the other hand, the integrands involve the unflowed four-point functions which, as discussed in Sec.~\ref{sec: unflowed factorization}, are not known in closed form. 
Nevertheless, below we will extend the discussion of the particular examples considered so far to four-point functions with arbitrary spectral flow charges, which leads to an important number of non-trivial results concerning the factorization properties of the expressions proposed in \cite{Dei:2021yom}. 

Let us now try to provide a heuristic intuition for why we have only been able to isolate exchanges with spacetime spins such as $h = h_1-h_4$ when dealing with flowed channels.  (Here we assume for simplicity that $\w_1\geq \w_4$). The main issue is that, due to the explicit form of the conformal dimension $\Delta$ given in Eq.~\eqref{def Delta w} the small $z$ limit interacts non-trivially with the sum over $h$. In order to describe the consequences of this observation it will be useful to massage a bit the formal expansion presented in Eqs.~\eqref{general factorization} and \eqref{general block small z}. Let us consider the contribution from the exchange of a state with a given spectral flow charge $\w$ and write the flowed structure constants appearing in \eqref{general factorization} in terms of their $y$-integrals. By using  Eqs.~\eqref{reflection w>0}, \eqref{reflection ybasis} and \eqref{def Cw(ji,hi)}, we get 
\begin{align}
    \fdot{\Ff}_{\boldsymbol{\w}}(x,z)  &\sim \int_{\frac{1}{2}+i \R} dj \sum_{h}  \frac{(2h-1)^2}{\pi^2}\int  d^2y' d^2y'' d^2x' d^2x'' \prod_{i=1}^4 d^2y_i \,  |y_i^{h_i^0-h_i-1}z^{\Delta-\Delta_1-\Delta_4}  
    \nn \\
    &\quad \times y'^{h^0-h-1}y''^{\tilde{h}^0-h-1}|^2 
    \tdot{\mathcal{F}}_{(\w_1,\w_4,\w)}(y_1,y_4,y') \tdot{\mathcal{F}}_{(\w,\w_2,\w_3)}^{(1-j)}(y'',y_2,y_3)   \\[1ex]
& \quad \times |x^{h-h_1-h_4} \of{x'-x''}^{2h-2}x'^{h_4-h_1-h}(x-x')^{h_1-h_4-h}(1-x'')^{h_3-h_2-h}|^2 \, , \nn
\end{align}
with $h_i^0 = j_i + \frac{k}{2}\w_i$, $h^0 = j + \frac{k}{2}\w$ and $\tilde{h}^0 = 1-j + \frac{k}{2}\w$. In the second three-point function the superscript indicates that one has to replace $j \to 1-j$. We can now remove the dependence of the overall power of $z$ in $h$ by rescaling 
\begin{equation}
    y_1 \to y_1z^{\w_1} \, , \qquad 
    y_4 \to y_4z^{\w_4}
    \, , \qquad 
    y' \to y'z^{-\w} \, , 
\end{equation}
which gives  
\begin{align}
    \fdot{\Ff}_{\boldsymbol{\w}}(x,z)  &\sim \int_{\frac{1}{2}+i \R} dj \sum_{h}  \frac{(2h-1)^2}{\pi^2}\int  d^2y' d^2y'' d^2x' d^2x'' \prod_{i=1}^4 d^2y_i \,  |y_i^{h_i^0-h_i-1}z^{\Delta_j^0-\Delta_1^0-\Delta_4^0}  
    \nn \\
    &\quad \times y'^{h^0-h-1}y''^{\tilde{h}^0-h-1}|^2 
    \tdot{\mathcal{F}}_{(\w_1,\w_4,\w)}\left(y_1 z^{\w_1},y_4z^{\w_4},\frac{y'}{z^{\w}}\right) \tdot{\mathcal{F}}_{(\w,\w_2,\w_3)}^{(1-j)}(y'',y_2,y_3)   \\[1ex]
& \quad \times |x^{h-h_1-h_4} \of{x'-x''}^{2h-2}x'^{h_4-h_1-h}(x-x')^{h_1-h_4-h}(1-x'')^{h_3-h_2-h}|^2 \, , \nn
\end{align}
where $\Delta^0_i = -\frac{j_i(j_i-1)}{k-2}-j_i \w_i - \frac{k}{4}\w^2_i$. This suggests that, at least in terms of scaling with $z$, the new $y_i$ should be identified with those appearing in Eqs.~\eqref{even4pt} and \eqref{odd4pt}. The issue is that the arguments of $\tdot{\mathcal{F}}_{(\w_1,\w_4,\w)}$ now scale either to zero or to infinity as we take $z\to 0$. This is only well defined for generic values of $h_i$ when it reduces to a collision limit analogous to the one in Eq.~\eqref{eq: Mobiusunfix}. As a consequence, we are able to  select channels for which $h = h_1-h_4$ and $\w = \w_1-\w_4+\delta$ with $\delta=0,\pm 1$ by implementing 
\begin{equation}
    y_1 \to y_1 z^{-\w_4+\delta} \, , \qquad y_4 \to y_4 z^{-\w_1-\delta}
    \label{collisionrescalings}
\end{equation}
In such cases, the form of $\tdot{\mathcal{F}}_{(\w_1,\w_4,\w)}$ reduces to that given in Eq.~\eqref{3ptcollision}. The integration over $x'$ and $x''$ simplifies drastically for this value of $h$, which also allows us to make use of  the identity \eqref{3ptproductId} derived in the Appendix. After the dust settles, we find that these particular contributions can be written as 
\begin{align}
\label{finalextremalfactrorization}
    \fdot{\Ff}_{\boldsymbol{\w}}(x,z)  &\sim \int_{\frac{1}{2}+i \R} dj  \int  \prod_{i=1}^4 d^2y_i \,  |y_i^{h_i^0-h_i-1}z^{\Delta-\Delta_1-\Delta_4}  x^{-2h_4}|^2
    \\
    &\quad \times \int d^2y \,  |y^{2j-2}|^2
    \tdot{\mathcal{F}}_{(\w_1,\w_4,\w_1-\w_4+\delta)}\left(y_1,y_4,y\right) \tdot{\mathcal{F}}_{(\w_1-\w_4+\delta,\w_2,\w_3)}^{(1-j)}(y^{-1},y_2,y_3)  \, .  \nn
\end{align}
Recall that the dependence of three-point functions with respect to the $y$-variables is much more complicated than that in $x$ or $z$. The latter is fixed in the usual way by conformal symmetry, while the former depends non-trivially on the choice of spectral flow charges, see Eq.~\eqref{3pt-final}. Nevertheless, the crucial point is that even though the factors $Z_I(y_i)$ are not simple differences such as $y_{ij}$, they remain linear in all the $y_i$, and their powers depend on the unflowed spins $j_i$ analogously to the unflowed case (up to shifts of the form $j_i \to \frac{k}{2}- j_i$ in the odd parity case). This implies that the integral appearing in the second line of Eq.~\eqref{finalextremalfactrorization} is again of the hypergeometric type and can be carried out following  App.~\eqref{sec: App A - Integrals}. The explicit result will be given below when discussing several cases in detail. 

Finally, let us also state that, depending on the values of the spectral flow charges, a similar analysis involving rescalings of $y_2$ and $y_3$ with factors of $z^{\w}$ where $\w= \pm(\w_3-\w_2)+\delta$ can be used to compute the contribution from channels where the three-point function on the RHS becomes extremal, i.e.~with either $h=h_3-h_2$ or $h= h_2-h_3$. The same goes for channels with $\w = \w_1+\w_4$, etc.

%%%%%%%%%%%%%%%%%%%%%%%%%%%%%%%%

\section{Edge cases and $m$-basis limits}
\label{Sec4}

In this section we move past particular cases and start testing the factorization properties of the SL(2,$\R$) four-point functions proposed in \cite{Dei:2021yom} for more general spectral flow configurations. As a stepping stone, we consider correlators which admit a well-defined $m$-basis limit. The latter were dubbed as edge cases in \cite{Dei:2021xgh,Dei:2021yom,Bufalini:2022toj}. 

%%%%%%%%%%%%%%%%%%%%%%%%%%%%

\subsection{Factorization in the $m$-basis}

In the $m$-basis approach \cite{Maldacena:2001km,Cagnacci:2015pka} one usually works with the so-called flowed primary states, denoted as $V_{jm}^{\w}(z)$ \cite{Iguri:2022eat}. They are obtained by considering the affine primary states and applying the spectral flow operation along the direction of the Cartan current $J^3(z)$. One can obtain them as limits of the $x$-basis operators we have been using so far:
\begin{equation}
    \lim_{x\to 0} V_{jh}^w(x,z) = V_{jm}^\w(z) \,,  \qquad
    \lim_{x\to \infty} x^{2h}V_{jh}^w(x,z) = V_{j,-m}^{-\w}(z) \,,  \qquad
    m = h - \frac{k}{2}\w\ .  
\end{equation}
The $m$-basis four-point function
\begin{equation}
    \langle V_{j_1 m_1}^{w_1}(0)V_{j_2 m_2}^{w_2}(1)V_{j_3,- m_3}^{-w_3}(\infty)V_{j_4 m_4}^{w_4}(z) \rangle
\end{equation}
thus corresponds to the collision  limit\footnote{One could also consider taking $x_4\to 0$ but $x_2 \to \infty$. We will consider such configurations later on.} $x_2,x_4\to 0$ of 
\begin{equation}
\langle V_{j_1}^{w_1}(0,y_1,z_1)V_{j_2}^{w_2}(x_2,y_2,1)V_{j_3}^{w_3}(\infty,y_3,\infty)V_{j_4}^{w_4}(x_4,y_4,z) \rangle \, .
 \label{eq: coincience limit 4pt}
\end{equation}
Since the global Ward identities \eqref{eq:global Ward identities solution 4pt function} imply that for generic $x_2$ and $x_4$ this correlator takes the form 
\begin{align}
|x_{2}|^{2(-h_1^0-h_2^0+h_3^0-h_4^0)}  
    \Bigg \langle V_{j_1}^{w_1} \left(0,\frac{y_1 }{x_{2}},0\right) V_{j_2}^{w_2} \left(1,\frac{y_2 }{x_{2} },1\right) V_{j_3}^{w_3}\left(\infty, y_3 \, x_{2},\infty\right)  V_{j_4}^{w_4} \left(\frac{x_{4}}{x_{2}},\frac{y_4 }{x_{2}}, z\right)\Bigg\rangle \nn \, ,
\end{align}
one finds that the limit in \eqref{eq: coincience limit 4pt} is only well-defined for spectral flow charges satisfying \cite{Dei:2021yom} \begin{equation}
    |w_3-w_1-w_2-w_4| \leq 2 \, .
\end{equation}
 The resulting formulae for the $y$-basis four-point functions  are drastically simplified in the above regime. We now study the factorization properties for the cases near the edge of the range allowed by the selection rules \eqref{Wselectionrules} separately. For simplicity, we will focus on a single flowed intermediate channel in each case.

\subsection*{Spectral flow conservation}

We first consider the case $\w_3=\w_1+\w_2+\w_4$, where the total spectral flow is conserved. After rescaling the $y_i$ variables by the appropriate factors of $z$ and $z-1$ and changing $y_3 \to -y_3^{-1}$ one gets \cite{Dei:2021yom} 
\begin{align}
&\langle V_{j_1 m_1}^{w_1}(0)V_{j_2 m_2}^{w_2}(1)V_{j_3,- m_3}^{-w_3}(\infty)V_{j_4 m_4}^{w_4}(z) \rangle=
\big|z^{-\frac{k w_1 w_4}{2}-w_1m_4-w_4m_1}(1-z)^{\frac{-k w_2 w_4}{2}-w_2m_4-w_4m_2}\big|^2\nonumber\\
&\qquad\times\int \prod_{i=1}^4 d^2 y_i \prod_{i\neq 3}| y_i^{j_i-m_i-1} y_3^{j_3+m_3-1} y_{12}^{-j_1-j_2+j_3-j_4}y_{13}^{-j_1+j_2-j_3+j_4} y_{23}^{j_1-j_2-j_3+j_4} y_{34}^{-2j_4}|^2\nonumber\\
&\qquad\qquad\times  \left\langle V_{j_1}^0(0;0) V_{j_2}^0(1;1) V_{j_3}^0(\infty;\infty) V_{j_4}^0\left(\frac{y_{32} \, y_{41}}{y_{21}\, y_{34}};z\right) \right \rangle \ . \label{eq:coincidence limit spectral flow preserving}
\end{align}
We see that, in this limit, the $y_i$ variables effectively play the role of the traditional $x_i$ for the unflowed correlator. The result is then simply the usual $m$-basis unflowed expression\footnote{This can be understood from parafermionic decomposition \cite{Maldacena:2000hw}, which implies that in the $m$-basis one only cares about the total spectral flow \cite{Fateev}, which vanishes in this particular case.}. 

The factorization analysis thus parallels the unflowed one. The selection rules for the relevant three-point functions allow for a channel with $\w = \w_1+\w_4 = \w_3-\w_2$. Further taking $x_2,x_4 \to 0$, the relevant expressions for the factorization limit take the form 
\begin{align}
    & \hspace{-0.3cm}
    \langle V_{j_1}^{w_1}(0,y_1,0)V_{j_4}^{w_4}(0,y_4,1)V_{j}^{w}(\infty,y,\infty)\rangle = 
    \frac{C(j_1,j_4,j)}{|y_{14}^{j_1+j_4-j}
    (1+y_1 y)^{j+j_1-j_4}
    (1+y_4 y)^{j+j_4-j_1}|^2}\,,  \\[1ex]
    & \hspace{-0.3cm}
    \langle V_{j}^{w}(0,y',0)V_{j_2}^{w_2}(0,y_2,1)V_{j_3}^{w_3}(\infty,y_3,\infty)\rangle 
    %\\ & \qquad \qquad 
    = 
    \frac{C(j,j_2,j_3) |(y' -y_2)^{j_3-j_2-j}|^2}{|(1+y_2y_3)^{j_2+j_3-j}
    (1+y_3y' )^{j+j_3-j_2}
    |^2}
    , 
\end{align}
again up to a few unimportant signs, and where $j$ is the intermediate spin. By taking $y \to - y^{-1}$ and $y_3 \to -y_3^{-1}$ one finds that, in this regime, the spectral flow preserving three-point functions take precisely the same form as the unflowed $x$-basis ones, but with $x_i \to y_i$. 

It follows that the integration over the intermediate $y$ thus parallels the derivation of \cite{Teschner:1999ug} leading to the hypergeometric function appearing in Eqs.~\eqref{unflowedblocks}-\eqref{unflowed 2F1}. More explicitly, proceeding in analogy to what was done in the previous section and using \eqref{reflection ybasis} and \eqref{3ptproductId} shows that we need to compute 
\begin{align}
   \int d^2y |(y_1-y)^{j_4-j_1-j}(y_4-y)^{j_1-j_4-j}
    (y y_2-1)^{j_3-j_2+j-1}
   (y y_3-1)^{j_2-j_3+j-1}|^2 \, , \nn
\end{align}
which is of the hypergeometric type. Combined with the overall factor of $y_{14}^{j-j_1-j_4} y_{23}^{j-j_2-j_3}$, this precisely matches the factorization derived by inserting the small $z$ expansion (at fixed $y_i$) of the unflowed correlator in Eq.~\eqref{eq:coincidence limit spectral flow preserving}.  Note that in this case the sum over the intermediate values of $h$ in Eq.~\eqref{general factorization} trivializes as a consequence of charge conservation \cite{Bufalini:2022toj}. In other words, given the values of the external $m_i$, the intermediate state must have $m=m_1+m_4=m_3-m_2$. Of course, the overall power of $z$ is consistent with our analysis: the factors coming from the expansion of the unflowed correlator add up to those in Eq.~\eqref{eq:coincidence limit spectral flow preserving}, giving 
\begin{equation}
    \Delta_j^{(0)}-\Delta_1^{(0)}-\Delta_4^{(0)} -\frac{k w_1 w_4}{2}-w_1m_4-w_4m_1 =  \Delta-\Delta_1-\Delta_4 \, .
\end{equation}
%%%%%%%%%%%%%%%%%%%%%%%%%%%%
\subsection*{Spectral flow violation in one unit}

We now consider the case $\w_3=\w_1+\w_2+\w_4+1$  (the analysis for the case with $\w_3=\w_1+\w_2+\w_4-1$ is analogous). Even tough spectral flow is not a conserved quantity, non-vanishing $m$-basis $n$-point functions for which the total spectral flow is non-zero are usually referred to as spectral flow \textit{violating} correlators. Manipulations similar to those of the previous section give   
\begin{align}
&\langle V_{j_1 m_1}^{w_1}(0)V_{j_2 m_2}^{w_2}(1)V_{j_3,- m_3}^{-w_3}(\infty)V_{j_4 m_4}^{w_4}(z) \rangle=
\Big|z^{-\frac{k w_1 w_4}{2}-w_1m_4-w_4m_1}(1-z)^{\frac{-k w_2 w_4}{2}-w_2m_4-w_4m_2}\Big|^2\nonumber\\
& \qquad \qquad\times \int \prod_{i=1}^4 d^2 y_i \ \prod_{i\neq 3} |y_i^{j_i-m_i-1}  y_3^{j_3+m_3-1} (y_1+y_2+y_3)^{\frac{k}{2}-j_1-j_2-j_3-j_4}  |^2\nonumber\\
&\qquad \qquad\times \left\langle V_{j_1}^0(0;0) V_{j_2}^0(1;1) V_{\frac{k}{2}-j_3}^0(\infty;\infty) V_{j_4}^0\left(\frac{y_1+z y_3- y_4}{y_1+y_2+y_3};z\right) \right \rangle\ .
\label{violationw1}
\end{align}
Taking the small $z$ limit and using the expansion for the unflowed correlator, we find that the $y$-integral becomes 
\begin{align}
&   \int_{\frac{1}{2}+i \R} dj \prod_{i=1}^4 d^2 y_i \ \prod_{i\neq 3}\Bigg| y_i^{j_i-m_i-1}  y_3^{j_3+m_3-1}  y_{14}^{j-j_1-j_4}(y_1+y_2+y_3)^{\frac{k}{2}-j-j_2-j_3} {}  \nonumber\\
&\qquad \qquad\times \pFq{2}{1}{j-j_1+j_4,j+j_2 +j_3-\frac{k}{2}}{2j}{\frac{y_{14}}{y_1+y_2 + y_3}}\Bigg|^2 \, .
\label{aaaa}
\end{align}
The latter can be obtained directly by integrating over the product of $y$-basis three-point functions  involved in the exchange of a state with $\w = \w_1+\w_4 = \w_3-\w_2-1$ (times the usual factor related to the corresponding reflection $j \to 1-j$) which, as before, is constrained by charge conservation. The only element we have not discussed so far is the three-point function on the RHS, which, up to signs, takes the form 
\begin{align}
    & \hspace{-0.3cm}
    \langle V_{j}^{w}(0,y',0)V_{j_2}^{w_2}(0,y_2,1)V_{j_3}^{w_3}(\infty,y_3,\infty)\rangle = \frac{{\cal{N}}(j_3)C\left(j,j_2,\frac{k}{2}-j_3 \right)}{|y_3^{\frac{k}{2}+j_3-j_2-j}(1+y_2 y_3 + y' y_3)^{j+j_2+j_3-\frac{k}{2}}|^2}
    \, .
\end{align}
Indeed, the integral
\begin{align} 
   \int d^2y  |(y_1y-1)^{j_4-j_1+j-1}(y_4y-1)^{j_1-j_4+j-1}
    (1+y_2y_3+y y_3)^{\frac{k}{2}-j-j_2-j_3}|^2
    \, , \nn
\end{align}
and the overall factors of $y_{14}^{1-j-j_1-j_4}y_3^{j+j_2-j_3-\frac{k}{2}}$ combine to give the integrand in \eqref{aaaa}. The product of the unflowed structure constants $C\left(j_1,j_4,j\right)$ and  ${\cal{N}}(j_3)C\left(j,j_2,\frac{k}{2}-j_3 \right)$, and the overall powers of $z$ then lead to the expected factorization.  

\subsection*{Spectral flow violation in two units}

Finally, we consider the case $\w_3= \w_1+\w_2+\w_4+2$. Correlators with $\w_3= \w_1+\w_2+\w_4-2$ can be treated similarly. This corresponds to the situation where spectral flow violation is maximal \cite{Fateev,Giribet:2011xf}. As discussed above, in this case, the conjecture of \cite{Dei:2021yom} looks slightly different. We get  
\begin{align}
& \langle V_{j_1 m_1}^{w_1}(0)V_{j_2 m_2}^{w_2}(1)V_{j_3,- m_3}^{-w_3}(\infty)V_{j_4 m_4}^{w_4}(z) \rangle = |z^{m_2-j_2-m_3-j_3}(1-z)^{m_1-j_1-m_3-j_3}|^2 f(z)\nonumber\\
&\qquad\times\int d^2 y_i \ \prod_{i=1,2,4} |y_i^{j_i-m_i-1}y_3^{j_3+m_3-1} (y_1+y_2+y_3+y_4)^{k-j_1-j_2-j_3-j_4} |^2\, ,
\label{violationw2}
\end{align}
where $f(z)$ is given in Eq.~\eqref{def f(z)} below. 

Only a single channel is allowed, for which the exchanged state has spectral flow charge $\w = \w_1+\w_4+1 = \w_3-\w_2-1$. After changing variables to $y_3 \to -y_3^{-1}$, the relevant integral can be written as  
\begin{align}
  & \hspace{-0.1cm} \int d^2y   |(y+y_1+y_4)^{\frac{k}{2}-j_1-j_4+j-1} y_3^{2j_3-2}(y+y_2+y_3)^{\frac{k}{2}-j_2-j_3-j}|^2  \\
  & \qquad = \pi\frac{\gamma(\frac{k}{2}-j_1-j_4+j)\gamma\left(j_1+j_2+j_3+j_4-k\right)}{\gamma\left(j+j_2+j_3-\frac{k}{2}\right)} |(y_1+y_2+y_3+y_4)^{k-j_1-j_2-j_3-j_4}|^2 \nn
  \, ,
\end{align}
where $j$ is the spin of the intermediate state. 
Comparing with \eqref{violationw2} at leading order in $z$, by means of \eqref{edgeflimit} and \eqref{Creflection}, we obtain a consistent factorization limit. 

%%%%%%%%%%%%%%%%%%%%%%%%%%%%%%%

\subsection{Edge cases beyond the collision limit} 
\label{sec: edge case}

Of course, the analysis performed above is quite restrictive, as the $m$-basis limit only captures a subset of four-point functions for which all spectral flows are, so to speak, along the same direction in $x$-space \cite{Eberhardt:2018ouy}. We would now like to relax this constraint, and work directly in the $x$-basis, i.e.~for generic values of $x_2$ and $x_4$.   

As it turns out, from the three choices of spectral flow charges considered in the previous section, the first two, namely those we have called spectral flow conserving and spectral flow violating by one unit, can be analyzed along the same lines as those with generic values of the $\w_i$, hence we will discuss them in the next section.
Here we focus on the edge case 
\begin{equation}
    \w_3 = \w_1+\w_2+\w_4+2,
\end{equation}
and set $(x_1,x_2,x_3,x_4) = (0,1,\infty,x)$ as usual. For such configurations, the quantity  $X_\emptyset$ appearing in the conjectured general formula \eqref{even4pt} vanishes. However, one also finds that the generalized cross-ratio behaves as $X \to z$. It is therefore necessary to make use of the identity 
\begin{equation}
    z-X = \frac{X_\emptyset X_{1234}}{X_{12}X_{34}} 
\end{equation}
to rewrite the $y$-basis correlator as 
\begin{align}
    & \hspace{-0.2cm} \fdot{\mathcal{F}}_{\rm edge}(x,y_i,z) = \big|X_{1234}^{k-j_1-j_2-j_3-j_4} X_{12}^{2j_3-k}X_{13}^{-j_1+j_2-j_3+j_4} X_{23}^{j_1-j_2-j_3+j_4}X_{34}^{j_1+j_2+j_3-j_4-k}\big|^2 f(z)\label{edge4pt},
\end{align}
where the function $f(z)$ is defined as a limit of the unflowed correlator, namely 
\begin{equation}
    f(z) = \lim_{x\rightarrow z}|x-z|^{2(j_1+j_2+j_3+j_4-k)}\fdot{\mathcal{F}}_{0}(x,z)\,.
    \label{def f(z)}
\end{equation}

As in the example considered in Sec.~\ref{sec: (0020)}, we will be able to carry out the integration over the $y_i$ variables explicitly. In order to show this we need to specify the relevant $X_I$ factors. For this choice of spectral flow charges, many of the polynomials $P_{\boldsymbol{\w}}(x,z)$ involved in the definition of the $X_I$ either vanish or trivialize. We have 
\begin{equation}
X_{12} = P_{\boldsymbol{\w}+e_1+e_2}\, , \quad X_{13} = P_{\boldsymbol{\w}+e_1-e_3}y_3\, , \quad X_{23} = P_{\boldsymbol{\w}+e_2-e_3}y_3\, , \quad X_{34} = P_{\boldsymbol{\w}-e_3+e_4}y_3 \, ,
\end{equation}
and
\begin{align}
    \frac{X_{1234}}{\sqrt{z(1-z)}} &= P_{\boldsymbol{\w}+e_1+e_2+e_3+e_4} + y_3 (P_{\boldsymbol{\w}+e_1+e_2-e_3+e_4}+P_{\boldsymbol{\w}-e_1+e_2-e_3+e_4} y_1 
    \nn \\
   & \qquad  +P_{\boldsymbol{\w}+e_1-e_2-e_3+e_4}y_2+P_{\boldsymbol{\w}+e_1+e_2-e_3-e_4}y_4)\, .
\end{align}
Here and in what follows we leave  the dependence of the $P_{\boldsymbol{\w}}(x,z)$ in $x$ and $z$ implicit. 
By rescaling the $y_i$-variables appropriately we can express the exact $x$-basis four-point function
\begin{equation}
    \fdot{\mathcal{F}}_{\text{edge}}(x,z) = \int d^2y_i \prod_{i=1}^{4} |y_i^{j_i + \frac{k}{2}\w_i-h_i-1}|^2 \fdot{\mathcal{F}}_{\text{edge}}(x,y_i,z),
\end{equation}
as 
\begin{align}
    & \fdot{\mathcal{F}}_{\text{edge}}(x,z) = h(x,z)\int d^2y_i \prod_{i \neq 3}\Bigg|y_i^{j_i + \frac{k}{2}\w_i-h_i-1}y_3^{h_1+h_2+h_4-h_3-1} \of{1+\sum_{i=1}^4 y_i}^{k-j_1-j_2-j_3-j_4}\Bigg|^2\nn\\
    & \qquad = \frac{\pi^4 h(x,z) \gamma(h_1+h_2+h_4-h_3) \gamma\left(j_3+h_3-\frac{k}{2}\w_3\right) \prod_{i\neq 3}\gamma\left(j_i + \frac{k}{2}\w_i-h_i\right)}{\gamma(j_1+j_2+j_3+j_4-k)},\label{edgeintegrated}
\end{align}
with
\begin{align}
    h(x,z) = f(z)\Bigg|\frac{P_{\boldsymbol{\w}+e_1-e_3}^{-j_3-j_1+j_2+j_4}P_{\boldsymbol{\w}+e_2-e_3}^{j_1-j_2-j_3+j_4}}{[z(1-z)]^{\frac{1}{2}(j_1+j_2+j_3+j_4-k)}P_{\boldsymbol{\w}+e_1+e_2}^{k-2j_3}P_{\boldsymbol{\w}-e_1+e_2-e_3+e_4}^{\alpha_1}} \times \nn\\
    \times \frac{P_{\boldsymbol{\w}-e_3+e_4}^{j_1+j_2+j_3-j_4-k}P_{\boldsymbol{\w}+e_1+e_2+e_3+e_4}^{j_1+j_2+j_4-h_3+\frac{k}{2}(\w_3-2)}}{P_{\boldsymbol{\w}+e_1-e_2-e_3+e_4}^{\alpha_2}P_{\boldsymbol{\w}+e_1+e_2-e_3-e_4}^{\alpha_4}P_{\boldsymbol{\w}+e_1+e_2-e_3+e_4}^{h_1+h_2+h_4-h_3}}\Bigg|^2 \, .
\end{align}

In order to analyze this result from the factorization perspective we consider the behaviour of the relevant $P_{\boldsymbol{\w}}(x,z)$ in the $z \to 0$ limit. This is done in Appendix \ref{sec: Pw at small z}, where we show that   
they reduce to a numerical prefactor of the form $N_{\boldsymbol{\w}}=n_{\boldsymbol{\w}}\tilde{P}_{\boldsymbol{\w}}(1,0)$, together with a number of powers of $x$ and $(1-x)$.  
Further using Eq.~\eqref{edgeflimit} leads to the following small $z$ behavior of the function $h(x,z)$: 
\begin{align}
    h(x,z) & \sim \Big|z^{\Delta_j - \Delta_1 -\Delta_4+\frac{k\w_1\w_4}{2}+\frac{k}{2}(\w_3-\w_2)-\frac{3k}{4}-(h_3-h_2)(\w_3-\w_2-1)+\w_1 h_1 + \w_4 h_4 }\nn\\
    &\times \of{N_{++-+}x}^{h_3-h_1-h_2-h_4}\Big|^2\frac{\gamma(2j)\gamma(j_1+j_2+j_3+j_4-k)}{\gamma(j_1+j_4+j-\frac{k}{2})\gamma(j_2+j_3+j-\frac{k}{2})}.
\end{align}
The overall powers of $x$ and $z$ we have obtained show that we have picked up the contribution coming from the exchange of an intermediate state with quantum numbers 
\begin{equation}
    \w = \w_3-\w_2-1 = \w_1 + \w_4 +1 \, , \qquad 
    h = h_3-h_2 \, .
    \label{edge intermediate w and h}
\end{equation}
Recall that this value of $\w$ is the only one allowed by the selection rules \eqref{Wselectionrules}. 
We conclude that, as long as the external spins $j_i$ are in the range \eqref{1010j1j3range}, the small $z$ limit of four-point functions saturating the bound on spectral flow can be written as 
\begin{align}
    \fdot{\mathcal{F}}_{\text{edge}} (x,z) &\sim \int_{\frac{1}{2}+i\R} dj \Big| z^{\Delta_{j}-\Delta_{1}-\Delta_{4}}\of{N_{++-+}x}^{h-h_1-h_4}\Big|^2 
    C\left(j_1,j_4,\frac{k}{2}-j\right)C\left(\frac{k}{2}-j,j_2,j_3\right) \nn \\
    &\times\frac{  \mathcal{N}(j)^2 \pi^4\gamma(2j) \gamma\left(j_3+h_3-\frac{k}{2}\w_3\right)\prod_{i\neq 3}\gamma\left(j_i+\frac{k}{2}\w_i-h_i\right)}{B(j)\gamma\left(j_1+j_4+j-\frac{k}{2}\right)\gamma\left(j_2+j_3+j-\frac{k}{2}\right)}.
\end{align}
This is consistent with the expected factorization structure. Indeed, for these values of $\w$ and $h$ we find that the constants in this  expression precisely match the product of three-point functions  appearing in Eq.~\eqref{general factorization}, namely 
\begin{equation}
 \frac{C_{(\w_1,\w_4,\w_1+\w_4+1)}(j_i,h_i)C_{(\w_3-\w_2-1,\w_2,\w_3)}(j_i,h_i)}{R(j,h,\w)}\,, 
\end{equation}  
thanks to the identity 
\begin{equation}
    N_{++-+} = n_{(\w_1+1,\w_2+1,\w_3-1,\w_4+1)}\tilde{P}_{(\w_1+1,\w_2+1,\w_3-1,\w_4+1)}(1,0) = Q_{(\w_3-\w_2,\w_2+1,w_3-1)} .\label{edgePidentity}
\end{equation}
This is a particular case of Eq.~\eqref{plimit}. It corresponds to a non-trivial relation between the (small $z$ limit of the) polynomials $P_{\boldsymbol{\w}}(x,z)$ involved in the conjecture of \cite{Dei:2021yom} for flowed four-point functions and the numerical factors $Q_{\boldsymbol{\w}}$ appearing in the flowed three-point functions of \cite{Dei:2021xgh,Bufalini:2022toj}. 

%%%%%%%%%%%%%%%%%%%%%%%%%%%%%%%%%%%%%%%%%

\section{Correlators with arbitrary spectral flow charges}
\label{Sec5}

We now discuss the general case, namely, four-point correlators with arbitrary assignments of spectral flow charges. We first show that, whenever it is allowed by the AdS$_3$ selection rules, the formula put forward in \cite{Dei:2021yom} for flowed four-point functions in terms of their unflowed counterparts consistently accounts for the exchange of unflowed states. We then move to arbitrary insertions and show that, in the small $z$ limit, the properties of correlation functions in the unflowed sector of the model lead to consistent contributions coming from the exchange of states with a non-zero spectral flow charge. The precise agreement with the structure anticipated in Eq.~\eqref{general factorization} we derive involves the full set of highly non-trivial spectrally flowed three-point functions obtained in \cite{Dei:2021xgh,Bufalini:2022toj}.   

\subsection{Contributions from the exchange of unflowed states}

As it follows from \eqref{Wselectionrules}, only four-point functions for which the external spectral flow charges satisfy 
\begin{equation}
    |\w_1-\w_4| \leq 1 \,, \qquad |\w_3-\w_2| \leq 1 \, , 
    \label{allowed unflowed exchanges}
\end{equation}
can contain contributions from the exchange of unflowed states along the $14\to 23$ channel. For concreteness, we first focus on the cases where 
$\w_1 = \w_4 = \w_L$ and $\w_2 = \w_3 = \w_R$, and then briefly discuss other situations. 
Consequently, the original four-point function and both three-point functions involved in the following discussion will be of the even-parity type. 

The relevant expression for the four-point function under consideration is given in Eq.~\eqref{even4pt}. The analysis of Sec.~\ref{sec: PwAnalysis} implies that, after the following rescaling 
\begin{equation}
    y_1\rightarrow (-1)^{\w_L}\frac{x}{z^{\w_L}}y_1,\qquad y_4\rightarrow \frac{x}{z^{\w_L}}y_4,
    \qquad y_2 \rightarrow (-1)^{\w_R}y_2,
\end{equation}
the relevant factors $X_I$, at leading order in $z$, take the form %\footnote{The relative signs are not determined by eq. \eqref{plimit} but were checked numerically.}
\begin{align}
    &X_{14} \rightarrow z^{-\frac{(\w_L+1)^2}{2}+\frac{1}{2}}x^{\w_L+1}(1+y_1 y_4), \nn\\
    &X_{23} \rightarrow z^{-\frac{\w_L^2}{2}}x^{\w_L}(1+y_2 y_3),\nn\\
    &X_{12} \rightarrow z^{-\frac{\w_L(\w_L+1)}{2}}x^{\w_L+1}(1+y_2+y_1(1-x+y_2)),\\
    &X_{13} \rightarrow z^{-\frac{\w_L(\w_L+1)}{2}}x^{\w_L+1}(1-y_3+y_1(1+(x-1)y_3)),\nn\\
    &X_{34} \rightarrow z^{-\frac{\w_L(\w_L+1)}{2}}x^{\w_L+1}(1-y_4+y_3(x-1+y_4)). \nn
\end{align}
In particular, this shows that for the generalized cross-ratio we have  
\begin{equation}
    \frac{X_{14}X_{23}}{X_{12}X_{34}} \rightarrow \frac{x(1+y_1y_4)(1+y_2y_3)}{(1+y_2+y_1(1-x+y_2))(1-y_4+y_3(x-1+y_4))} \, .
\end{equation}
Therefore,  after using \eqref{unflowed 2F1} under the assumption that the $j_i$ are in the range \eqref{unflowedrange}, we obtain a contribution to the factorized expansion of the form 
\begin{align}
    \fdot{\mathcal{F}}_{\ww}(x,z)&\sim \int_{\frac{1}{2}+i\RR}dj\mathcal{C}(j) |x^{j-h_1-h_4}z^{\Delta^{(0)}_j-\Delta_1-\Delta_4}|^2\int d^2y_i \prod_{i=1}^{4}\Bigg|y_i^{j_i+\frac{k}{2}\w_i-h_i-1}\label{wwexpansion}\\
    &\hspace{-1cm}\times(1+y_2+y_1(1-x+y_2))^{-j-j_2+j_3}(1-y_4+y_3(x-1+y_4))^{j_1-j-j_4}\nn\\
    &\hspace{-1cm}\times(1-y_3+y_1(1+(x-1)y_3))^{-j_1+j_2-j_3+j_4} 
    (1+y_1y_4)^{j-j_1-j_4}(1+y_2 y_3)^{j-j_2-j_3}\nn\\
    &\hspace{-1cm}\times\pFq{2}{1}{j-j_1+j_4,j-j_3+j_2}{2j}{\frac{x(1+y_1y_4)(1+y_2y_3)}{(1+y_2+y_1(1-x+y_2))(1-y_4+y_3(x-1+y_4))}}\Bigg|^2.\nn
\end{align}
Here we have written the leading order expression for the unflowed correlator  as in Eq.~\eqref{unflowedzexp} and used the shorthand $\mathcal{C}(j) = C(j_1,j_4,j)C(j,j_2,j_3)/B(j)$.
The overall powers of $x$ and $z$ we have obtained in \eqref{wwexpansion} show that we have indeed isolated the contribution associated with the  exchange of an unflowed state of spin $j$. 

In order to show  that this expression  is consistent with the expectation set up in Sec.~\ref{sec: lessons} we proceed by reverse engineering. For the case at hand, the relevant contribution to the RHS of \eqref{general factorization} reads 
\begin{align}\label{expected fact unflowed exchage}
    &\fdot{\Ff}_{\boldsymbol{\w}}(x,z) \sim  \int_{\frac{1}{2}+i \R} dj \frac{C_{(\w_L,\w_L,0)}(j_i,h_i)C_{(0,\w_R,\w_R)}(j_i,h_i)}{B(j)} |F^{0}_{jh}(x,z)|^2 \, , 
\end{align}
where, at leading order in $z$,  
\begin{equation}
    F^{0}_{jh}(x,z) = z^{\Delta^{(0)}_j-\Delta_1-\Delta_4} x^{j-h_1-h_4} \pFq{2}{1}{j-h_1+h_4,j-h_3+h_2}{2j}{x} \, .
    \label{unflowed channel h-2F1}
\end{equation}
This can be re-written in terms of an integral over two auxiliary variables $x'$ and $x''$, interpreted as the locations of the intermediate vertex operators. This is analogous to 
\eqref{2F1-integral-Teschner}, but with the replacements $j_i \to h_i$ for the external spins. On the other hand, inserting the integral expressions for the flowed structure constants obtained from the $y$-basis analysis, we have 
\begin{align}
    \frac{C_{(\w_L,\w_L,0)}(j_i,h_i)C_{(0,\w_R,\w_R)}(j_i,h_i)}{B(j)} &= 
    {\cal{C}}(j)  \int d^2y_i \prod_{i=1}^{4}|y_i^{h_i^0-h_i-1} (1+y_1y_4)^{j-j_1-j_4}  \\
    &\hspace{-4cm} \times (1+y_2y_3)^{j-j_2-j_3}(y_4+1)^{j_1-j_4-j}(y_1-1)^{j_4-j_1-j}(y_3-1)^{j_2-j_3-j}(y_2+1)^{j_3-j_2-j}|^2 \, . \nn 
\end{align}
with $h_i^0 = j_i+\frac{k}{2}\w_i$.
Plugging this back into  Eq.~\eqref{expected fact unflowed exchage}, we see that the full dependence on the $h_i$  quantum numbers coming from the hypergeometric integral can be eliminated by rescaling 
\begin{align}
    y_1 \rightarrow \frac{x-x'}{x'}y_1,\qquad y_2\rightarrow \frac{y_2}{1-x''},\qquad y_3\rightarrow (1-x'')y_3, \qquad y_4 \rightarrow \frac{x'}{x-x'}y_4\, .
\end{align}
 Indeed, as could have been anticipated from the global Ward identity \eqref{ybasisx1x2x3fixing}, this effectively replaces the $h_i$ appearing in the powers of $x'$, $(x-x')$ and $(1-x'')$ by the corresponding $h_i^0$. Moreover, for the spectral flow charges under consideration we have that $h_1^0-h_4^0 = j_1-j_4$ and $h_2^0-h_3^0 = j_2-j_3$, hence the resulting factors nicely combine with those involving the $y_i$ variables, thus leading to 
\begin{align}
    \fdot{\Ff}_{\boldsymbol{\w}}(x,z) & \sim \int_{\frac{1}{2}+i\R} dj {\cal{C}}(j)  \int d^2y_id^2x' d^2x'' \prod_{i=1}^{4}|y_i^{h_i^0-h_i-1}  \of{xy_1+x'(1-y_1)}^{j_4-j_1-j}\\
    &\hspace{-1cm} \times \of{x-x'(-y_4+1)}^{j_1-j_4-j}\of{y_3-1-y_3 x''}^{j_2-j_3-j}\of{y_2+1-x''}^{j_3-j_2-j}   \of{x'-x''}^{2j-2}|^2 \, .\nn
\end{align}
We have landed on an integral over $x'$ and $x''$ which is again of the hypergeometric type, with all powers written in terms of the unflowed spins $j_i$. This showcases how the presence of the integrals over the $y_i$ associated to the external states allows one to perform changes of variables that, roughly speaking, end up rewriting flowed quantities in terms of the unflowed ones. After integrating out $x'$ and $x''$, we finally re-obtain the expression derived from the conjectured formula for the flowed four-point function, namely Eq.~\eqref{wwexpansion}. 

For completeness, let us provide an additional example involving an odd three-point function. We set $\w_1 = \w_4+1 = \w_L+1$ and $\w_2 = \w_3= \w_R$. After taking $y_1 \rightarrow (-z)^{-\w_1}x y_1$ and $y_4 \rightarrow z^{-\w_4}x y_4$, the corresponding generalized cross-ratio behaves at small $z$ as  
\begin{align}
    \frac{X_{2}X_{134}}{X_{4}X_{123}} \rightarrow  \frac{z(1+y_2)(1-y_3+y_1(1-y_4+y_3(x-1+y_4)))}{x y_1 (1+y_2y_3)} \,.
\end{align}
Therefore, after using the expansion \eqref{zxexpansion}, the relevant contribution to the four-point becomes  
\begin{align}
    &\fdot{\mathcal{F}}_{\ww}(x,z) \sim  \int_{\frac{1}{2}+i\R} dj \mathcal{C}(j)|z^{\Delta_j^{(0)}-\Delta_1-\Delta_4}x^{j-h_1-h_4}|^2\int \prod_{i=1}^{4}d^2y_i \Bigg|y_i^{h^0_i-h_i-1} y_1^{j+j_4-j_1-\frac{k}{2}} \nn\\
    & \quad \times (1+y_2y_3)^{j-j_2-j_3}(1+y_2)^{j_3-j_2-j}(1-y_3)^{j_1+j_2-j_3+j_4-\frac{k}{2}} \label{unflowed2limit}\\
    & \quad \times (1-y_3+y_1(1-y_4+y_3(x-1+y_4)))^{\frac{k}{2}-j-j_1-j_4}\nn\\
    & \quad \times \pFq{2}{1}{j+j_1+j_4-\frac{k}{2},j-j_3+j_2}{2j}{\frac{x y_1 (1+y_2y_3)}{(1+y_2)(1-y_3+y_1(1-y_4+y_3(x-1+y_4)))}}\Bigg|^2, \nn
\end{align}
where $\mathcal{C}(j)$ now stands for 
\begin{equation}
    \mathcal{C}(j) =  \frac{\mathcal{N}(j_1)C(\frac{k}{2}-j_1,j_4,j)C(j,j_2,j_3)}{B(j)} \, .
\end{equation}
As before, the overall dependence in $z$ and $x$ matches with the desired unflowed intermediate channel. 

In order to re-derive this expression from the expected factorized form \eqref{general factorization} we proceed analogously to the previous case. In other words, we combine the hypergeometric integral \eqref{2F1-integral-Teschner} with all external $j_i$ replaced by $h_i$ with the $y$-basis expression
\begin{align}
    \frac{C_{(\w_L+1,\w_L,0)}(j_i,h_i)C_{(0,\w_R,\w_R)}(j_i,h_i)}{B(j)}&=\mathcal{C}(j)\int \prod_{i=1}^{4}d^2y_i| y_i^{h^0_i-h_i-1} y_1^{j+j_4-j_1-\frac{k}{2}}(1+y_2)^{j_3-j_2-j}\nn\\
    & \hspace{-2cm} (1-y_1(1+y_4))^{\frac{k}{2}-j_1-j_4-j}(1-y_3)^{j_2-j_3-j}(1+y_2y_3)^{j-j_2-j_3}|^2 \, ,
\end{align}
perform the appropriate rescalings, use that $h_1^0-h_4^0 = j_1-j_4+\frac{k}{2}$, and finally compute the integral over $x'$ and $x''$. The rest of the combinations allowed by \eqref{allowed unflowed exchanges} can be analyzed similarly. 

%%%%%%%%%%%%%%%%%%%%%%%%%%
\subsection{Contributions from the exchange of spectrally flowed states}

We now move to the exchange of states with non-zero spectral flow charges, which allows us to work with arbitrary values of the external $\w_i$. For concreteness, we first focus on four-point functions of odd-parity type, i.e. those described by Eq.~\eqref{odd4pt}, and assume that the $\w_i$ satisfy the following inequalities:
\begin{equation}
   \w_1-\w_4 > |\w_2-w_3|\,. \label{wCaseOdd1}
\end{equation}
This is the first option considered in Eq.~\eqref{pCases}, which showcases how the corresponding polynomials $P_{\ww}(x,z)$ behave at small $z$. The identity \eqref{plimit} then allows us to derive  the leading order expressions for each of the terms in the relevant $X_{I}$, see Eq.~\eqref{def XI 4pt}. Roughly speaking, we find that all $X_I$ reduce to one of the  $Z_I$ factors appearing in the flowed three-point functions \eqref{3pt-final} that are expected to appear in a given contribution to the factorization limit. 
Indeed, at small $z$, we get
\begin{align}
    &X_1 \sim z^{-\frac{(\w_1+1)\w_4}{2}}x^{\w_4}\off{Q_{(\w_1-\w_4+1,\w_2,\w_3)}+Q_{(\w_1-\w_4-1,\w_2,\w_3)}(-z)^{\w_4}y_1}  \nn \, ,
    \\[1ex] &X_{2}\sim z^{-\frac{\w_1\w_4}{2}}x^{\w_4}\off{Q_{(\w_1-\w_4,\w_2+1,w_3)}+Q_{(\w_1-\w_4,\w_2-1,w_3)}y_2}  \, , \\[1ex]
    & X_{3} \sim 
    z^{-\frac{\w_1\w_4}{2}}x^{\w_4}
    \off{Q_{(\w_1-\w_4,\w_2,w_3+1)}+Q_{(\w_1-\w_4,\w_2,w_3-1)}y_3} 
    \, , \nn\\[1ex]
    &X_4 \sim z^{-\frac{\w_1(\w_4+1)}{2}}x^{\w_4+1}\off{Q_{(\w_1-\w_4-1,\w_2,\w_3)}-Q_{(\w_1-\w_4+1,\w_2,\w_3)}z^{\w_1} x^{-2} y_4} \, , \nn
\end{align}
with the $Q_{\ww}$ defined in \eqref{Qw-definition}, while 
\begin{align}
   & X_{123}\sim z^{-\frac{(\w_1+1)\w_4}{2}}x^{\w_4} \left[
    Q_{(\w_1-\w_4+1,\w_2+1,\w_3+1)}
    + Q_{(\w_1-\w_4-1,\w_2+1,\w_3+1)} (-z)^{\w_4}y_1 \right.  \nn \\
    & \quad 
    + Q_{(\w_1-\w_4+1,\w_2-1,\w_3+1)} y_2
    + Q_{(\w_1-\w_4+1,\w_2+1,\w_3-1)} y_3
    + Q_{(\w_1-\w_4-1,\w_2-1,\w_3+1)} 
    (-z)^{\w_4}y_1 y_2 \nn \\
    & \quad + Q_{(\w_1-\w_4+1,\w_2-1,\w_3-1)}
    y_2 y_3
    + Q_{(\w_1-\w_4-1,\w_2+1,\w_3-1)}
    (-z)^{\w_4}y_1 y_3 \nn \\
    & \quad + \left. Q_{(\w_1-\w_4+1,\w_2+1,\w_3+1)}
    (-z)^{\w_4}y_1 y_2 y_3\right]
    \, ,
\end{align}
and 
\begin{align}
    &X_{134}\sim z^{-\frac{(\w_1+1)(\w_4+1)}{2}}x^{\w_4+1} \Big[ \left(1+(-z)^{\w_4}y_1 z^{\w_1}x^{-2}y_4\right) 
    \left(Q_{(\w_1-\w_4,\w_2,\w_3+1)}\right. \nn \\
    & \quad + \left. Q_{(\w_1-\w_4,\w_2,\w_3-1)} y_3 \right)+ z^{\w_1+1} x^{-2}y_4 \of{Q_{(\w_1-\w_4+2,\w_2,\w_3+1)}+Q_{(\w_1-\w_4+2,\w_2,\w_3-1)}y_3}\nn\\
    & \quad - (-z)^{\w_4+1}y_1\of{Q_{(\w_1-\w_4-2,\w_2,\w_3+1)}+Q_{(\w_1-\w_4-2,\w_2,\w_3-1)}y_3}\Big].
\end{align}
In what follows we will show that, as anticipated in Sec.~\ref{sec: lessons}, from this expansion we can read off contributions to the factorized expansion of the flowed four-point functions associated with the exchange of flowed states with charges $\w= \w_1-\w_4 + \delta$ with $\delta = -1,0,+1$.

Let us start with the case where the three-point function on the left of Fig.~\ref{fig: Basic diagram} is, in the language of  \cite{Dei:2021xgh,Bufalini:2022toj}, of the even edge type, namely when the intermediate state has $\w= \w_1-\w_4$. Importantly, the three-point function on the RHS remains unconstrained, except for the fact that it must be an odd-parity one. We will make use of the notation in Eq.~\eqref{3pt-final} for the $y_i$-dependent factors involved in the latter three-point function. By rescaling 
\begin{equation}
y_1\rightarrow (-z)^{-\w_4}y_1,\qquad y_4\rightarrow z^{-\w_1}x^2 y_4\, ,
\end{equation}
and working at first non-trivial order in $z$, we find that the above expressions give 
\begin{equation}
    \left[X_1,X_2,X_3,X_4\right] \to 
     z^{-\frac{\w_1\w_4}{2}}x^{\w_4}
       \left[ z^{-\frac{\w_4}{2}}Z_{1}(y_1),Z_{2}(y_2),Z_{3}(y_3), - x z^{-\frac{\w_1}{2}} y_4 Z_{1}(-y_4^{-1})\right] \, , 
\end{equation}
while 
\begin{equation}
    X_{123} \to z^{-\frac{(\w_1+1)\w_4}{2}} x^{\w_4+1} Z_{123}(y_1,y_2,y_3) \, , 
\end{equation}
and 
\begin{equation}
       X_{134} \to z^{-\frac{(\w_1+1)(\w_4+1)}{2}} x^{\w_4+1} (1+ y_1 y_4) Z_3 (y_3)\, . 
\end{equation}
The overall powers of $x$ and $z$ cancel out in the generalized cross-ratio, giving 
\begin{equation}
    \frac{X_2X_{134}}{X_4X_{123}} \to \frac{(-1)^{\w+\w_2+1}\of{1+y_1y_4} Z_2(y_2)Z_3(y_3)}{y_4 Z_1(-y_4^{-1}) Z_{123}(y_1,y_2,y_3)} \, .
\end{equation}
Here, the $Z_I$ functions are those corresponding to a spectrally flowed three-point function with charges $(\w,\w_2,\w_3)$. Note that, somewhat surprisingly , either $y_1$ or $y_4^{-1}$ seem to play a role analogous to a putative $y$-variable associated to the intermediate state. Finally, combining these results with the leading order expression for the unflowed correlator appearing in \eqref{odd4pt}, we get
\begin{equation}
    \fdot{\mathcal{F}}_{\ww}(x,z) \sim |x^{-2h_4}z^{h_1\w_4+h_4\w_1-\frac{k}{2}\w_1\w_4}|^2\int \prod_i d^2y_i |y_i^{j_i + \frac{k}{2}\w_i-h_i-1}|^2\mathcal{G}(y_i,z) \,, 
    \label{F general odd 1}
\end{equation}
where, at small $z$ and for external spins in the range \eqref{0010j3range}, we have 
\begin{equation}
    \mathcal{G}(y_i,z) \sim \int_{\frac{1}{2}+i\R} dj {\cal{C}}(j)|z^{\Delta_j^{(0)}-\Delta_{1}^{(0)}-\Delta_{4}^{(0)}} \hat{\mathcal{G}}(y_i)|^2 \, , 
\end{equation}
with
\begin{align}
 \hat{\mathcal{G}}(y_i ) &=  Z_1(y_1)^{-j_1+j_2+j_3+j_4-\frac{k}{2}}Z_{2}(y_2)^{j-j_2+j_3-\frac{k}{2}} Z_{3}(y_3)^{j+j_2-j_3-\frac{k}{2}}\nn\\
    &\quad \times Z_{123}(y_1,y_2,y_3)^{\frac{k}{2}-j-j_2-j_3} \off{y_4Z_{1}(-y_4^{-1})}^{j_1-j-j_4}(1+y_1y_4)^{j-j_1-j_4} 
    \label{def Ghat odd}\\
    & \quad  \times \pFq{2}{1}{j-j_1+j_4,j+j_2+j_3-\frac{k}{2}}{2j}{\frac{(-1)^{s}(1+y_1y_4)}{y_4Z_{1}(-y_4^{-1})}\frac{Z_{2}(y_2)Z_3(y_3)}{Z_{123}(y_1,y_2,y_3)}} \, , \nn
\end{align}
and 
\begin{equation}
    {\cal{C}}(j) = \frac{C(j_1,j_4,j) \mathcal{N}(j_3) C(j,j_2,\frac{k}{2}-j_3)}{B(j)} \, .
\end{equation}
As it happened in the sample cases studied in Sec.~\ref{sec: unflowedcases}, the overall power of $x$ shows that the $z\to 0$ limit has fixed the spacetime spin of the exchanged state to be 
\begin{equation}
    h = h_1-h_4 \, .  
\end{equation}
Moreover, the remaining integral is independent of $x$, which is consistent with the fact that the hypergeometric function in \eqref{general factorization} trivializes for this particular value of $h$. 
This is also consistent with the power of $z$ in Eq.~\eqref{F general odd 1} since,   
\begin{equation}
    \Delta_j^{(0)}-\Delta_{1}^{(0)}-\Delta_{4}^{(0)}+h_1\w_4+h_4\w_1-\frac{k}{2}\w_1\w_4 = \Delta_j\Big|^{\w=\w_1-\w_4}_{h=h_1-h_4}-\Delta_{1}-\Delta_{4} \, .
\end{equation}

In order to argue that this is consistent with the expected factorization structure in Eq.~\eqref{general factorization} we need to show that it matches the relevant product of three-point functions. 
As before, we proceed by reverse engineering. In the present context, the starting point corresponds to 
\begin{equation}
 \frac{C_{(\w_1,\w_4,\w_1-\w_4)}(j_i,h_i)C_{(\w_1-\w_4,\w_2,\w_3)}(j_i,h_i)}{R(j,h_1-h_4,\w_1-\w_4)}\,.  
\end{equation}    
Following the analysis of Sec.~\ref{sec: lessons}, and by means of the identity given in Eq.~\eqref{3ptproductId}, it suffices to show that 
\begin{align}
{\cal{I}}_{\text{odd}}     & \equiv \int d^2y \Big|(1+y y_4)^{j_1+j-j_4-1}(y-y_1)^{j_4+j-j_1-1}(1+y_1y_4)^{1-j_1-j_4-j}\times \label{oddIntegral}\\
    &\times Z_{123}^{\frac{k}{2}-j-j_2-j_3}\of{y,y_2,y_3}Z_{1}^{j_3+j_2-j-\frac{k}{2}}\of{y}Z_{2}^{j+j_3-j_2-\frac{k}{2}}(y_2)Z_{3}^{j+j_2-j_3-\frac{k}{2}}(y_3)\Big|^2  \nn 
\end{align}
integrates to $\hat{\mathcal{G}}(y_i)$ up to the appropriate multiplicative factor depending only on the unflowed spins. Roughly speaking, in Eq.~\eqref{oddIntegral} we have included the $y$-variable of the intermediate state. As discussed in Sec.~\eqref{sec: lessons}, the $Z_I$ do not take the form of the usual differences $y_{ij}$ and contain complicated numerical factors coming from the relevant holomorphic covering maps  \cite{Dei:2021xgh,Bufalini:2022toj}, but they remain at most linear functions of each of the $y_i$. Since their powers involve the usual combinations of the external spins $j_i$ and of the  intermediate spin $j$ (up to shifts of the form $j \to \frac{k}{2}-j$), this allows us to carry out the integral over $y$ as in App.~\ref{sec: App A - Integrals}, leading once again to a result that can be expressed in terms of hypergeometric functions of the ${}_2F_1$ type. More precisely, denoting 
\begin{align}
    Z_{123}(y,y_2,y_3) \equiv A^+(y_2,y_3) + A^-(y_2,y_3) y\qqquad Z_{1}(y) \equiv B^++B^-y,
\end{align}
we find that the resulting hypergeometric function is evaluated at 
\begin{equation}
    \frac{(1+y_1y_4)}{y_4Z_1\of{-y_4^{-1}}}\frac{\off{A^-(y_2,y_3)B^+-A^+(y_2,y_3)B^-}}{Z_{123}\of{y_1,y_2,y_3}} \, .
\end{equation}
Finally, from the explicit expressions for $A^{\pm}(y_2,y_3)$ and $B^{\pm}$ one can check that
\begin{equation}
    A^-(y_2,y_3)B^+-A^+(y_2,y_3)B^- = (-1)^{\w+\w_2}Z_2\of{y_2}Z_3\of{y_3} \,, 
\end{equation}
hence reproducing the (small $z$ limit of the) generalized cross-ratio appearing in \eqref{def Ghat odd}. The precise matching is given by 
\begin{equation}
    C\left(j,j_2,\frac{k}{2}-j_3\right)\mathcal{N}(j_3)C(j_1,j_4,1-j){\cal{I}}_{\text{odd}} = {\cal{C}}(j)|\hat{\mathcal{G}} (y_i)|^2 \, . 
\end{equation}
We have thus recovered the expected factorization structure. Moreover, one of the relevant three-point functions, namely that on the RHS of Fig.~\ref{fig: Basic diagram}, involves the full complexity of the exact odd-parity spectrally flowed three-point functions of the SL(2,$\R$) WZW model, derived recently in \cite{Dei:2021xgh,Iguri:2022eat,Bufalini:2022toj}. 

For the four-point function under consideration, we can also isolate the contribution coming from the exchange of a flowed state with $\w = \w_1 - \w_4 - 1$. Indeed, by rescaling 
\begin{equation}
    y_1 \rightarrow (-z)^{-\w_4-1} y_1, \qquad y_4 \rightarrow z^{-\w_1+1}x^2 y_4\, ,
\end{equation} 
we find that generalized cross-ratio goes to
\begin{align}
    \frac{X_2 X_{134}}{X_4 X_{123}} \to z \frac{A^+(y_2)\off{(1+y_1y_4)B^+(y_3)-y_1B^-(y_3)}}{(-1)^{\w+\w_2+1} y_1 Z_{\emptyset}Z_{23}(y_2,y_3)},
\end{align}
where now the $Z_I$ are those corresponding to an even spectrally flowed three-point function with charges $(\w,\w_2,\w_3)$. The factors $A^{\pm}$ and $B^{\pm}$ are defined accordingly by
\begin{equation}
    Z_{12}(y,y_2) = A^{+}(y_2)+A^-(y_2)y, \qquad Z_{13}(y,y_3) =B^{+}(y_3)+B^-(y_3)y \,.
\end{equation}
This time the leading order cross-ratio is proportional to $z$, hence the factorization limit is governed by Eq.~\eqref{zxexpansion} whenever the external spins are in the range \eqref{0010j1range}. Therefore the corresponding contribution to the factorized form of the four-point function is given by 
\begin{align}
    \fdot{\mathcal{F}}_{\ww}(x,z)\sim |x^{-2h_4}z^{h_1(\w_4-1)+h_4(\w_1+1)-\frac{k}{2}(\w_1\w_4+\w_1-\w_4)-j_1-j_4}|^2\int \prod_i d^2y_i |y^{j_i+\frac{k}{2}\w_i-h_i-1}|^2\mathcal{G}(y_i,z),
\end{align}
where
\begin{equation}
    \mathcal{G}(y_i,z) \sim \int_{\frac{1}{2}+i\R} dj \mathcal{C}(j)|z^{\Delta^{(0)}_j-\Delta^{(0)}_1-\Delta^{(0)}_4-j_1-j_4+\frac{k}{4}}\hat{\mathcal{G}}(y_i)|^2
    \,,
\end{equation}
with
\begin{align}
 \hat{\mathcal{G}}(y_i) &=  Z_{23}(y_2,y_3)^{j-j_2-j_3}A^+(y_2)^{j_3-j_2-j}B^+(y_3)^{j_1+j_4+j_2-j_3-\frac{k}{2}}Z_{\emptyset}^{j+j_2+j_3-k}\nn\\
    &\quad \times \off{(1+y_1y_4)B^+(y_3)-y_1B^-(y_3)}^{\frac{k}{2}-j-j_1-j_4} y_1^{j+j_4-j_1-\frac{k}{2}}
    \label{def Ghat odd 2}\\
    & \quad  \times \pFq{2}{1}{j+j_1+j_4-\frac{k}{2},j-j_3+j_2}{2j}{\frac{(-1)^{\w+\w_2+1} y_1 Z_{\emptyset}Z_{23}(y_2,y_3)}{A^+(y_2)\off{(1+y_1y_4)B^+(y_3)-y_1B^-(y_3)}}} \, . \nn
\end{align}
and 
\begin{equation}
    \mathcal{C}(j) = \frac{\mathcal{N}(j_1) C(\frac{k}{2}-j_1,j_4,j)C(j,j_2,j_3)}{B(j)} \,  . 
\end{equation}
The overall $x$-power again shows that we have picked up an intermediate spacetime spin $h = h_1-h_4$, and the remaining $z$-dependence is also consistent since
\begin{equation}
    \Delta^{(0)}_j-\Delta^{(0)}_1-\Delta^{(0)}_4+h_1(\w_4-1)+h_4(\w_1+1)-\frac{k}{2}(\w_1\w_4+\w_1-\w_4)+\frac{k}{4} = \Delta_j\Big|_{h=h_1-h_4}^{\w=\w_1-\w_4-1} -\Delta_{1} - \Delta_{4}\, . 
\end{equation}
As in the previous case, it it possible to show that the entire function matches the product
\begin{equation}
    \frac{C_{(\w_1,\w_4,\w_1-\w_4-1)}(j_i,h_i)C_{(\w_1-\w_4-1,\w_2,\w_3)}(j_i,h_i)}{R(j,h_1-h_4,\w_1-\w_4-1)} 
\end{equation}
by integrating an expression analogous to Eq.~\eqref{oddIntegral}. Since the four-point function spectral flow parity has not changed, the three-point function on the RHS must now be of the even-parity type. The corresponding integral takes the form
\begin{align}
    \mathcal{I}_{\rm odd}' = &\int d^2y\Big| y_1^{j+j_4-j-\frac{k}{2}}(1+y_1y_4+y_1y)^{\frac{k}{2}-j-j_1-j_4} Z_{\emptyset}^{1-j+j_2+j_3-k}\\
   & \quad \times y^{2j-2} Z_{23}(y_2,y_3)^{1-j-j_2-j_3} Z_{13}(y^{-1},y_3)^{j_2+j-j_3-1}Z_{12}(y^{-1},y_2)^{j_3+j-j_2-1}\Big|^2 \, .\nn
\end{align}
Indeed, this leads exactly to the $\hat{\mathcal{G}}(y_i)$ in Eq.~\eqref{def Ghat odd 2} by means of the identity
\begin{equation}
    (-1)^{\w+\w_2+1}Z_{\emptyset}Z_{23}(y_2,y_3) = A^{+}(y_2)B^-(y_3)-A^{-}(y_2)B^+(y_3) \, .
\end{equation}

The necessary steps to isolate the $\w = \w_1-\w_4+1$ intermediate channel also follow analogously by rescaling 
\begin{equation}
y_1 \rightarrow (-z)^{-\w_4+1}y_1,\qquad y_4 \rightarrow z^{-\w_1-1}x^2 y_4\,.     
\end{equation}
In the small $z$ limit, the generalized cross-ratio now gives
\begin{align}
    \frac{X_{2}X_{134}}{X_{4}X_{123}} \to z \frac{A^-(y_2)\off{(1+y_1y_4)B^-(y_3)+y_4B^+(y_3)}}{y_4 Z_{\emptyset}Z_{23}(y_2,y_3)},
\end{align}
where the $Z_I$ corresponds to the factors involved in the three-point function with charges $\ww = (\w_1-\w_4+1,\w_2,\w_3)$, and with  
\begin{equation}
    Z_{12}(y,y_2)A^{+}(y_2)+A^-(y_2)y,\qquad Z_{13}(y,y_3) = B^+(y_3) + B^-(y_3)y.
\end{equation}
The four-point function limit then becomes 
\begin{align}
    \fdot{\mathcal{F}}_{\ww}(x,z) &= \int_{\frac{1}{2}+i\R} dj \mathcal{C}(j)|z^{\Delta_j-\Delta_1-\Delta_4}x^{-2h_4}|^2\int \prod_{i=1}^4d^2y_i \Bigg|y_i^{h_i^0-h_i-1}\nn\\
    &\times Z_{\emptyset}^{j+j_2+j_3-k}Z_{23}^{j-j_2-j_3}A^-(y_2)^{j_3-j-j_2}B^-(y_3)^{j_1+j_2-j_3+j_4-\frac{k}{2}}\\
    &\times y_4^{j_1+j-j_4-\frac{k}{2}}\of{(1+y_1y_4)B^-(y_3)+y_4B^+(y_3)}^{\frac{k}{2}-j_1-j_4-j}\nn\\
    &\times \pFq{2}{1}{j+j_1+j_4-\frac{k}{2},j-j_3+j_2}{2j}{\frac{y_4 Z_{\emptyset}Z_{23}(y_2,y_3)}{A^-(y_2)\of{(1+y_1y_4)B^-(y_3)+y_4B^+(y_3)}}}\Bigg|^2,\nn
    \end{align}
where $\Delta_j = \Delta_j\Big|^{\w=\w_1-\w_4+1}_{h=h_1-h_4}$. Again, the normalization matches the product
\begin{equation}
    \frac{C_{(\w_1,\w_4,\w_1-\w_4+1)}(j_i,h_i)C_{(\w_1-\w_4+1,\w_2,\w_3)}(j_i,h_i)}{R(j,h_1-h_4,\w_1-\w_4+1)},
\end{equation}
which is now computed from the integral
\begin{align}
    \mathcal{I}_{\text{odd}}=&\int d^2y \Big| y_4^{j+j_1-j_4-\frac{k}{2}}y^{j_1+j_4-j-\frac{k}{2}}(y_4+(1+y_1y_4)y)^{\frac{k}{2}-j-j_1-j_4} Z_{\emptyset}^{1-j+j_2+j_3-k}\\
   & \quad \times y^{2j-2} Z_{23}(y_2,y_3)^{1-j-j_2-j_3} Z_{13}(y^{-1},y_3)^{j_2+j-j_3-1}Z_{12}(y^{-1},y_2)^{j_3+j-j_2-1}\Big|^2 \, .\nn
\end{align}
Let us stress that the matchings we have obtained for $\w = \w_1-\w_4\pm 1$ involves all the even-parity  spectrally flowed three-point functions. 
This ends our analysis of four-point functions satisfying \eqref{wCaseOdd1}. 

For completeness, let us now consider spectral flow charges satisfying 
\begin{equation}
\label{wCaseOdd2}
    \w_2-\w_3>|\w_1-\w_4| \, . 
\end{equation}
Following a similar path as in the previous cases and using the rescalings 
\begin{alignat}{2}
&y_1 \rightarrow \frac{xy_1 }{z^{\w_1}(1-x)} , \qquad && y_2 \rightarrow (-1)^{\w_3}\frac{(1-x)z^{\w}y_2}{x} , \\
&y_3 \rightarrow -\frac{x y_3}{(1-x)z^{\w}} , \qquad &&y_4 \rightarrow \frac{x(1-x) y_4}{z^{\w_4}} , \nn
\end{alignat} 
now leads to a small $z$ expression of the form 
\begin{align}
    \fdot{\mathcal{F}}_{\ww}(x,z) \sim & \, |x^{h-h_1-h_4}(1-x)^{h_1-h_4-h} z^{j_1+j_4-\frac{k}{4}-h\w+h_1\w_1+h_4\w_4+\frac{k}{4}\of{\w^2-\w_1^2-\w_4^2}}|^2\nn\\
    &\times\int \prod_i d^2y_i |y_i^{j_i + \frac{k}{2}\w_i-h_i-1}|^2\mathcal{G}(y_i,z) \, .
\end{align}
The overall powers of $x$ and $z$ now  correspond to the exchange of an intermediate state with 
\begin{equation}
    \w = \w_2-\w_3 \, , \qquad  h = h_2-h_3 \, . 
\end{equation}
Note that one important difference is that there is also a factor of $(1-x)^{h_1-h_4-h}$. This is consistent with the fact that for $h = h_2-h_3$ the hypergeometric function analogous to \eqref{unflowed 2F1} but with the external $j_i$ replaced by $h_i$ simplifies, but it does not trivialize completely. 
Moreover, in this case the generalized cross-ratio becomes 
\begin{equation}
 \frac{X_2 X_{134}}{X_4 X_{123}} \to  z (-1)^{\w+\w_1}\frac{y_3Z_{3}(y_2)Z_{123}(y_1,y_4,y_3^{-1})}{Z_{2}(y_4)Z_{1}(y_1)(1-y_2y_3)} \, , 
\end{equation}
with $Z_I(y_i)$ the factors involved in the LHS three-point function with $\ww = (\w_1,\w_4,\w)$. 
For spins in the range \eqref{0010j1range} we can thus  use Eq.~\eqref{zxexpansion} to provide the leading order expression for the relevant unflowed correlator, leading to
\begin{equation}
    \mathcal{G}(y_i,z) \sim \int_{\frac{1}{2}+i\R} dj {\cal{C}}(j)|z^{\Delta_j^{(0)}-\Delta_{1}^{(0)}-\Delta_{4}^{(0)}-j_1-j_4+\frac{k}{4}} \hat{\mathcal{G}}(y_i)|^2 \, , 
\end{equation}
which gives the expected power of $z$, where 
\begin{align}
   \hat{\mathcal{G}}(y_i) & = \of{1-y_2y_3}^{j-j_2-j_3}Z_{1}(y_1)^{j_4+j-j_1-\frac{k}{2}}\of{y_3Z_{123}(y_1,y_4,y_3^{-1})}^{\frac{k}{2}-j-j_1-j_4}\nn\\
    &\times Z_{2}(y_4)^{j_1+j-j_4-\frac{k}{2}}Z_3(y_2)^{j_3-j_2-j}\of{y_3Z_3\of{y_3^{-1}}}^{j_1+j_2-j_3+j_4-\frac{k}{2}}\\
    &\times \pFq{2}{1}{j_1+j_4+j-\frac{k}{2},j-j_3+j_2}{2j}{ (-1)^{\w+\w_1}\frac{Z_{2}(y_4)Z_{1}(y_1)(1-y_2y_3)}{y_3Z_{3}(y_2)Z_{123}(y_1,y_4,y_3^{-1})}}.\nn
\end{align}
and 
\begin{equation}
    \mathcal{C}(j) = \frac{\mathcal{N}(j_1) C(\frac{k}{2}-j_1,j_4,j)C(j,j_2,j_3)}{B(j)} \,  . 
\end{equation}
After taking $y_2 \rightarrow -y_2^{-1}$ and $y_3 \rightarrow y_3^{-1}$, the $\hat{\mathcal{G}}(y_i)$ we have obtained takes exactly the same form as in Eq.~\eqref{def Ghat odd}, hence it can be obtained from the product of the $y$-basis three-point functions in an analogous way. As before, one can also derive the contributions associated with exchanged states of winding $\w = \w_2 - \w_3 \pm 1$ (and $h = h_2-h_3$) in a similar manner. This completes our study of the small $z$ behavior and  factorization properties of spectrally flowed four-point functions of the odd-parity type.

The even-parity cases, i.e.~the four-point functions described by \eqref{even4pt}, can be studied in the same way. 
Let us briefly discuss one example for completeness. We assume that 
\begin{align}
     \w_3-\w_2>|\w_1-\w_4| \, .
\end{align}
After taking
\begin{equation}
y_1 \rightarrow \frac{x}{z^{\w_1}} y_1\, , \qquad 
y_2 \rightarrow (-1)^{\w}\frac{x}{z^{\w}} y_2
\, , \qquad 
y_3 \rightarrow \frac{z^{\w}}{x} y_3 \, , \qquad 
y_4 \rightarrow \frac{x}{z^{\w_4}} y_4 \, ,
\end{equation}
the relevant small $z$ contribution describes the exchange of a state with 
\begin{equation}
    \w = \w_3-\w_2 \, , \qquad h = h_3 - h_2 \, .
\end{equation}
More explicitly, this  takes the form 
\begin{equation}
    \fdot{\mathcal{F}}_{\ww}(x,z)  \sim \int_{\frac{1}{2}+i\R} dj  |x^{h-h_1-h_4}z^{\Delta_{j}-\Delta_{j_1}-\Delta_4}|^2 \mathcal{C}(j)  \int \prod_i d^2y_i |y_i^{j_i + \frac{k}{2}\w_i-h_i-1}\hat{\mathcal{G}}(y_i)|^2,\label{4ptoddLimit3}
\end{equation}
with 
\begin{align}
    \hat{\mathcal{G}}(y_i)&=Z_{\emptyset}^{j+j_1+j_4-k}(1+y_2y_3)^{j-j_2-j_3}\off{y_2Z_{13}\of{y_1,-y_2^{-1}}}^{j_3-j_2-j}\nn\\
    &\times Z_{13}^{j_4-j_1-j_3+j_2}(y_1,y_3)Z_{23}(y_4,y_3)^{j_1-j-j_4}Z_{12}(y_1,y_4)^{j-j_1-j_4} \\
    &\times \pFq{2}{1}{j-j_1+j_4,j-j_3+j_2}{2j}{\frac{(-1)^{\w+\w_1}(1+y_2y_3)Z_{\emptyset}Z_{12}(y_1,y_4)}{y_2Z_{13}(y_1,-y_2^{-1})Z_{23}(y_1,y_3)}} \, . \nn
\end{align}
and 
\begin{equation}
    \mathcal{C}(j) = \frac{C(j_1,j_4,j)C(j,j_2,j_3)}{B(j)}
    \, .
\end{equation}
As usual, this can be obtained from the integral associated to the relevant product of three-point functions, namely 
\begin{align}
    {\cal{I}}_{\text{even}} = &\int d^2y \Big|(1+y y_2)^{j_3+j-j_2-1}(y-y_3)^{j_2+j-j_3-1}(1+y_2y_3)^{1-j_2-j_3-j} \label{evenIntegral}\\
    &\times Z_{23}^{j_1-j-j_4}\of{y_4,y}Z_{13}^{j_4-j_1-j}\of{y_1,y}Z_{12}^{j-j_1-j_4}(y_1,y_4)Z_{\emptyset}^{j+j_1+j_4-k}\Big|^2 \, . \nn
\end{align}
The matching holds thanks to the identity  
\begin{align}
    A^-(y_4)B^{+}(y_1)- A^+(y_4)B^{-}(y_1) = (-1)^{\w+\w_1}Z_{\emptyset}Z_{12}(y_1,y_4) \, ,
\end{align}
where 
\begin{align}
    Z_{23}(y_4,y) = A^+(y_4)+A^-(y_4)y\, , \qquad  Z_{13}(y_1,y) = B^+(y_1)+B^-(y_1)y \, . 
\end{align}
The remaining contributions for these spectral flow charges and the rest of the cases in Eq.~\eqref{pCases} can be studied similarly.

\section{Concluding remarks and outlook}
\label{Sec7}

Let us briefly recapitulate what we have achieved in this paper. The WZW model based on the universal cover of SL(2,$\R$) at level $k$ describes string propagation in an AdS$_3$ background of radius  $\sqrt{k}$ in string units.  The importance of spectral flow in this context  was established a long time ago in \cite{Maldacena:2000hw} by solving a number of puzzles regarding the spectrum of the theory. Despite the fact that its role at the level of correlation functions was initially discussed shortly after in \cite{Maldacena:2001km}, structure constants involving states in the flowed sectors were derived in full generality only recently \cite{Dei:2021xgh,Iguri:2022eat,Bufalini:2022toj}. Four-point functions are even harder to study due to the intricate dependence on the worldsheet and boundary cross-ratios, usually denoted as $z$ and $x$, respectively. In \cite{Dei:2021yom} the authors conjectured a formula relating flowed four-point functions with arbitrary spectral flow charges with their unflowed counterparts \cite{Teschner:1999ug,Maldacena:2001km} by means of a complicated integral transform, see Eqs.~\eqref{even4pt} and \eqref{odd4pt}.

Unflowed four-point functions are not known in closed form, but only in a formal expansion in powers of $z$. This is the so-called factorization expansion, which accounts for the exchange of unflowed states along the $14 \to 23$ channel in the language of Fig.~\ref{fig: Basic diagram}. It thus makes sense to ask whether this can be used to show that the proposal of \cite{Dei:2021yom} leads to a consistent factorization picture in the flowed sectors of the theory, where the structure of the OPE has not been explored so far. This paper constitutes a first approach to this important question, which, if answered positively, would lend conclusive support to the above conjecture. 

On general grounds, the factorization structure is expected to be of the form discussed in Sec.~\eqref{sec: lessons}, see Eqs.~\eqref{general factorization} and \eqref{general block small z}. The difficulties we have encountered so far can be understood as follows. Flowed vertex operators have worldsheet conformal dimensions $\Delta$ that depend not only on their unflowed spin $j$ but also on the spacetime spin $h$ and the spectral flow charge. It follows that, for a given  value of the intermediate $\w$, the small $z$ limit we have used to study the flowed four-point functions interacts non-trivially with the sum over the quantum number $h$ associated to the exchanged state. In terms of the integral transform from the $y$-basis to the $x$-basis which captures the effect of spectral flow, the consequence is that different scalings of the integration variables $y_i$ with respect to $z$ correspond to different channels in the conformal block decomposition. 

Nevertheless, we have been able to consider four-point functions with arbitrary values of the external $\w_i$ and study in detail all exchanges of unflowed states, together with an important number of flowed channels. We have shown that, at small $z$, in all cases one can isolate several contributions to the factorization expansion that are precisely consistent with the structure anticipated in Eq.~\eqref{general factorization}. Let us stress that this matching involves the full complexity of the spectrally flowed  three-point functions of the model. 
We see this as providing substantial evidence for the general formulae in Eqs.~\eqref{even4pt} and \eqref{odd4pt}. 

Of course, this is not the full story. We leave for the near future the possibility of computing the total sum over all allowed intermediate states. Further integrating over $z$, at least at small $x$, would allow us to study the problem at hand from the point of view of the OPE structure and conformal block decomposition of the holographic CFT. This could help establishing the duality proposed in \cite{Eberhardt:2021vsx} beyond the perturbative analysis developed recently in \cite{Dei:2022pkr,Knighton:2023mhq,Knighton:2024qxd}.%
 It would also be interesting to extend the present analysis to the supersymmetric AdS$_3 \times S^3 \times T^4$ model \cite{Dabholkar:2007ey,Giribet:2007wp,Pakman:2009zz,Cardona:2010qf,Kirsch:2011na,Iguri:2022pbp,Iguri:2023khc}. 
Finally, let us highlight the fact that the ability to compute the worldsheet correlators in the SL(2,$\R$) model also leads to applications for black hole phenomenology \cite{Witten:1991yr,Banados:1992wn,Maldacena:2000kv}, most notably in the context of the Fuzzball program and holography beyond AdS \cite{Lunin:2001jy,Martinec:2017ztd,Martinec:2018nco,Martinec:2019wzw,Martinec:2020gkv,Martinec:2022okx,Bufalini:2021ndn,Bufalini:2022wyp,Bufalini:2022wzu,Kutasov:2001uf,Israel:2003ry,Giveon:2017nie,Asrat:2017tzd,Giveon:2017myj,Giribet:2017imm,Chakraborty:2019mdf,Apolo:2019zai,Cui:2023jrb,Georgescu:2024iam}.

\acknowledgments

It is a pleasure to thank Emiliano Barone, Silvia Georgescu, Bob Knighton and Vit Sriprachyakul for discussions. The work of S.I.~and J.H.T.~is supported by CONICET.  
The work of N.K.~was partly supported by the ERC Consolidator Grant 772408-Stringlandscape.

\appendix 
\section{Useful identities}
\label{sec: app A}
%%%%%%%%%%%%%%%%%%%%%%%%
\subsection{Complex integrals}
\label{sec: App A - Integrals}
Here we record some identities related to complex integrals. We start with 
\begin{align}
    \int d^2 y y^{a-1} \bar{y}^{\bar{a}-1} (1-y)^{b-1}(1-\bar{y})^{\bar{b}-1}=\frac{\pi \gamma(a)\gamma(b)}{\gamma(a+b)}\,,
\end{align}
where 
\begin{align}
    \gamma(x) = \frac{\Gamma(x)}{\Gamma(1-\bar{x})}.
\end{align}
Along the paper, we also make use of the following generalized version of this formula: 
\begin{equation}
    \int \prod_{i=1}^n d^2 y_i y_i^{a_i-1} \bar{y}_i^{\bar{a}_i-1} \left(1-\sum_{j=1}^n y_j \right)^{b-1}\left(1-\sum_{j=1}^n \bar{y}_j \right)^{\bar{b}-1}=\frac{\pi^n \gamma(b)\prod_{i=1}^n \gamma(a_i)}{\gamma\left(b+\sum_{i=1}^n a_i\right)}\, .
\end{equation}
Another useful identity is given by \cite{Maldacena:2001km}  
\begin{align}
    \Bigg|\pFq{2}{1}{a,b}{c}{x}\Bigg|^{2}+|x^{1-c}\lambda &\Bigg|\pFq{2}{1}{1-c+a,1-c+b}{2-c}{x}\Bigg|^{2}=\nn\\
    &\frac{\pi \gamma(c)}{\gamma(b)\gamma(c-b)}\int d^2t|t^{b-1}(1-t)^{c-b-1}(1-xt)^{-a}|^2,
\end{align}
where
\begin{equation}
    \lambda = - \frac{\gamma(c)^2\gamma(a-c+1)\gamma(b-c+1)}{(1-c)^2\gamma(a)\gamma(b)}.
\end{equation}
For $a= j+\alpha$, $b = j+\beta$ and $c = 2j$, the above expression becomes symmetric under $j\rightarrow 1-j$, hence for any function $g(j)$ invariant under this reflection we can write 
\begin{align}
    \int_{\frac{1}{2}+i \RR} dj & g(j)\Bigg|\pFq{2}{1}{j+\alpha,j+\beta}{2j}{x}\Bigg|^{2} =\nn\\ &\int_{\frac{1}{2}+i \RR} dj g(j) \frac{\pi \gamma(2j)}{\gamma(j+\beta)\gamma(j-\beta)}\int d^2t |t^{j+\beta-1}(1-t)^{j-\beta-1}(1-xt)^{-j-\alpha}|^2.
\end{align}
A related identity is the Appell function-type integral, given by 
\begin{align}
    \int d^2u |u^{\alpha-1}&(1-u)^{\beta+\beta'-\alpha-1}(1-qu)^{-\beta}(1-pu )^{-\beta'}|^2=\nn\\
    &|(1-p)|^{-2\alpha}\int d^2t \Bigg|t^{\alpha-1}(1-t)^{\beta+\beta'-\alpha-1}\left(1-\frac{q-p}{1-p}t\right)^{-\beta}\Bigg|^2\label{Appell},
\end{align}
which follows by taking 
\begin{equation}
    u = \frac{t}{1+p(t-1)}.
\end{equation}
Assuming $g(j) = g(1-j)$ as before, this implies 
\begin{align}
    \int_{\frac{1}{2}+i \RR} dj g(j)\Bigg|\pFq{2}{1}{j+\alpha,j+\beta}{2j}{\frac{q-p}{1-p}}\Bigg|^{2} =\\ \int_{\frac{1}{2}+i \RR} dj \frac{\pi \gamma(2j)|(1-p)|^{2(j+\beta)}}{\gamma(j+\beta)\gamma(j-\beta)}\int d^2u |u^{j+\beta-1}&(1-u)^{j-\beta-1}(1-qu)^{-(j+\alpha)}(1-pu )^{-(j-\alpha)}|^2\nn \, .
\end{align}
 
%%%%%%%%%%%%%%%%%%%%%%%%%
\subsection{The behaviour of $P_{\ww}(x,z)$ near $z=0$}\label{sec: PwAnalysis}
\label{sec: Pw at small z}

Here we discuss the small $z$ behaviour of the polynomials $P_{\ww}(x,z)$ involved in the integral formulas for spectrally flowed four-point functions \eqref{even4pt} and \eqref{odd4pt}. The corresponding definitions, provided in Eq.~\eqref{pfunction}, show that, at small $z$, there is an overall power-law dependence, accompanied by a factor $\tilde{P}_{\ww}(x,z)$ which is regular at $z=0$, 
\begin{equation}
    \tilde{P}_{\ww}(x,0) = x^{\Lambda_{\ww}}\tilde{P}_{\ww}(1,0),
\end{equation}
with $\Lambda_\w$ as in \eqref{lambda}.
More explicitly, we find 
\begin{align}
    P_{\ww}(x,z) &  \sim n(\boldsymbol{\w})\tilde{P}_{\boldsymbol{\w}}(1,0)\of{1-x}^{\frac{1}{2}s(\w_2+\w_4-\w_1-\w_3)} x^{\frac{1}{2}s(\w_1+\w_4-\w_2-\w_3)+\Lambda_{\ww}}\nn\\
    &\qquad \times z^{\frac{1}{4}s((\w_1+\w_2-\w_3-\w_4)(\w_2+\w_4-\w_1-\w_3))-\frac{1}{2}\w_1\w_4} \, .
\end{align}
The overall $z$ power is completely determined in terms of $\max{\off{|\w_1-\w_4|,|\w_3-\w_2|}}$ since 
\begin{align}
    s((\w_1+\w_2-\w_3-\w_4)(\w_2+\w_4-\w_1-\w_3)) = s\of{\of{\w_2-\w_3}^2-\of{\w_1-\w_4}^2}.
\end{align}
Once this is fixed, one further condition is necessary in order to fix the $x$-dependence. This can be written as follows:
\begin{align}
   \frac{P_{\ww}(x,z)}{n(\boldsymbol{\w})\tilde{P}_{\boldsymbol{\w}}(1,0)} \sim \left\{
   \begin{array}{ccc}
    z^{-\frac{\w_1\w_4}{2}} x^{\w_4}  & {\rm if} & \w_1-\w_4>|\w_3-\w_2| \, , \\[1ex]
    z^{-\frac{\w_1\w_4}{2}} x^{\w_1}\of{1-x}^{\frac{\w_2-\w_3-\w_1+\w_4}{2}} & {\rm if} & \w_4-\w_1>|\w_3-\w_2| \, , \\[1ex]
    z^{\frac{(\w_2-\w_3)^2}{4}-\frac{\w_1^2}{4}-\frac{\w_4^2}{4}} 
    x^{\frac{\w_2-\w_3+\w_1+\w_4}{2}} & {\rm if} & |\w_1-\w_4|<\w_3-\w_2 \, , \\[1ex]
   z^{\frac{(\w_2-\w_3)^2}{4}-\frac{\w_1^2}{4}-\frac{\w_4^2}{4}} 
    x^{\frac{\w_3-\w_2+\w_1+\w_4}{2}}\of{1-x}^{\frac{\w_2-\w_3-\w_1+\w_4}{2}} 
    & {\rm if} & |\w_1-\w_4|<\w_2-\w_3 \, .
   \end{array}
   \right.
\label{pCases}
\end{align}
Moreover, we have checked numerically  with the help of the ancillary \texttt{Mathematica} file provided in \cite{Dei:2021yom} that, up to an unimportant overall sign, the normalization $n(\boldsymbol{\w})\tilde{P}_{\boldsymbol{\w}}(1,0)$ can be expressed as 
\begin{equation}
    |n(\boldsymbol{\w})\tilde{P}_{\boldsymbol{\w}}(1,0)| = |Q_{\max{\off{|\w_1-\w_4|,|\w_3-\w_2|}},\w_2,\w_3}Q_{\max{\off{|\w_1-\w_4|,|\w_3-\w_2|}},\w_1,\w_4}| \,\label{plimit} \, .
\end{equation}
where $Q_{\boldsymbol{\w}}$ are the numbers involved in the $y$-basis flowed three-point functions \cite{Dei:2021xgh,Bufalini:2022toj}, see Eq.~\eqref{Qw-definition}. This relation is crucial when considering the factorization properties of the spectrally flowed four-point functions. 

%%%%%%%%%%%%%%%%%%%%%%%%%%%%%%%%%%%%%%%%%%%%%%%%%%%%%

\subsection{Useful properties of flowed three-point functions}
\label{Useful properties}

Here we will derive an integral representation of the product of three-point functions given by 
\begin{equation}
    \frac{\braket{V^{\w_1}_{j_1,h_1}(0,0)
    V^{\w_4}_{j_4,h_4}(x_4,1)
    V^{\w}_{j,h}(\infty,\infty)}\braket{
    V^{\w}_{j,h}(0,0)
    V^{\w_2}_{j_2,h_2}(1,1)
    V^{\w_3}_{j_3,h_3}(\infty,\infty)}}{\braket{
    V^{\w}_{j,h}(0,0)
    V^{\w}_{j,h}(\infty,\infty)}}  
    % \\ &\qquad \qquad \qquad = \frac{C_{(\w_1,\w_4,\w)}(j_1,j_4,j,h_1,h_4,h)C_{(\w,\w_2,\w_3)}(j,j_2,j_3,h,h_2,h_3)}{R(j,h,\w)}
    \, ,
\end{equation}
when one of the involved correlators is in the so-called collision limit, namely $x_4 \to 0$.

Let us  consider three-point functions satisfying either $\w_3 = \w_1+\w_2$ or 
$\w_3 = \w_1+\w_2\pm 1$. For such correlators taking $x_2 \rightarrow x_1$ leads to the conservation equation $h_3 = h_1+h_2$. In the even-parity case one then gets structure constants of the form  \cite{Dei:2021xgh,Bufalini:2022toj}
\begin{align}
    &\braket{
    V^{\w_1}_{j_1,h_1}(0,0)
    V^{\w_2}_{j_2,h_2}(0,1)
    V^{\w_3}_{j_3,h_3}(\infty,\infty)  }  \,  \\
    & = C(j_1,j_2,j_3) \int \prod_{i=1}^3 d^2y_i \, |y_i^{\alpha_i-1}( y_1 -y_2)^{j_3 -j_1 - j_2 }
    (1 -y_2 y_3 )^{j_1 - j_2 - j_3} 
    (1 -y_1 y_3)^{ j_2  - j_1  - j_3}|^2\, , \nn 
\end{align}
up to an overall sign, and where we have introduced the shorthand $\alpha_i= j_i - h_i +\frac{k}{2}\w_i$. By defining the function  
\begin{equation}
    \varphi_{0}(y_1)\equiv \int d^2y_2d^2y_3 |y_2^{\alpha_2-1} y_3^{\alpha_3-1}( y_1 -y_2)^{j_3 -j_1 - j_2 }
    (1 -y_2 y_3 )^{j_1 - j_2 - j_3} 
    (1 -y_1 y_3)^{ j_2  - j_1  - j_3}|^2\label{phi+}\,  ,
\end{equation}
one immediately notices that 
\begin{align}
    \varphi_0(y_1) = |y_1^{-j_1+h_3-h_2-\frac{k}{2}\w_1}|^2\varphi_0(1) \, .
    \label{prophomogeneity}
\end{align}
It follows that the structure constant can be expressed as 
\begin{align}
    &\braket{
    V^{\w_1}_{j_1h_1}(0,0)
    V^{\w_2}_{j_2h_2}(0,1)
    V^{\w_3}_{j_3h_3}(\infty,\infty)  }_{\rm even} = i(2\pi)^2 C(j_1,j_2,j_3) \delta^{(2)}(h_3-h_1-h_2)\varphi_0(1).
\end{align} 
Similarly, in the odd-parity case we have 
\begin{equation}
    \braket{
    V^{\w_1}_{j_1,h_1}(0,0)
    V^{\w_2}_{j_2,h_2}(0,1)
    V^{\w_3}_{j_3,h_3}(\infty,\infty)  }_{\rm odd} =   {\cal{N}}(j_1)C\left(\frac{k}{2}-j_1,j_2,j_3\right) \int d^2y_1 \,|y_1^{\alpha_1-1}|^2 \varphi_\pm(y_1) \, , \nn 
\end{equation}
with 
\begin{align}
    &\varphi_+(y_1) = \int d^2y_2d^2y_3 \,| y_2^{\alpha_2-1} y_3^{\alpha_3-1} \, y_3^{j_1 +j_2 - j_3-\frac{k}{2}}
    (1 -y_1 y_3 - y_2 y_3)^{\frac{k}{2}-j_1 - j_2 - j_3}|^2 \, , \\
    &\varphi_{-}(y_1) =\int d^2y_2d^2y_3 \, |y_2^{\alpha_2-1} y_3^{\alpha_3-1} \, y_2^{j_1+j_3-j_2-\frac{k}{2}}y_1^{j_2+j_3-j_1-\frac{k}{2}}(y_2+(1+y_3y_2)y_1)^{\frac{k}{2}-j_1-j_2-j_3}|^2.
\end{align}
Since $\varphi_\pm(y_1)$ satisfy \eqref{prophomogeneity} as well, we get 
\begin{align}
    &\braket{
    V^{\w_1}_{j_1h_1}(0,0)
    V^{\w_2}_{j_2h_2}(0,1)
    V^{\w_3}_{j_3h_3}(\infty,\infty)  }_{\rm odd} =\\
    &\hspace{5cm}i(2\pi)^2{\cal{N}}(j_1)C\left(\frac{k}{2}-j_1,j_2,j_3\right) \delta^{(2)}(h_3-h_1-h_2)\varphi_\pm(1).\nn
\end{align} 

From this we derive a useful identity involving the normalization of certain $x$-basis spectrally flowed conformal blocks. This reads
\begin{align}
 &  \frac{\braket{V^{\w_1}_{j_1,h_1}
    V^{\w_4}_{j_4,h_4}
    V^{\w}_{j,h_3-h_2}}\braket{
    V^{\w}_{j,h_3-h_2}
    V^{\w_2}_{j_2,h_2}
    V^{\w_3}_{j_3,h_3}}}{\braket{
    V^{\w}_{j,h_3-h_2}
    V^{\w}_{j,h_3-h_2}}}  \label{3ptproductId0} \\ &  
    \qquad = 
    B(1-j)\int d^2y d^2y' |(1-yy')^{2j-2}|^2\braket{V^{\w_1}_{j_1,h_1}
    V^{\w_4}_{j_4,h_4}
    V^{\w}_{j}(y)}\braket{
    V^{\w}_{j}(y')
    V^{\w_2}_{j_2,h_2}
    V^{\w_3}_{j_3,h_3}} \nn \, ,
\end{align}
where on both sides the  three-point  function on the left is evaluated at $(x_1,x_4,x)=(z_1,z_4,z)=(0,0,\infty)$ while the one on the right one is evaluated at $(x,x_2,x_3)=(0,1,\infty)$ and $(z,z_2,z_3)=(0,1,\infty)$.
Indeed, by using the reflection properties in Eqs.~\eqref{reflection w>0} and \eqref{reflection ybasis} we get 
\begin{align}
 &  \frac{\braket{V^{\w_1}_{j_1,h_1}
    V^{\w_4}_{j_4,h_4}
    V^{\w}_{j,h_3-h_2}}\braket{
    V^{\w}_{j,h_3-h_2}
    V^{\w_2}_{j_2,h_2}
    V^{\w_3}_{j_3,h_3}}}{\braket{
    V^{\w}_{j,h_3-h_2}
    V^{\w}_{j,h_3-h_2}}}  \label{sergito}\\[1ex] 
    &  \qquad = \braket{V^{\w_1}_{j_1,h_1}
    V^{\w_4}_{j_4,h_4}
    V^{\w}_{j,h_3-h_2}}\braket{
    V^{\w}_{1-j,h_3-h_2}
    V^{\w_2}_{j_2,h_2}
    V^{\w_3}_{j_3,h_3}} \nn \\[1ex]
    &\qquad = \int d^2y |y^{\alpha-2j}|^2 \braket{V^{\w_1}_{j_1,h_1}
    V^{\w_4}_{j_4,h_4}
    V^{\w}_{j,h_3-h_2}}\braket{
    V^{\w}_{1-j}(y)
    V^{\w_2}_{j_2,h_2}
    V^{\w_3}_{j_3,h_3}} \nn \\[1ex]
    &\qquad = B(1-j)\int d^2yd^2y' |y^{\alpha-2j} (y-y')^{2j-2}|^2 \braket{V^{\w_1}_{j_1,h_1}
    V^{\w_4}_{j_4,h_4}
    V^{\w}_{j,h_3-h_2}}\braket{
    V^{\w}_{j}(y')
    V^{\w_2}_{j_2,h_2}
    V^{\w_3}_{j_3,h_3}} \nn \\[1ex]
    &\qquad = B(1-j)\int d^2yd^2y' |y^{-\alpha} (1-yy')^{2j-2}|^2 \braket{V^{\w_1}_{j_1,h_1}
    V^{\w_4}_{j_4,h_4}
    V^{\w}_{j,h_3-h_2}}\braket{
    V^{\w}_{j}(y')
    V^{\w_2}_{j_2,h_2}
    V^{\w_3}_{j_3,h_3}} \nn \, ,
\end{align}
where $\alpha = j + \frac{k}{2}\w-h$, and in the last step we have inverted $y \to y^{-1}$. The final expression in \eqref{sergito} reduces to the RHS of Eq.~\eqref{3ptproductId0} by virtue of \eqref{prophomogeneity}. 
Eq.~\eqref{3ptproductId0} can also be expressed as 
\begin{align}
    \int d^2y |y^{2j-2}|^2\braket{V^{\w_1}_{j_1,h_1}
    V^{\w_4}_{j_4,h_4}
    V^{\w}_{1-j}\of{\frac{1}{y}}}&\braket{
    V^{\w}_{j}(y)
    V^{\w_2}_{j_2,h_2}
    V^{\w_3}_{j_3,h_3}}=\label{3ptproductId1}\\
    &\frac{\braket{V^{\w_1}_{j_1,h_1}
    V^{\w_4}_{j_4,h_4}
    V^{\w}_{j,h_3-h_2}}\braket{
    V^{\w}_{j,h_3-h_2}
    V^{\w_2}_{j_2,h_2}
    V^{\w_3}_{j_3,h_3}}}{\braket{
    V^{\w}_{j,h_3-h_2}
    V^{\w}_{j,h_3-h_2}}} \, , \nn
\end{align}
or, equivalently, as
\begin{align}
    \int d^2y |y^{-2j}|^2\braket{V^{\w_1}_{j_1,h_1}
    V^{\w_4}_{j_4,h_4}
    V^{\w}_{j}\of{\frac{1}{y}}}&\braket{
    V^{\w}_{1-j}(y)
    V^{\w_2}_{j_2,h_2}
    V^{\w_3}_{j_3,h_3}}= \label{3ptproductId}\\
    &\frac{\braket{V^{\w_1}_{j_1,h_1}
    V^{\w_4}_{j_4,h_4}
    V^{\w}_{j,h_3-h_2}}\braket{
    V^{\w}_{j,h_3-h_2}
    V^{\w_2}_{j_2,h_2}
    V^{\w_3}_{j_3,h_3}}}{\braket{
    V^{\w}_{j,h_3-h_2}
    V^{\w}_{j,h_3-h_2}}} \, . \nn
\end{align}
These identities are used multiple times along the paper while studying the factorization properties of different examples of spectrally flowed four-point functions.

\bibliographystyle{JHEP}
\bibliography{refs}

\end{document}